\documentclass[12pt,reqno]{amsart}

\usepackage{amsmath}
\usepackage{amsthm}
\usepackage{amssymb}
\usepackage{graphicx}
\usepackage{fullpage}
\usepackage{color}

\newcommand{\N}{\mathbb{N}}
\newcommand{\R}{\mathbb{R}}
\renewcommand{\S}{\mathbb{S}}
\newcommand{\C}{\mathbb{C}}

\newcommand{\B}{\mathbb{B}}
\renewcommand{\H}{\mathbb{H}}
\newcommand{\M}{\mathbb{M}}
\renewcommand{\Re}{\operatorname{Re}}
\renewcommand{\Im}{\operatorname{Im}}
\newcommand{\supp}{\operatorname{supp}}
\newcommand{\dist}{\operatorname{dist}}

\newcommand{\mc}{\mathcal}

\newcommand{\rg}{\operatorname{rg}}

\newcommand{\I}{i}

\newtheorem{lemma}{Lemma}[section]
\newtheorem{proposition}[lemma]{Proposition}
\newtheorem{theorem}[lemma]{Theorem}
\newtheorem{corollary}[lemma]{Corollary}
\newtheorem{hypothesis}[lemma]{Hypothesis}
\theoremstyle{remark}
\newtheorem{remark}[lemma]{Remark}

\theoremstyle{definition}
\newtheorem{definition}[lemma]{Definition}

\numberwithin{equation}{section}
\numberwithin{table}{section}

\title{Existence and stability of Schr\"odinger solitons on noncompact
  manifolds}
\author{David Borthwick}
\address{Department of Mathematics, Emory
  University, Atlanta, GA 30322, USA} 
\email{dborthw@emory.edu}

\author{Roland Donninger}
\thanks{R.D.~is supported by the Austrian Science Fund FWF, Project P 30076.}
\address{Universit\"at Wien, Fakult\"at f\"ur Mathematik,
  Oskar-Morgenstern-Platz 1, 1090 Vienna, Austria}
\email{roland.donninger@univie.ac.at}

\author{Enno Lenzmann}
\address{University of Basel, Department of Mathematics and Computer
  Science, Spiegelgasse 1, CH-4051 Basel, Switzerland}
\email{enno.lenzmann@unibas.ch}

\author{Jeremy L. Marzuola}
\thanks{J.L.M. was supported in part by  NSF Applied Math Grant DMS-1312874 and NSF CAREER Grant DMS-1352353.}
\address{Department of Mathematics,
University of North Carolina, Chapel Hill,
CB \#3250 Phillips Hall,
Chapel Hill, NC 27599,
USA}
\email{marzuola@math.unc.edu}

\begin{document}
\maketitle
\begin{abstract}
We consider the focusing nonlinear Schr\"odinger equation on a large
class of rotationally symmetric, noncompact manifolds. We prove the existence of a
solitary wave by perturbing off the flat Euclidean case. Furthermore,
we study the stability of the solitary wave under radial perturbations
by
analyzing 
spectral properties of the associated
linearized operator.
Finally, in the $L^2$-critical case, by considering the Vakhitov-Kolokolov criterion (see also results of Grillakis-Shatah-Strauss), we provide numerical evidence
showing that the introduction of a nontrivial geometry destabilizes the
solitary wave in a wide variety of cases, regardless of the curvature of the manifold. In particular, the parameters of the metric corresponding to standard hyperbolic space will lead to instability consistent with the blow-up results of Banica-Duyckaerts (2015).  We also provide numerical evidence for geometries under which it would be possible for the Vakhitov-Kolokolov condition to suggest stability, provided certain spectral properties hold in these spaces.  
  \end{abstract}

\section{Introduction}
\noindent The focusing nonlinear Schr\"odinger equation
\begin{equation}
  \label{eq:NLS} i\partial_t u(t,x)+\Delta_x u(t,x)+u(t,x)|u(t,x)|^{p-1}=0,\quad
  p>1,
  \end{equation}
for an unknown $u: \R\times\R^d\to\C$, is a prototypical dispersive
partial differential equation
that arises in various situations in
 physics, e.g., in nonlinear optics or as an effective equation in many
 particle quantum mechanics. We refer the reader to the standard monograph
 \cite{SulSul99} for the general background.
 It is a classical result that in the parameter range
 $1<p<1+\frac{4}{d-2}$ ($d\geq 2$, no upper bound if $d=2$),
 Eq.~\eqref{eq:NLS} possesses \emph{solitary waves} or \emph{solitons},
 i.e., solutions of the form
 \begin{equation}
   \label{eq:flatsoliton}
   u_\alpha^*(t,x)=e^{i\alpha^2 t}Q_{\R^d,\alpha}(x), \quad \alpha>0,
   \end{equation}
 where the \emph{profile function} $Q_{\R^d,\alpha}\in H^1(\R^d)$ is radial,
 smooth, positive, and exponentially decaying, see \cite{Cof72, Str77, ColGlaMar78, Est80,
   GidNiNir81, BerLio83, BerGalKav83}.
Note that $Q_{\R^d,\alpha}$ satisfies the elliptic equation
 \begin{equation}
   \label{eq:Q}
   -\Delta Q_{\R^d,\alpha}+\alpha^2
   Q_{\R^d,\alpha}-Q_{\R^d,\alpha}|Q_{\R^d,\alpha}|^{p-1}=0.
   \end{equation}
The upper bound $p=1+\frac{4}{d-2}$ has an interpretation in terms of
scaling.
Observe that if $u$ is a solution to Eq.~\eqref{eq:NLS}, then so is
the rescaled function
 \[u_\lambda(t,x):=\lambda^{-\frac{2}{p-1}}u(t/\lambda^2,x/\lambda) \] for
 any $\lambda>0$.
 When measured in homogeneous Sobolev spaces, the rescaled solution satisfies
 \[ \|u_\lambda(t,\cdot)\|_{\dot
     H^s(\R^d)}=\lambda^{\frac{d}{2}-s-\frac{2}{p-1}}\|u(t/\lambda^2,\cdot)\|_{\dot
     H^s(\R^d)} \]
 and thus, if $p=1+\frac{4}{d-2}$, the $\dot H^1(\R^d)$-norm is
 invariant under the scaling.
 This is called the \emph{energy-critical case}.
Similarly, $p=1+\frac{4}{d}$ is called the \emph{mass-critical} or
  \emph{$L^2$-critical case} 
as it leaves the $L^2(\R^d)$-norm invariant.
 The scaling symmetry also shows that it is
   enough to consider $\alpha=1$ in Eq.~\eqref{eq:flatsoliton}, and in
   this case, the solution $u_1^*$ is unique \cite{Cof72, McLSer87, Kwo89} and called
 the \emph{ground state}.

The ground state has a variational characterization which is
closely related to stability properties. More precisely, this refers
to the notion of \emph{orbital stability}. Roughly
speaking, $u_1^*$ is orbitally stable if any solution $u$ that starts out
close to $u_1^*$ stays close to $u_1^*$ for all times,
modulo symmetries of the equation. It is known that the ground state
$u_1^*$ is orbitally stable in the $L^2$-subcritical case
$p<1+\frac{4}{d}$
and unstable otherwise \cite{BerCaz81, CazLio82, Wei83, ShaStr85,
  Wei86, GriShaStr87, Gri88, GriShaStr90}.

The stronger notion of \emph{asymptotic stability} of $u_1^*$ refers to the
property that
 all solutions $u$ starting out
sufficiently close to $u_1^*$ converge to $u_1^*$ as $t\to\infty$,
modulo symmetries of the equation. Proving asymptotic stability is
challenging as it presupposes a detailed knowledge of the spectrum of
the nonself-adjoint operator that arises upon linearization of the
equation
at the ground state. Unfortunately, the mathematical understanding of
this operator is still unsatisfactory and one has to rely in part on
numerical evidence. Consequently, asymptotic stability is known only
in special cases or under suitable spectral assumptions, see e.g.~\cite{SofWei90, SofWei92, BusPer95,
  PilWay97, Cuc01, Cuc03, BusSul03, Per04, CucMiz08, Sch09, Bec08,
  Cuc11, Bec12, NakSch12} for an incomplete selection of available results.

\subsection{Main results}
In the present paper we change the geometry and investigate the existence of solitary waves and
their spectral stability for Schr\"odinger equations on manifolds.
More precisely, let $\M^d=(0,\infty)\times_A \S^{d-1}$, $d\geq 2$, be
a warped product manifold
with warping function $A:\R\to\R$ and $\S^{d-1}$ equipped with the
standard round metric, see e.g.~\cite{Pet16}. For the sake of concreteness, we
use the stereographic projection $\psi: \R^{d-1}\to \S^{d-1}$,
\[ \psi(y):=\left (\frac{2y}{|y|^2+1},\frac{|y|^2-1}{|y^2|+1}\right
  ), \]
to parametrize the sphere.
Then we have
\[ \partial_a \psi^j(y)\partial_b
  \psi_j(y)=\frac{4}{(|y|^2+1)^2}\delta_{ab} \]
and the components $g_{jk}$ of the
Riemannian
metric on the warped product $\M^d$ are given by
\[
  g_{jk}(r,y)=\delta_{1j}\delta_{1k}+\frac{4A(r)^2}{(|y|^2+1)^2}\delta_{jk}(1-\delta_{1j}\delta_{1k}) \]
for $j,k\in \{1,2,\dots,d\}$.
We also remark that the sectional curvatures of $\M^d$ are given by
\[ K(\partial_r,\partial_{y^a})(r,y)=-\frac{A''(r)}{A(r)},\qquad
K(\partial_{y^a},
\partial_{y^b})(r,y)=\frac{1-A'(r)^2}{A(r)^2}(1-\delta_{ab}), \]
for $a,b\in \{1,2,\dots,d-1\}$, see \cite{Pet16}.
\begin{hypothesis}
  \label{hyp:A}
  We make the following assumptions on the warping function $A$.
\begin{itemize}
\item $A: \R\to\R$ is smooth and odd with $A'(0)=1$.
 \item $A(r)\gtrsim r$ for all $r>0$.
\item There exists a constant $V_{0,d}\in \R$ such that
  \[ 
      \frac{d-1}{2}\frac{A''(r)}{A(r)}+\frac{(d-1)(d-3)}{4}\left [
    \frac{A'(r)^2}{A(r)^2}-\frac{1}{r^2}\right ]=V_{0,d}[1+O(\langle
  r\rangle^{-2})]  \]
for all $r>0$.
  \end{itemize}
\end{hypothesis}

\begin{remark}
A classical example covered by Hypothesis \ref{hyp:A} is $A(r)=\sinh(r)$ so that $\M^d$ is the
hyperbolic space.  
\end{remark}

As usual, we denote by $(g^{jk})$ the matrix inverse of $(g_{jk})$ and
$\det g$ is the determinant of the latter matrix. Explicitly, we have
\[ \sqrt{\det g(r,y)}=A(r)^{d-1}\left (\frac{2}{|y|^2+1}\right )^{d-1},
\]
and for the inner product $(\cdot|\cdot)_{L^2(\M^d)}$ on $L^2(\M^d)$
we obtain the expression
\[ (f|g)_{L^2(\M^d)}=\int_0^\infty
  \int_{\R^{d-1}}f(r,y)\overline{g(r,y)}A(r)^{d-1}\left
    (\frac{2}{|y|^2+1}\right )^{d-1}dy \>dr. \]
Furthermore, the Laplace-Beltrami
operator $\Delta_{\M^d}$ on $\M^d$ is given by
\[ \Delta_{\M^d}:=\frac{1}{\sqrt{\det g}} \partial_j\left (\sqrt{\det
      g}\,g^{jk}\partial_k\right ), \]
where $\partial_1=\partial_r$ and $\partial_j=\partial_{y^{j-1}}$ for $j=2,3,\dots,d$.
We consider the focusing nonlinear Schr\"odinger equation
\begin{equation}
\label{eq:NLSM}
\I\partial_t u(t,\cdot)+\Delta_{\M^d}u(t,\cdot)+u(t,\cdot)|u(t,\cdot)|^{p-1}=0
\end{equation}
on $\M^d$ for a function $u: \R\times \M^d \to \C$.
Our first result concerns the existence of solitary waves or solitons.

\begin{theorem}[Existence of solitary waves]
  \label{thm:ex}
  Assume Hypothesis \ref{hyp:A} and $1<p<1+\frac{4}{d-2}$ (no upper
  bound in the case $d=2$).
  Then there exists an $\alpha_0>0$ such that
  for any $\alpha\geq\alpha_0$, there exists a real-valued function
  $Q_{\M^d,\alpha}\in C^2(\M^d)$ for which
  $u_\alpha^*: \R\times \M^d\to \C$, given by
\[ u_\alpha^*(t,r,y):=e^{i\alpha^2 t}Q_{\M^d,\alpha}(r,y), \] is a solution
  to Eq.~\eqref{eq:NLSM} for all $t\in \R$. More precisely, we have
  \[ Q_{\M^d,\alpha}(r,y)=\alpha^\frac{2}{p-1}\left
        (\frac{r}{A(r)}\right )^{\!\frac{d-1}{2}} \left [Q_{\R^d,1}(\alpha
        r e_1)+\rho_\alpha(\alpha re_1)\right ], \]
    where $\rho_\alpha \in C^2(\R^d)$ satisfies
    $
      \|\rho_\alpha\|_{H^2(\R^d)}+\|\rho_\alpha\|_{L^\infty(\R^d)}\lesssim
      \alpha^{-1} $
    for all $\alpha\geq \alpha_0$.
    In particular, $Q_{\M^d,\alpha}$ is radial. 
\end{theorem}

\begin{remark}
  The soliton profile on the manifold is a perturbation of
  the Euclidean profile. The heuristic behind this fact is that for large
  $\alpha$ the soliton is supposed to concentrate near the origin and one expects
  the curvature to become negligible. This effect is quantified in
  Theorem \ref{thm:ex}. 
\end{remark}

We continue by investigating the linear stability of the solitary
wave from Theorem \ref{thm:ex}. By plugging the ansatz
$u(t,r,y)=e^{i\alpha^2 t}[Q_{\M^d,\alpha}(r,y)+w(t,r,y)]$ into
Eq.~\eqref{eq:NLSM}, one obtains, upon dropping the nonlinear terms,
the evolution equation
\begin{equation}
\label{eq:tildew}
  \partial_t
  \begin{pmatrix}
    \Re w(t,\cdot) \\ \Im w(t,\cdot) 
  \end{pmatrix}
  =\widetilde{\mc L}_{\M^d,\alpha}  \begin{pmatrix}
    \Re w(t,\cdot) \\ \Im w(t,\cdot) 
  \end{pmatrix},
\end{equation}
with the operator
\[ \widetilde{\mc L}_{\M^d,\alpha}:=
  \begin{pmatrix}
    0 & \widetilde{\mc L}_{\M^d,\alpha,-} \\
    -\widetilde{\mc L}_{\M^d,\alpha,+} & 0
  \end{pmatrix},
  \]
  where
  \begin{align*}
    \widetilde{\mc
    L}_{\M^d,\alpha,-}&:=-\Delta_{\M^d}+\alpha^2-|Q_{\M^d,\alpha}|^{p-1}
    \\
    \widetilde{\mc L}_{\M^d,\alpha,+}&:=-\Delta_{\M^d}+\alpha^2-p|Q_{\M^d,\alpha}|^{p-1}.
  \end{align*}
  Evidently, the linear stability of $u_\alpha^*$ is encoded in the
  spectral properties of (a closed realization of) the operator $\widetilde{\mc L}_{\M^d,\alpha}$. We
  restrict our attention to the radial case and consider $\widetilde{\mc
  L}_{\M^d,\alpha}$ on
  the space $L^2_\mathrm{rad}(\M^d,\C^2)$ with domain
  \[ \mc D(\widetilde{\mc L}_{\M^d,\alpha}):=\{(f_1,f_2)\in
    C^\infty_c(\M^d,\C^2): f_1,f_2
    \mbox{ radial}\}. \]
  Accordingly, we equip the scalar operators $\widetilde{\mc
    L}_{\M^d,\alpha,\pm}$ with the domains
  \[ \mc D(\widetilde{\mc L}_{\M^d,\alpha,\pm}):=\{f\in C^\infty_c(\M^d): f \mbox{
      radial}\}. \]

  \begin{theorem}[Structure of the spectrum of the linearized operator]
    \label{thm:spec}
    Assume Hypothesis \ref{hyp:A} and $1<p<1+\frac{4}{d-2}$ (no upper
    bound in the case $d=2$).
    There exists an $\alpha_0>0$ such that for any $\alpha\geq
    \alpha_0$, the operator $\widetilde{\mc L}_{\M^d,\alpha}:
   \mc D(\widetilde{\mc L}_{\M^d,\alpha})\subset L^2_\mathrm{rad}(\M^d,\C^2)\to
    L^2_\mathrm{rad}(\M^d,\C^2)$ is closable.  Its closure $\mc L_{\M^d,\alpha}$
has the following properties:
    \begin{itemize}
    \item The spectrum of $\mc L_{\M^d,\alpha}$ is a subset of $\R\cup
      i\R$.
      \item If $\lambda\in \sigma(\mc L_{\M^d,\alpha})$ then
        $-\lambda\in \sigma(\mc L_{\M^d,\alpha})$.
        \item The essential spectrum\footnote{There are various (in
            general inequivalent) definitions of the essential
            spectrum of a closed operator. For us, the essential
            spectrum is the part of the spectrum that is invariant
            under relatively compact perturbations, see Definition
            \ref{def:essspec} and Remark \ref{rem:essspec} below.}  of $\mc L_{\M^d,\alpha}$ is given
          by
          \[ \sigma_e(\mc L_{\M^d,\alpha})=\{z\in \C: \Re z=0, |\Im
            z|\geq \alpha^2+V_{0,d}\} .\]
\item The set $\sigma(\mc L_{\M^d,\alpha})\setminus \sigma_e(\mc
  L_{\M^d,\alpha})$ is free of accumulation points and consists of
  eigenvalues with finite algebraic multiplicities.
\item We have $0\in \sigma_p(\mc L_{\M^d,\alpha})$ and
  \[ \ker\mc L_{\M^d,\alpha}=\left \langle \begin{pmatrix}0 \\
        Q_{\M^d,\alpha}\end{pmatrix}\right \rangle. \]
    \end{itemize}
  \end{theorem}

For $p\not= 1+\frac{4}{d}$ we obtain a very clear picture concerning the
linear stability which is analogous to the Euclidean situation.
  
\begin{theorem}[Spectral stability in the noncritical case]
  \label{thm:stabnoncrit}
    Assume Hypothesis \ref{hyp:A} and $1<p<1+\frac{4}{d-2}$ (no upper
    bound in the case $d=2$). Then there exists an $\alpha_0>0$ such
    that for all $\alpha\geq \alpha_0$ the following holds.
    \begin{itemize}
    \item If $p\not= 1+\frac{4}{d}$, the algebraic multiplicity of
      $0\in \sigma_p(\mc L_{\M^d,\alpha})$ equals $2$.
      \item If $p<1+\frac{4}{d}$, there are no positive eigenvalues of
        $\mc L_{\M^d,\alpha}$.
        \item If $p>1+\frac{4}{d}$, there exists precisely one
          positive eigenvalue $\lambda_\alpha\in \sigma_p(\mc
          L_{\M^d,\alpha})$ and the eigenvalues $\pm \lambda_\alpha\in
          \sigma_p(\mc L_{\M^d,\alpha})$
          are simple.
    \end{itemize}
  \end{theorem}

  In the critical case $p=1+\frac{4}{d}$, the stability of the
  solitary wave is more involved and depends on finer properties of the
  underlying geometry.
The corresponding condition is formulated in terms of $\mc
L_{\M^d,\alpha,+}^{-1}$, where $\mc L_{\M^d,\alpha,+}$ is the closure
of $\widetilde{\mc L}_{\M^d,\alpha,+}$.

  \begin{theorem}[Spectral stability in the critical case]
    \label{thm:stabcrit}
    Assume Hypothesis \ref{hyp:A} and $1<p<1+\frac{4}{d-2}$ (no upper
    bound in the case $d=2$). Then
    there exists an $\alpha_0>0$ such that for all $\alpha\geq
    \alpha_0$ the
    operator $\widetilde{\mc L}_{\M^d,\alpha,+}: \mc D(\widetilde {\mc L}_{\M^d,\alpha,+})\subset
    L^2_\mathrm{rad}(\M^d)
    \to L^2_\mathrm{rad}(\M^d)$ is essentially self-adjoint and its
    closure $\mc L_{\M^d,\alpha,+}$ is bounded
    invertible. If $p=1+\frac{4}{d}$
    then for all $\alpha\geq \alpha_0$ the following holds:

    \begin{itemize}
    \item If $(\mc L_{\M^d,\alpha,+}^{-1}
      Q_{\M^d,\alpha}|Q_{\M^d,\alpha})_{L^2(\M^d)}>0$ then $\mc
      L_{\M^d,\alpha}$ has precisely one positive eigenvalue
      $\lambda_\alpha$ and the eigenvalues $\pm\lambda_\alpha\in
      \sigma_p(\mc L_{\M^d,\alpha})$ are simple.

      \item If $(\mc L_{\M^d,\alpha,+}^{-1}
        Q_{\M^d,\alpha}|Q_{\M^d,\alpha})_{L^2(\M^d)}\leq 0$ then $\mc
        L_{\M^d,\alpha}$ has no positive eigenvalues.
    \end{itemize}
  \end{theorem}

  \begin{remark}
    \label{rem:stabcrit}
    If $\partial_\alpha Q_{\M^d,\alpha}$ is sufficiently smooth and
    belongs to the domain of $\mc L_{\M^d,\alpha,+}$,
    the (in)stability condition in Theorem \ref{thm:stabcrit} can be
    simplified. Indeed, by differentiating
\[ -\Delta_{\M^d} Q_{\M^d,\alpha}+\alpha^2
  Q_{\M^d,\alpha}-Q_{\M^d,\alpha}|Q_{\M^d,\alpha}|^{p-1}=0 \]
with respect to $\alpha$ we find
\[ \mc L_{\M^d,\alpha,+}\partial_\alpha Q_{\M^d,\alpha}=-2\alpha
  Q_{\M^d,\alpha} \]
and thus,
\[ \big (\mc L_{\M^d,\alpha,+}^{-1}Q_{\M^d,\alpha}\big |
  Q_{\M^d,\alpha}\big )_{L^2(\M^d)}=-\tfrac{1}{2\alpha} \big (\partial_\alpha
  Q_{\M^d,\alpha}\big | Q_{\M^d,\alpha}\big )_{L^2(\M^d)}
=-\tfrac{1}{4\alpha}\partial_\alpha\|Q_{\M^d,\alpha}\|_{L^2(\M^d)}^2. \]
  \end{remark}

  \begin{remark}
\label{rem:stabilize}
  Theorem \ref{thm:stabcrit} raises the intriguing question of whether it is
  possible to ``stabilize'' the borderline unstable soliton in
  Euclidean space by changing the background geometry.
 Unfortunately, we cannot
  answer this question in the affirmative as we are unable to provide
  a \emph{sufficient} criterion for stability in the critical
  case. This appears to be challenging, as it requires a
  good understanding of eigenvalues and resonances on the imaginary
  axis, a question which is still largely open even in the purely Euclidean
  setting. However, Theorem \ref{thm:stabcrit}
  provides a sufficient criterion for (linear)
  \emph{in}stability.
Using this, we provide numerical evidence that there exists a large
class of negatively curved manifolds
such that the soliton becomes (linearly) unstable, see Section \ref{sec:stabcurv}.
This fits well with the blow-up instability for the $L^2$ critical
(and super-critical) nonlinear Schr\"odinger equation on the
hyperbolic space $\H^d$ computed via virial identities in
\cite{BanDuy15}.
Blow-up was also established in \cite{boulenger2012blow} in the $L^2$
critical setting with a Riemannian manifold that is locally like
$\H^d$ and asymptotically like $\R^d$.
\end{remark}

\subsection{Further related results}
Unfortunately, there is still no general satisfactory understanding of the
linearized operator even in the Euclidean
case, and as a consequence, this classical problem remains a topic of
contemporary research.
For instance, see \cite{CucPelVou05} for an analysis of embedded eigenvalues in the
essential spectrum and
\cite{ChaGusNakTsa07} for a modern account of the general theory and new numerical results.
Furthermore, decay properties of eigenfunctions are investigated in
\cite{HunLee07}. The paper \cite{Sch09} is concerned with asymptotic
stability but also contains a thorough analysis of the linearized
operator. In \cite{CosHuaSch12}, a novel computer-assisted method is
introduced to prove the absence of eigenvalues in the essential
spectral gap.  In addition, in \cite{marzuola2010spectral}, the
authors give a numerically assisted proof for the absence of embedded
eigenvalues in a variety of settings on $\R^d$.  In the case of
potential perturbations, stability analysis in both the small and
large mass limits have been studied through both dispersive techniques
as well as bifurcation theory on $\R^d$ for a range of nonlinearities
in many works, for a small sampling see
e.g. \cite{tsai2002asymptotic,cuccagna2003asymptotic,gustafson2004asymptotic,kirr2008symmetry,kirr2011symmetry,marzuola2017nonlinear,nakanishi2017global,nakanishi2017global1}
and the references captured within.

Needless to say, the literature on Schr\"odinger equations on
manifolds is vast and we just mention some closely related recent works.
There is a number of papers devoted to the study of the focusing nonlinear
Schr\"odinger equation on hyperbolic space. A recurring theme, compare
Remark \ref{rem:stabilize}, is the question whether the negative curvature may
improve the situation compared to the Euclidean case and
stabilize the evolution, see, e.g.,~\cite{Ban07, AnkPie09,
   BanDuy15}. Spectral properties in hyperbolic space are studied in
 \cite{BorMar15} and
the existence of ground states on noncompact manifolds is investigated
in, e.g.,~\cite{ChrMar10, TerTzvVis14, ChrMarMetTay14}.   See also the recent works \cite{chen2014resolvent,chen2016resolvent,chen2018resolvent} for advances on the spectral measure for asymptotically hyperbolic manifolds, the analysis of which is required for good dispersive estimates that can lead to results on asymptotic stability when understood with perturbations and for the linearized operator.   The literature on spectral measures for the asymptotically Euclidean and conic cases is quite vast, but see \cite{marzuola2007strichartz,bouclet2011low,hassell2016global} and references therein.

\section{Preliminary transformations}

\noindent We proceed by transforming the radial case of Eq.~\eqref{eq:NLSM} 
to a standard nonlinear Schr\"odinger equation on $\R^d$ with a
potential.
This is a well-known reduction, see e.g.~\cite{BanDuy07, ChrMar10}.

\subsection{The Laplace-Beltrami operator}

The Laplace-Beltrami operator $\Delta_{\M^d}$ is given by
\[ \Delta_{\M^d}f=\frac{1}{\sqrt{\det g}}\partial_j\left (\sqrt{\det g}\,g^{jk}\partial_k f \right ). \]
We now assume that $f(r,y)=f(r)$, i.e., we restrict ourselves to the
radial case. Then,
\begin{align*} \Delta_{\M^d}f(r)&=\frac{1}{\sqrt{\det g(r,y)}}\partial_r \left (\sqrt{\det g(r,y)}\,\partial_r f(r) \right ) 
=\frac{1}{A(r)^{d-1}}\partial_r \left [A(r)^{d-1}\partial_r f(r)\right ] \\
                                &=\left
                                  [\partial_r^2+(d-1)\frac{A'(r)}{A(r)}\partial_r
                                  \right ]f(r) \\
                                  &=:\Delta_{\M^d}^\mathrm{rad}f(r).
\end{align*}
Obviously, $\Delta_{\M^d}^\mathrm{rad}$ is formally self-adjoint on
$L^2_{A^{d-1}}(0,\infty)$.
Eq.~\eqref{eq:NLSM} for radial functions  reduces to
\begin{equation}
  \label{eq:NLSMchart}
  i\partial_t \widetilde
  u(t,\cdot)+\Delta_{\M^d}^\mathrm{rad}\widetilde
  u(t,\cdot)+\widetilde u(t,\cdot)|\widetilde u(t,\cdot)|^{p-1}=0
\end{equation}
for $\widetilde u: \R\times (0,\infty)\to\C$.

\subsection{Conjugation to Euclidean} 
In order to perturb off the Euclidean case,
we would like to compare the Laplace-Beltrami operator $\Delta_{\M^d}$ to the ordinary Laplace operator on $\R^d$, henceforth denoted by 
$\Delta_{\R^d}$.  The restriction of the Euclidean operator to radial functions yields the operator 
\[ \Delta_{\R^d}^\mathrm{rad}:=\partial_r^2+\frac{d-1}{r}\partial_r, \]
acting on $L^2_{|\cdot|^{d-1}}(0,\infty)$.
To compare the two operators, we need to conjugate by the unitary map that relates
the radial function spaces.

Let
$\mc U_d: L^2_{|\cdot|^{d-1}}(0,\infty)\to L^2_{A^{d-1}}(0,\infty)$ be defined by
\[ \mc U_d f(r):=\left (\frac{r}{A(r)}\right )^{\frac{d-1}{2}}f(r), \]
so that $\|\mc U_d f\|_{L^2_{A^{d-1}}(0,\infty)}=\|f\|_{L^2_{|\cdot|^{d-1}}(0,\infty)}$.
Now we consider the operator $\mc
U_d^{-1}\Delta_{\M^d}^\mathrm{rad}\mc U_d$ on $L^2_{|\cdot|^{d-1}}(0,\infty)$.
Explicitly, we have
\[ \mc U_d^{-1}\Delta_{\M^d}^\mathrm{rad}\mc U_df(r)=\left ( \Delta_{\R^d}^\mathrm{rad}
-\frac{d-1}{2}\frac{A''(r)}{A(r)}-\frac{(d-1)(d-3)}{4}\frac{A'(r)^2}{A(r)^2}
+\frac{(d-1)(d-3)}{4r^2}\right )f(r),
\]
By setting
\[ \widetilde u(t,r)=\mc U_d\left (\widetilde v(t,\cdot)\right )(r)=\left
    (\frac{r}{A(r)}\right )^\frac{d-1}{2}\widetilde v(t,r), \] for a function
$\widetilde v: \R\times
(0,\infty)\to \C$, Eq.~\eqref{eq:NLSMchart} can now be written as
\[ \I \partial_t \widetilde
  v(t,\cdot)+\mc U_d^{-1}\Delta_{\M^d}^\mathrm{rad}\mc U_d \widetilde
  v(t,\cdot)+\widetilde v(t,\cdot)|\mc U_d \widetilde
  v(t,\cdot)|^{p-1}=0. \]
In fact, we find it more convenient to formulate this equation in terms
of the auxiliary function $v: \R\times \R^d\to\C$, given by $v(t,x):=\widetilde
v(t,|x|)$. This yields
\begin{equation}
  \label{eq:NLSv}
  i\partial_t v(t,\cdot)+\Delta_{\R^d}v(t,\cdot)-V_d
  v(t,\cdot)+\varphi_{d,p}v(t,\cdot)|v(t,\cdot)|^{p-1}=0,
  \end{equation}
with
\[ \varphi_{d,p}(x):=\left (\frac{|x|}{A(|x|)}\right
  )^{\frac{(d-1)(p-1)}{2}} \]
and
\[ V_d(x):=\frac{d-1}{2}\frac{A''(|x|)}{A(|x|)}+\frac{(d-1)(d-3)}{4}\frac{A'(|x|)^2}{A(|x|)^2}
-\frac{(d-1)(d-3)}{4|x|^2}. \]
We keep in mind that $v(t,\cdot)$ is radial.
Note that Eq.~\eqref{eq:NLSv} resembles a standard nonlinear
Schr\"odinger equation on Euclidean
space with a potential $V_d$.

To look for solitons, we plug the ansatz $v(t,x)=e^{\I\alpha^2
  t}R_\alpha(x)$ into Eq.~\eqref{eq:NLSv} with $R_\alpha$ radial.
This yields the elliptic equation
\begin{equation}
\label{eq:R} \Delta_{\R^d}R_\alpha-\alpha^2
R_\alpha-V_d R_\alpha+\varphi_{d,p}R_\alpha|R_\alpha|^{p-1}=0.
\end{equation}
In terms of the rescaled profile $\widetilde R_\alpha$, defined by $R_\alpha(x)=\alpha^{\frac{2}{p-1}}\widetilde R_\alpha(\alpha x)$, Eq.~\eqref{eq:R} reads
\begin{equation}
\label{eq:Ralpha} \Delta_{\R^d}\widetilde R_\alpha(x) -\widetilde
R_\alpha(x)
-\alpha^{-2}V_d(\alpha^{-1}x) \widetilde R_\alpha(x)
+\varphi_{d,p}(\alpha^{-1}x)F_p\left (\widetilde R_\alpha(x)\right )=0, 
\end{equation}
where $F_p(s):=s|s|^{p-1}$.
We intend to solve Eq.~\eqref{eq:Ralpha} by perturbing off the
Euclidean situation and hence
insert the ansatz $\widetilde R_\alpha(x)=Q_{\R^d}(x)+\rho(x)$ into
Eq.~\eqref{eq:Ralpha}, where $Q_{\R^d}:=Q_{\R^d,1}$. In view of Eq.~\eqref{eq:Q}, we obtain
\begin{equation}
  \label{eq:rho}
  \begin{split}
-\mc A_\alpha\rho(x)=&
q_\alpha(x)F_p'(Q_{\R^d}(x))\rho(x)+[q_\alpha(x)-1]\mc N(\rho)(x) \\
&+\alpha^{-2} V_d(\alpha^{-1}x)Q_{\R^d}(x)+q_\alpha(x)F_p(Q_{\R^d}(x)),
\end{split}
\end{equation}
where
\begin{align*}
\mc A_\alpha \rho(x)&:=-\Delta_{\R^d}\rho(x)+\rho(x)-
F_p'(Q_{\R^d}(x))\rho(x)+\alpha^{-2} V_d(\alpha^{-1}x)\rho(x), \\
q_\alpha(x)&:=1-\varphi_{d,p}(\alpha^{-1}x), \\
\mc N(\rho)(x)&:=F_p(Q_{\R^d}(x)+\rho(x))-F_p(Q_{\R^d}(x))-F_p'(Q_{\R^d}(x))\rho(x).
\end{align*}

\section{Existence of a soliton}

\noindent In this section we show that Eq.~\eqref{eq:rho} has a
solution $\rho=\rho_\alpha$, provided $\alpha\geq 1$ is sufficiently
large. This way, we obtain a soliton solution
\[ v_\alpha^*(t,x):=\alpha^{\frac{2}{p-1}}e^{i\alpha^2
    t}\left [Q_{\R^d}(\alpha x)+\rho_\alpha(\alpha x)\right ] \]
to Eq.~\eqref{eq:NLSv}.

\subsection{Analysis of the linear operator}
Initially, we define the operator $\mc A_\alpha$ 
as a classical differential operator acting on $C^\infty_c(\R^d)$.
Recall that $Q_{\R^d}\in C^\infty(\R^d)$, $Q_{\R^d}>0$, and $V_d\in
C^\infty(\R^d)$ by Hypothesis \ref{hyp:A}.
As a consequence, $\mc A_\alpha$ is a continuous map from $\mc D(\R^d)$ to $\mc D(\R^d)$.
Furthermore, $\mc A_\alpha$ is formally self-adjoint on $L^2(\R^d)$
and thus,
$\mc A_\alpha$ extends to $\mc D'(\R^d)$ by
\[ \mc A_\alpha u(\varphi):=u(\mc A_\alpha \varphi) \]
for $u\in \mc D'(\R^d)$ and $\varphi\in \mc D(\R^d)$.
In the limit $\alpha\to\infty$, $\mc A_\alpha$ formally
reduces to $\mc L_+$, given by
\[ \mc L_+
  f(x)=-\Delta_{\R^d}f(x)+f(x)-F_p'(Q_{\R^d}(x))f(x). \]
This is a well-known operator in the Euclidean setting that occurs in
the linearization about solitary waves.

Note that both $\mc A_\alpha$ and $\mc L_+$ map radial distributions
to radial distributions since $Q_{\R^d}$ and $V_d$ are radial. 
Consequently, $\mc A_\alpha$ and $\mc L_+$ may
be viewed as unbounded operators on $L^2_\mathrm{rad}(\R^d)$.

\begin{lemma}
  \label{lem:basicLinf}
  The operator $\mc L_+: H^2_\mathrm{rad}(\R^d)\subset
  L^2_\mathrm{rad}(\R^d)\to L^2_\mathrm{rad}(\R^d)$ is self-adjoint. Furthermore, $\mc L_+$ is
  invertible and we have the smoothing estimate
  \[ \|\mc L_+^{-1} g\|_{H^2(\R^d)}\lesssim \|g\|_{L^2(\R^d)} \]
  for all $g\in L^2_\mathrm{rad}(\R^d)$.
\end{lemma}

\begin{proof}
  By the exponential decay of $Q_{\R^d}$ and \cite{Tes14}, p.~258, Theorem 10.2, we see that
$\mc L_+$ is self-adjoint with domain
$H^2_\mathrm{rad}(\R^d)$ and essential spectrum $\sigma_e(\mc
L_+)=[1,\infty)$. Consequently, $0\notin \sigma(\mc L_+)$
follows from \cite{ChaGusNakTsa07}, Lemma 2.1. Thus, it remains to prove the smoothing
estimate.
To this end, let $\mc L_0: H^2_{\mathrm{rad}}(\R^d)\subset L^2_{\mathrm{rad}}(\R^d)\to
L^2_\mathrm{rad}(\R^d)$ be given by $\mc L_0 f=-\Delta_{\R^d} f+f$.
For $f\in \mc S(\R^d)$ we have
\[ \mc F\mc L_0 f(\xi)=\mc F(-\Delta_{\R^d}f+f)(\xi)=(4\pi^2|\xi|^2+1)\mc
  Ff(\xi), \]
where $ \mc F$ denotes the Fourier transform
\[ \mc F f(\xi):=\int_{\R^d}e^{-2\pi i \xi x}f(x)dx. \]
Thus, on the Fourier side, the equation $\mc L_0 f=g$ reads
\[ (1+4\pi^2|\xi|^2)\mc Ff(\xi)=\mc Fg(\xi). \]
Consequently, by Plancherel,
\begin{align*} \|\mc
  L_0^{-1}g\|_{H^2(\R^d)}
  &=\|f\|_{H^2(\R^d)}\simeq\|\langle\cdot\rangle^2
  \mc F
  f\|_{L^2(\R^d)}=\left
    \|\langle\cdot\rangle^2(1+4\pi^2|\cdot|^2)^{-1}\mc Fg\right
  \|_{L^2(\R^d)}
  \lesssim \|\mc Fg\|_{L^2(\R^d)} \\
  &\simeq \|g\|_{L^2(\R^d)}
\end{align*}
for $g\in \mc S(\R^d)$. By approximation, this bound holds for all
$g\in L^2_\mathrm{rad}(\R^d)$.
Let $\mc B:
L^2_{\mathrm{rad}}(\R^d)\to L^2_{\mathrm{rad}}(\R^d)$ be given by $\mc
B f(x)=-F_p'(Q_{\R^d}(x))f(x)$.
By definition, we have the identity
\[ \mc L_+=(1+\mc B\mc L_0^{-1})\mc L_0, \]
and this shows that $\mc L_+ \mc L_0^{-1}$ is a bounded operator
on $L^2_{\mathrm{rad}}(\R^d)$. By the open mapping theorem, its
inverse $\mc L_0\mc L_+^{-1}$ is also bounded.
Consequently, the smoothing property of $\mc L_0^{-1}$ implies the bound
\[ \|\mc L_+^{-1}g\|_{H^2(\R^d)}=\|\mc L_0^{-1}\mc L_0\mc
  L_+^{-1} g\|_{H^2(\R^d)}\lesssim \|\mc L_0\mc
  L_+^{-1}g\|_{L^2(\R^d)}\lesssim \|g\|_{L^2(\R^d)} \]
for all $g\in L^2_\mathrm{rad}(\R^d)$.
\end{proof}

\begin{lemma}
\label{lem:Lalpha}
There exists an $\alpha_0>0$ such that,
for any $\alpha\geq \alpha_0$, the operator $\mc A_\alpha:
H^2_{\mathrm{rad}}(\R^d)\subset L^2_\mathrm{rad}(\R^d)\to
L^2_\mathrm{rad}(\R^d)$ is self-adjoint and invertible. 
Furthermore, we have the smoothing estimate
\[ \|\mc A_\alpha^{-1}f\|_{H^2(\R^d)}\lesssim \|f\|_{L^2(\R^d)} \]
for all $f\in L^2_\mathrm{rad}(\R^d)$ and all $\alpha \geq \alpha_0$.
\end{lemma}

\begin{proof}
For any $\alpha>0$ we define a bounded operator $\mc B_\alpha$ on
$L^2_\mathrm{rad}(\R^d)$ by setting
\[ \mathcal B_\alpha f(x):=-V_d(\alpha^{-1}x)f(x). \]
Since $V_d\in L^\infty(\R^d)$ by Hypothesis \ref{hyp:A}, we infer $\|\mc B_\alpha f\|_{L^2(\R^d)}\lesssim \|f\|_{L^2(\R^d)}$ for all $\alpha>0$.
Consequently, a Neumann series argument shows the existence of the operator
$(1-\alpha^{-2}\mathcal B_\alpha \mc L_+^{-1})^{-1}$ with the bound
\[ \|(1-\alpha^{-2}\mathcal B_\alpha \mc L_+^{-1})^{-1}f\|_{L^2(\R^d)}\lesssim 
\|f\|_{L^2(\R^d)} \]
for all $\alpha\geq \alpha_0$, provided $\alpha_0>0$ is sufficiently
large. 
Thus, from the identity $\mc A_\alpha=(1-\alpha^{-2}\mc B_\alpha\mc L_+^{-1})\mc L_+$, we obtain
the existence of the operator
\[ \mc A_\alpha^{-1}=\mc L_+^{-1}(1-\alpha^{-2}\mathcal B_\alpha \mc L_+^{-1})^{-1}, \]
with the bound
\begin{align*}
 \|\mc A_\alpha^{-1}f\|_{H^2(\R^d)}&\lesssim 
 \|\mc L_+^{-1}(1-\alpha^{-2}\mathcal B_\alpha \mc L_+^{-1})^{-1}f\|_{H^2(\R^d)}
 \lesssim \|(1-\alpha^{-2}\mathcal B_\alpha \mc L_+^{-1})^{-1}f\|_{L^2(\R^d)} \\
 &\lesssim \|f\|_{L^2(\R^d)}
 \end{align*}
 for all $f\in L^2_\mathrm{rad}(\R^d)$ and $\alpha\geq \alpha_0$.
\end{proof}

As a consequence of Lemma \ref{lem:Lalpha}, we can now reformulate
Eq.~\eqref{eq:rho} as the fixed point problem
\begin{equation}
  \label{eq:rhofix}
  \rho=-\mc A_\alpha^{-1}\left [
    q_\alpha F_p'(Q_{\R^d}(\cdot))\rho+(q_\alpha-1)\mc N(\rho) 
+\alpha^{-2} V_d(\alpha^{-1}(\cdot))Q_{\R^d}+q_\alpha F_p(Q_{\R^d}(\cdot))
    \right ].
\end{equation}

\subsection{Refined bounds for $\mc A_\alpha^{-1}$}

Next, we prove an $L^\infty$ bound for $\mc A_\alpha^{-1}$, again by first proving the corresponding result for 
$\mc L_+^{-1}$.
  
\begin{lemma}
  \label{lem:Linfsup}
We have the bound
\[ \|\mc L_+^{-1}g\|_{L^\infty(\R^d)}\lesssim
  \|g\|_{L^2(\R^d)}+\|g\|_{L^\infty(\R^d)} \]
for all $g\in L^2_\mathrm{rad}(\R^d)\cap L^\infty(\R^d)\cap
C(\R^d)$.
\end{lemma}

\begin{proof}
  By Sobolev embedding and Lemma \ref{lem:basicLinf} the result is
  immediate in the case $d=2$, i.e.,
  \[ \|\mc L_+^{-1}g\|_{L^\infty(\R^2)}\lesssim \|\mc
    L_+^{-1}g\|_{H^2(\R^2)}\lesssim \|g\|_{L^2(\R^2)}. \]
  Thus, we may restrict ourselves to $d\geq 3$.
  Since all functions are radial, problems occur only at the origin.
  Indeed, 
  by the one-dimensional Sobolev embedding and Lemma \ref{lem:basicLinf},
  we have
  \[ \|\mc L_+^{-1}g\|_{L^\infty(\R^d\setminus \B^d)}\lesssim
    \|\mc L_+^{-1}g\|_{H^1(\R^d)}\lesssim \|g\|_{L^2(\R^d)} \]
  for all $g\in L^2_\mathrm{rad}(\R^d)$.
Consequently, it
suffices to prove the estimate
\[ \|\mc L_+^{-1}g\|_{L^\infty(\B^d)}\lesssim
  \|g\|_{L^2(\R^d)}+\|g\|_{L^\infty(\R^d)} \] for all $g\in
  L_\mathrm{rad}^2(\R^d)\cap L^\infty(\R^d)\cap C(\R^d)$.

  Let $f=\mc L_+^{-1}g$. Then $f\in H^2_\mathrm{rad}(\R^d)$, and by the
  radial Sobolev embedding we infer that $f\in
  C(\R^d\setminus\{0\})\cap L^1_\mathrm{loc}(\R^d)$.  
The equation $\mc L_+ f=g$ implies
  $\Delta_{\R^d} f^\sharp=h^\sharp$ in $\mc D'(\R^d)$ with
  \[ h(x):=-g(x)+f(x)-F_p'(Q_{\R^d}(x))f(x), \]
see Definition \ref{def:sharp} for the
notation.
Evidently, $h\in C(\R^d\setminus\{0\})\cap L^1_\mathrm{loc}(\R^d)$ and thus,
 Lemma \ref{lem:raddist} shows that the function
  $\widehat f(r):=r^\frac{d-1}{2}f(re_1)$ belongs to $C^2(0,\infty)$
  and satisfies
  \begin{equation}
    \label{eq:Linfrad}
    \widehat f''(r)-\frac{(d-1)(d-3)}{4r^2}\widehat f(r)-\widehat
  f(r)+F_p'(Q_{\R^d}(re_1))\widehat
  f(r)=-r^\frac{d-1}{2}g(re_1)
  \end{equation}
  for all $r>0$.
Now we consider the homogeneous version of Eq.~\eqref{eq:Linfrad},
i.e.,
\begin{equation}
  \label{eq:Linfradhom}\phi''(r)-\frac{(d-1)(d-3)}{4r^2}\phi(r)-\phi(r)
  +F_p'(Q_{\R^d}(re_1))\phi(r)=0.
\end{equation}
Eq.~\eqref{eq:Linfradhom} has a fundamental system
$\{\phi_0,\phi_\infty\}$ with the asymptotic behavior
\begin{align*}
  |\phi_0(r)|
  &\simeq r^\frac{d-1}{2}
  \mbox{ for } r\in [0,1],
  &
|\phi_0(r)|&\simeq e^r
  \mbox{ for }
             r\geq 1, \\
  |\phi_0'(r)|&\simeq r^\frac{d-3}{2} \mbox{ for }r\in [0,1],
  &
|\phi_0'(r)|&\simeq e^r \mbox{ for }r\geq 1, \\
  |\phi_\infty(r)|&\simeq r^{-\frac{d-3}{2}} \mbox{ for }r\in (0,1],
  &
    |\phi_\infty(r)|&\simeq e^{-r} \mbox{ for }r\geq 1, \\
  |\phi_\infty'(r)|&\simeq r^{-\frac{d-1}{2}}\mbox{ for }r\in (0,1],
& |\phi_\infty'(r)|&\simeq e^{-r} \mbox{ for }r\geq 1,                    
\end{align*}
and we may normalize so that $W(\phi_0,\phi_\infty)=1$, see Lemma
\ref{lem:phi0phiinf} below.
Consequently, the variation of
constants formula yields the existence of constants $a,b\in \C$ such
that
\[ \widehat f(r)=a\phi_0(r)+b\phi_\infty(r)-\phi_0(r)\int_r^\infty
  \phi_\infty(s)s^\frac{d-1}{2}g(se_1)ds
  -\phi_\infty(r)\int_0^r \phi_0(s)s^\frac{d-1}{2}g(se_1)ds, \]
and $f\in H_\mathrm{rad}^1(\R^d)$ implies that $a=b=0$.
Furthermore, we have the bounds
\begin{align*}
  |\phi_0(r)|\int_r^\infty
  \left |\phi_\infty(s)s^\frac{d-1}{2}g(se_1)\right |ds
  &\lesssim
    r^\frac{d-1}{2}\|g\|_{L^\infty(\R^d)}\left [\int_0^1
    s^{-\frac{d-3}{2}}s^\frac{d-1}{2}ds+\int_1^\infty
    e^{-s}s^{\frac{d-1}{2}}ds \right ] \\
  &\lesssim r^\frac{d-1}{2}\|g\|_{L^\infty(\R^d)}
\end{align*}
and
\begin{align*}
  |\phi_\infty(r)|\int_0^r \left
  |\phi_0(s)s^\frac{d-1}{2}g(se_1)\right |ds
  &\lesssim
  r^{-\frac{d-3}{2}}\|g\|_{L^\infty(\R^d)}\int_0^r s^{d-1}ds \\
  &\lesssim r^{\frac{d+3}{2}}\|g\|_{L^\infty(\R^d)}
\end{align*}
for all $r\in (0,1]$.
Consequently,
\[ |f(re_1)|=\left |r^{-\frac{d-1}{2}}\widehat f(r)\right |\lesssim
  \|g\|_{L^\infty(\R^d)} \]
for all $r\in (0,1]$, which 
implies the desired bound
\[ \|\mc
L_+^{-1}g\|_{L^\infty(\B^d)}=\|f\|_{L^\infty(\B^d)}\lesssim
\|g\|_{L^\infty(\R^d)}. \]
\end{proof}

By a simple perturbative argument, we obtain an analogous $L^\infty$
bound for the operator
$\mc A_\alpha^{-1}$.

\begin{corollary}
  \label{cor:Lalphasup}
  There exists an $\alpha_0>0$ such that
  \[ \|\mc A_\alpha^{-1}g\|_{L^\infty(\R^d)}\lesssim
    \|g\|_{L^2(\R^d)}+\|g\|_{L^\infty(\R^d)} \]
  for all $g\in L^2_\mathrm{rad}(\R^d)\cap L^\infty(\R^d)\cap C(\R^d)$ and $\alpha\geq\alpha_0$.
\end{corollary}

\begin{proof}
  Let $X:=L^2_\mathrm{rad}(\R^d)\cap L^\infty(\R^d)\cap C(\R^d)$ and write
  \[ \|g\|_X:=\|g\|_{L^2(\R^d)}+\|g\|_{L^\infty(\R^d)}. \]
  As in the proof of Lemma \ref{lem:Lalpha}, we set
  \[ \mc B_\alpha g(x):=-V_d(\alpha^{-1}x)g(x). \]
Note that $\|\mc B_\alpha g\|_X \lesssim
  \|g\|_X$ for all $g\in X$ and $\alpha>0$
  by Hypothesis \ref{hyp:A}. Consequently, the operator
  $(1-\alpha^{-2}\mc B_\alpha): X\to X$ is bounded invertible on $X$ for any
  $\alpha\geq\alpha_0$ by a Neumann series argument, provided
  $\alpha_0$ is sufficiently large. Furthermore,
  \[ \|(1-\alpha^{-2}\mc B_\alpha)^{-1}g\|_X\lesssim \|g\|_X \]
  for all $g\in X$ and $\alpha\geq\alpha_0$. By Lemmas
  \ref{lem:basicLinf} and \ref{lem:Linfsup}, we have $\|\mc
  L_+^{-1}g\|_X\lesssim \|g\|_X$ for all $g\in X$, and thus,
  \begin{align*} \|\mc A_\alpha^{-1}g\|_{L^\infty(\R^d)}
    &\leq \|\mc A_\alpha^{-1}g\|_X=\|\mc L_+^{-1}(1-\alpha^{-2}\mc
    B_\alpha\mc L_+^{-1})^{-1}g\|_X
  \lesssim  \|(1-\alpha^{-1}\mc
      B_\alpha\mc L_+^{-1})^{-1}g\|_X\lesssim \|g\|_X  \\
    &\lesssim \|g\|_{L^2(\R^d)}+\|g\|_{L^\infty(\R^d)}
  \end{align*}
  for all $g\in X$ and $\alpha\geq\alpha_0$, as desired.
\end{proof}

\subsection{Bounds on the right-hand side of Eq.~\eqref{eq:rhofix}}
Next, we provide suitable estimates for the terms appearing on the
right-hand side of Eq.~\eqref{eq:rhofix}. 

\begin{lemma}
  \label{lem:boundsrhslin}
We have the bounds
   \begin{align*}
    |q_\alpha(x)|&\lesssim 1, \\
    |q_\alpha(x)F_p'(Q_{\R^d}(x))|&\lesssim \alpha^{-1}, \\
    |\alpha^{-2}V_d(\alpha^{-1}x)Q_{\R^d}(x)|&\lesssim \alpha^{-2}\langle
                                               x\rangle^{-d}, \\
    |q_\alpha(x)F_p(Q_{\R^d}(x))|&\lesssim \alpha^{-1}\langle x\rangle^{-d},
  \end{align*}
  for all $x\in \R^d$ and all $\alpha\geq 1$.
\end{lemma}

\begin{proof}
  Recall that we assume Hypothesis \ref{hyp:A}.
  For $r\in [0,\alpha^\frac12]$ we have
  \[
    \frac{\alpha^{-1}r}{A(\alpha^{-1}r)}=\frac{\alpha^{-1}r}{\alpha^{-1}r[1+O(\alpha^{-2}r^2)]}=1+O(\alpha^{-1}), \]
  and thus,
  \[ |q_{\alpha}(x)|=\left |1-\varphi_{d,p}(\alpha^{-1}x)\right |=
    \left |1-\left (\frac{|\alpha^{-1}x|}{A(|\alpha^{-1}x|)}\right
      )^{\frac{(d-1)(p-1)}{2}}\right | \lesssim \alpha^{-1}\]
  for all $|x|\leq \alpha^\frac12$. For $|x|\geq
  \alpha^\frac12$ we trivially estimate
  \[ |q_\alpha(x)|\lesssim
    1+\left (\frac{|\alpha^{-1}x|}{A(|\alpha^{-1}x|)} \right )^\frac{(d-1)(p-1)}{2}\lesssim 1, \]
  since $A(|\alpha^{-1}x|)\gtrsim |\alpha^{-1}x|$.
  This yields the first statement. 
  
  For the second one we recall that
  $F_p'(Q_{\R^d}(x))=p|Q_{\R^d}(x)|^{p-1}$ decays exponentially as
  $|x|\to\infty$. In
  particular, $|F_p'(Q_{\R^d}(x))|\lesssim \langle x\rangle^{-2}$ and
  thus,
  \[ |q_\alpha(x)F_p'(Q_{\R^d}(x))|\lesssim \langle
    \alpha^\frac12\rangle^{-2}\lesssim \alpha^{-1} \]
  provided $|x|\geq \alpha^\frac12$.
  In the case $|x|\leq \alpha^\frac12$ we use the bound
  $|q_\alpha(x)|\lesssim \alpha^{-1}$ from above.
  This proves the second bound, and the fourth bound follows analogously.
Finally, the third estimate is obvious from $V_d\in L^\infty(\R^d)$ and the
  exponential decay of $Q_{\R^d}$.
  \end{proof}

Next, we provide Lipschitz estimates for the nonlinearity from  Eq.~\eqref{eq:rhofix}.
  
\begin{lemma}
  \label{lem:boundsrhsnlin}
    We have the
    bound
    \[ \|\mc N(f)-\mc N(g)\|_{L^2(\R^d)}\lesssim \left
        (\|f\|_{H^2(\R^d)}^{p-1}+\|f\|_{H^2(\R^d)}+\|g\|_{H^2(\R^d)}^{p-1}+\|g\|_{H^2(\R^d)}\right
      )\|f-g\|_{H^2(\R^d)} \]
    for all $f,g\in H^2(\R^d)$.
    Furthermore,
    \[ \|\mc N(f)-\mc N(g)\|_{L^\infty(\R^d)}\lesssim \left
        (\|f\|_{L^\infty(\R^d)}^{p-1}+\|f\|_{L^\infty(\R^d)}+\|g\|_{L^\infty(\R^d)}^{p-1}
        +\|g\|_{L^\infty(\R^d)}\right
      )\|f-g\|_{L^\infty(\R^d)} \]
    for all $f,g\in L^\infty(\R^d)$.
  \end{lemma}

  \begin{proof}
    Recall that we assume $p\in (1,\frac{d+2}{d-2})$ and $d\geq 2$.
    Let $N(t_0,t):=F_p(t_0+t)-F_p(t_0)-F_p'(t_0)t$. Then we have
    \[ \mc
      N(\rho)(x)=N(Q_{\R^d}(x), \rho(x)), \]
and the fundamental theorem of calculus yields
\begin{equation}
  \label{eq:ftN}
  \begin{split}
      N(t_0,t)-N(t_0,s)&=\int_0^1 \partial_u N(t_0, s+u(t-s))du \\
      &=(t-s)\int_0^1 [F_p'(t_0+s+u(t-s))-F_p'(t_0)]du.
      \end{split}
    \end{equation}
    Now we distinguish the cases $p\in (1,2]$ and $p> 2$.
    We proceed with the former and note the elementary estimate
    \begin{equation}
      \label{eq:el1}
     \left ||t_0+t|^{p-1}-|t_0|^{p-1}\right |\lesssim
      |t|^{p-1}
    \end{equation}
    for all $t_0,t\in \R$. 
    Since $F_p'(s)=p|s|^{p-1}$, we obtain from Eq.~\eqref{eq:ftN} the bound
    \begin{align*}
      |N(t_0,t)-N(t_0,s)|
      &\lesssim |t-s|\int_0^1 |s+u(t-s)|^{p-1} du \\
      &\lesssim \left (|t|^{p-1}+|s|^{p-1}\right )|t-s|
    \end{align*}
    for all $t_0,t,s\in \R$.
    Consequently, by H\"older's inequality and Sobolev embedding,
    \begin{align*}
      \|\mc N(f)-\mc N(g)\|_{L^2(\R^d)}
      &\lesssim
      \||f|^{p-1}(f-g)\|_{L^2(\R^d)}
        + \||g|^{p-1}(f-g)\|_{L^2(\R^d)} \\
      &\lesssim \|f\|_{L^{2p}(\R^d)}^{p-1}\|f-g\|_{L^{2p}(\R^d)}
        +\|g\|_{L^{2p}(\R^d)}^{p-1}\|f-g\|_{L^{2p}(\R^d)} \\
      &\lesssim \left
        (\|f\|_{H^2(\R^d)}^{p-1}+\|g\|_{H^2(\R^d)}^{p-1}\right )\|f-g\|_{H^2(\R^d)}.
    \end{align*}
    
    In the case $p>2$ (which only occurs if $d\leq 5$), we use the bound
    \begin{equation}
      \label{eq:el2}\left ||t_0+t|^{p-1}-|t_0|^{p-1}\right |\lesssim
      |t|^{p-1}+|t_0|^{p-2}|t|,
      \end{equation}
    which yields
    \begin{align*}
      |N(t_0,t)-N(t_0,s)|
      &\lesssim |t-s|\int_0^1 \left
        (|s+u(t-s)|^{p-1}+|t_0|^{p-2}|s+u(t-s)|\right )du \\
      &\lesssim \left
        (|t|^{p-1}+|t_0|^{p-2}|t|+|s|^{p-1}+|t_0|^{p-2}|s|\right )|t-s|
    \end{align*}
    for all $t_0,t,s\in \R$. Consequently,
    \begin{align*}
      \|\mc N(f)-\mc N(g)\|_{L^2(\R^d)}
      &\lesssim \||f|^{p-1}(f-g)\|_{L^2(\R^d)}+
        \|Q_{\R^d}\|_{L^\infty(\R^d)}^{p-2}\|f(f-g)\|_{L^2(\R^d)} \\
      &\quad +\||g|^{p-1}(f-g)\|_{L^2(\R^d)}+
        \|Q_{\R^d}\|_{L^\infty(\R^d)}^{p-2}\|g(f-g)\|_{L^2(\R^d)} \\
      &\lesssim \|f\|_{L^{2p}(\R^d)}^{p-1}\|f-g\|_{L^{2p}(\R^d)}+
        \|f\|_{L^4(\R^d)}\|f-g\|_{L^4(\R^d)} \\
      &\quad +\|g\|_{L^{2p}(\R^d)}^{p-1}\|f-g\|_{L^{2p}(\R^d)}+
        \|g\|_{L^4(\R^d)}\|f-g\|_{L^4(\R^d)} \\
      &\lesssim \left (\|f\|_{H^2(\R^d)}^{p-1}+\|f\|_{H^2(\R^d)}
        +\|g\|_{H^2(\R^d)}^{p-1}+\|g\|_{H^2(\R^d)} \right )\|f-g\|_{H^2(\R^d)},
    \end{align*}
    by the Sobolev embeddings $H^2(\R^d)\hookrightarrow L^{2p}(\R^d)$
    and $H^2(\R^d)\hookrightarrow L^4(\R^d)$ (recall that $d\leq 5$).
    The $L^\infty$ bound is immediate from the above.
      \end{proof}

\subsection{Existence of the soliton}
Now we are ready to prove the existence of the soliton profile $Q_{\M^d,\alpha}$.

\begin{proposition}
  \label{prop:exrho}
  There exists an $\alpha_0>0$ such that
  Eq.~\eqref{eq:rhofix} has a real-valued solution $\rho=\rho_\alpha \in
  H_\mathrm{rad}^2(\R^d)\cap C(\R^d)$
  for any $\alpha\geq \alpha_0$. Furthermore, $\rho_\alpha$ satisfies
  \[
    \|\rho_\alpha\|_{H^2(\R^d)}+\|\rho_\alpha\|_{L^\infty(\R^d)}\lesssim
    \alpha^{-1} \]
  for all $\alpha\geq\alpha_0$.
\end{proposition}

\begin{proof}
  Let $X:=H^2_\mathrm{rad}(\R^d)\cap C(\R^d)$ with norm
  \[ \|f\|_X:=\|f\|_{H^2(\R^d)}+\|f\|_{L^\infty(\R^d)} \]
  and set
  $X_\delta:=\{f\in X: \|f\|_X\leq \delta\}$.
 Similarly, we define $Y:=L^2_\mathrm{rad}(\R^d)\cap L^\infty(\R^d)$
  with
  \[ \|f\|_Y:=\|f\|_{L^2(\R^d)}+\|f\|_{L^\infty(\R^d)}. \]
Note that Lemmas \ref{lem:Lalpha}, \ref{lem:boundsrhsnlin} and Corollary
\ref{cor:Lalphasup} imply the estimates
\begin{align*}
  \|\mc A_\alpha^{-1}f\|_X&\lesssim \|f\|_Y \\
\|\mc N(f)-\mc N(g)\|_Y&\lesssim \left
                         (\|f\|_X+\|f\|_X^{p-1}+\|g\|_X+\|g\|_X^{p-1}\right )\|f-g\|_X                           
\end{align*}
for all $f,g\in X\subset Y$ and $\alpha\geq \alpha_0$, provided
$\alpha_0>0$ is sufficiently large.
In view of Eq.~\eqref{eq:rhofix}, we define a map $\mc K_\alpha$ on $X_\delta$ by
  \begin{align*} \mc K_\alpha(f):=-\mc A_\alpha^{-1}\left [
    q_\alpha F_p'(Q_{\R^d}(\cdot))f+(q_\alpha-1)\mc N(f) 
+\alpha^{-2} V_d(\alpha^{-1}(\cdot))Q_{\R^d}+q_\alpha F_p(Q_{\R^d}(\cdot))
    \right ].
  \end{align*}
  Then, by Lemma \ref{lem:boundsrhslin}, 
  \begin{align*}
    \|\mc K_\alpha(f)\|_X 
    &\lesssim \|q_\alpha
    F_p'(Q_{\R^d}(\cdot))f\|_Y
    +\|(q_\alpha-1)\mc N(f)\|_Y \\
      &\quad +\alpha^{-2} \|V_d(\alpha^{-1}(\cdot))Q_{\R^d}\|_Y+
        \|q_\alpha F_p(Q_{\R^d}(\cdot))\|_Y \\
    &\lesssim \alpha^{-1}\|f\|_X
      +\|f\|_X^p+\|f\|_X^2+\alpha^{-2}+\alpha^{-1}
    \\
    &\lesssim \alpha^{-1}\delta+\delta^p+\delta^2+\alpha^{-2}+\alpha^{-1}
  \end{align*}
  for all $f\in X_\delta$.  Thus, $\mc K_\alpha(f)\in X_\delta$ for
  all $f\in X_\delta$ and $\alpha\geq \alpha_0$, provided $\delta>0$ is sufficiently small
  and $\alpha_0\geq 1$ is sufficiently large.
  Similarly,
  \begin{align*}
    \|&\mc K_\alpha(f)-\mc K_\alpha(g)\|_X \\
    &\lesssim \|q_\alpha
      F_p'(Q_{\R^d}(\cdot))(f-g)\|_Y+\|(q_\alpha-1)(\mc
      N(f)-\mc N(g))\|_Y \\
    &\lesssim \alpha^{-1}\|f-g\|_X
      +\left (\|f\|_X^{p-1}+\|f\|_X
      +\|g\|_X^{p-1}+\|g\|_X\right
      )\|f-g\|_X \\
    &\lesssim (\alpha^{-1}+\delta^{p-1}+\delta)\|f-g\|_X.
  \end{align*}
  Thus, $\mc K_\alpha$ is a contraction on $X_\delta$ for all
  $\alpha\geq \alpha_0$, provided $\delta>0$ is small enough and
  $\alpha_0\geq 1$ is large enough.
  Consequently, the contraction mapping principle yields the existence
  of a fixed point $\rho_\alpha\in X_\delta\subset
  H^2_\mathrm{rad}(\R^d)\cap C(\R^d)$ of $\mc
  K_\alpha$ which, by construction, is a solution to
  Eq.~\eqref{eq:rhofix}. Finally, for the stated estimate on $\rho_\alpha$,
it suffices to note that
  \begin{align*}
    \|\rho_\alpha\|_X
&=\|\mc K_\alpha(\rho_\alpha)\|_X\lesssim
                        \alpha^{-1}\|\rho_\alpha\|_X+\delta^{p-1}\|\rho_\alpha\|_X
+\delta\|\rho_\alpha\|_X+\alpha^{-2}+\alpha^{-1},
  \end{align*}
by the above estimate for $\mc K_\alpha(f)$.
\end{proof}

\subsection{Decay and regularity}
From now on we denote by $\rho_\alpha$ the solution constructed in
Proposition \ref{prop:exrho}.
Note that the radiality of $\rho_\alpha$ immediately implies a pointwise
decay estimate. To see this, we recall the classical Strauss estimate.

\begin{lemma}
  \label{lem:Strauss}
We have the bound
\[ \left \||\cdot|^\frac{d-1}{2} f\right
  \|_{L^\infty(\R^d)}\lesssim \|f\|_{H^1(\R^d)} \]
for all $f\in H^1_\mathrm{rad}(\R^d)$.
\end{lemma}

\begin{proof}
  It suffices to prove the bound for real-valued $f$.
  First, we assume that $f\in C^\infty_c(\R^d)$.
  Then $f$ is given by $f(x)=\widehat f(|x|)$ for some $\widehat f\in
  C^\infty_c(\R)$.
  By the fundamental theorem of calculus and Cauchy-Schwarz, we obtain
  \begin{align*}
    r^{d-1}\widehat f(r)^2
    &=-\int_r^\infty \partial_s \left
      [s^{d-1}\widehat f(s)^2\right ]ds
      =-(d-1)\int_r^\infty s^{d-2}\widehat f(s)^2 ds
      -2\int_r^\infty s^{d-1}\widehat f'(s)\widehat f(s)ds \\
    &\lesssim \|f\|_{L^2(\R^d)}\|\nabla f\|_{L^2(\R^d)}\lesssim \|f\|_{H^1(\R^d)}^2
  \end{align*}
  for all $r\geq 0$, which implies the desired estimate.
By
  approximation, the bound extends to all $f\in H^1_\mathrm{rad}(\R^d)$.
\end{proof}

 Lemma \ref{lem:Strauss} implies the decay
\begin{equation}
\label{eq:decayrho}
 |\rho_\alpha(x)|\lesssim \langle x\rangle^{-\frac{d-1}{2}} 
\end{equation}
for all $x\in \R^d$.

\begin{lemma}
  \label{lem:rhoC2}
  We have $\rho_\alpha\in C^2(\R^d)$. In particular, the function
  $\widetilde R_\alpha(x)=Q_{\R^d}(x)+\rho_\alpha(x)$ satisfies
  \[ \Delta_{\R^d}\widetilde R_\alpha(x)-\widetilde R_\alpha(x)
    -\alpha^{-2}V_d(\alpha^{-1}x)\widetilde
    R_\alpha(x)+\varphi_{d,p}(\alpha^{-1}x)F_p\left (\widetilde
      R_\alpha(x)\right)=0 \]
  for all $x\in \R^d$, in the classical sense.
\end{lemma}

\begin{proof}
  Let
  \begin{align*} g_\alpha(x):=-\big [ &q_\alpha(x) F_p'(Q_{\R^d}(x))\rho_\alpha(x)+[q_\alpha(x)-1]\mc N(\rho_\alpha)(x) \\
&+\alpha^{-2} V_d(\alpha^{-1}x)Q_{\R^d}(x)+q_\alpha(x)
    F_p(Q_{\R^d}(x))\big ].
  \end{align*}
  Then, by Lemma \ref{lem:boundsrhsnlin}, we have $g_\alpha\in
  L^2_\mathrm{rad}(\R^d)\cap L^\infty(\R^d)\cap C(\R^d)$ and by
  construction,
$\mc A_\alpha \rho_\alpha^\sharp=g_\alpha^\sharp$.
  Equivalently,
  $\Delta_{\R^d}\rho_\alpha^\sharp=h_\alpha^\sharp$
  with
  \[
    h_\alpha(x):=\rho_\alpha(x)-F_p'(Q_{\R^d}(x))\rho_\alpha(x)+\alpha^{-2}V_d(\alpha^{-1}x)\rho_\alpha(x)
    -g_\alpha(x). \]
  Since $\rho_\alpha,h_\alpha\in C(\R^d)$ are radial, the claim
  follows from Lemma \ref{lem:regLaplace}.
\end{proof}

\begin{proof}[Proof of Theorem \ref{thm:ex}]
For $\alpha>0$ sufficiently large, let
 \[ Q_{\M^d,\alpha}(r,y):=\alpha^\frac{2}{p-1}\left(
      \frac{r}{A(r)}\right )^{\frac{d-1}{2}}[Q_{\R^d}(\alpha
    re_1)+\rho_\alpha(\alpha r e_1)]. \]
  By Lemma \ref{lem:rhoC2} and Hypothesis \ref{hyp:A}, $Q_{\M^d,\alpha}\in C^2(\M^d)$
and by construction, $u_\alpha^*(t,r,y)=e^{i\alpha^2
    t}Q_{\M^d,\alpha}(r,y)$ solves Eq.~\eqref{eq:NLSM} for all $t\in
  \R$. The remaining properties follow from Proposition \ref{prop:exrho}.
\end{proof}

\section{Spectral stability of the soliton}

\noindent In this section we investigate the \emph{linear} stability of the soliton
\[ v_\alpha^*(t,x)=e^{i\alpha^2 t}R_\alpha(x)=\alpha^\frac{2}{p-1}e^{i\alpha^2 t}\left
    [Q_{\R^d}(\alpha x)+\rho_\alpha(\alpha x)\right] \]
as a solution to the nonlinear Schr\"odinger equation \eqref{eq:NLSv}.
More precisely, we study spectral properties of the linearized
operator $\mc L_\alpha$ associated to the soliton $v_\alpha^*$.
We will see that the qualitative behavior is very similar to the Euclidean case.

\subsection{The linearized operator}

The notion of spectral stability derives from spectral properties of
the operator that is obtained by linearizing Eq.~\eqref{eq:NLSv} at
the soliton $v_\alpha^*$.
More precisely, we insert the ansatz
\[ v(t,x)=v_\alpha^*(t,x)+e^{i\alpha^2 t}w(t,x)=e^{i\alpha^2 t}\left
    [R_\alpha(x)+w(t,x)\right] \]
into Eq.~\eqref{eq:NLSv}. This yields
\begin{equation}
  \label{eq:NLSpert}
  \begin{split}
  i\partial_t w(t,\cdot)
  &+\Delta_{\R^d}w(t,\cdot)-\alpha^2
  w(t,\cdot)-V_dw(t,\cdot) \\
  &+\varphi_{d,p}F_p\left (
      R_\alpha(\cdot)+w(t,\cdot)\right
    )-\varphi_{d,p}F_p(R_\alpha(\cdot))=0,
    \end{split}
\end{equation}
where we have used Eq.~\eqref{eq:R}, i.e., 
\[ \Delta_{\R^d}R_\alpha-\alpha^2 R_\alpha-V_d
  R_\alpha=-\varphi_{d,p}F_p(R_\alpha(\cdot)). \]
Now note that for all $a_0,a,b\in \R$,
\begin{align*}
  |a_0+a+ib|^{p-1}
  &=(a_0^2+2a_0 a+a^2+b^2)^{\frac{p-1}{2}} \\
    &=|a_0|^{p-1}+\tfrac{p-1}{2}(a_0^2)^{\frac{p-1}{2}-1}(2a_0
      a+a^2+b^2)+N_1(a_0,a,b) \\
  &=|a_0|^{p-1}+(p-1)a_0|a_0|^{p-3}a+N_2(a_0,a,b),
\end{align*}
where $N_1(a_0,a,b)$ and $N_2(a_0,a,b)$ are quadratic in $a$ and $b$.
Hence,
\begin{align*}
  F_p(a_0+a+ib)&=(a_0+a+ib)|a_0+a+ib|^{p-1} \\
  &=a_0|a_0|^{p-1}+p|a_0|^{p-1}a+i|a_0|^{p-1}b+N(a_0,a,b),
\end{align*}
where $N(a_0,a,b)$ is quadratic in $a$ and $b$.
This yields
\begin{align*}
  F_p\left (R_\alpha(x)+w(t,x)\right )
  &=F_p\left (R_\alpha(x)+\Re w(t,x)+i\Im w(t,x)\right ) \\
  &=F_p(R_\alpha(x))+p|R_\alpha(x)|^{p-1}\Re
    w(t,x)+i|R_\alpha(x)|^{p-1}\Im w(t,x) \\
  &\quad +N\left (R_\alpha(x),\Re w(t,x),\Im w(t,x)\right ).
\end{align*}
By dropping the nonlinear terms, we obtain from
Eq.~\eqref{eq:NLSpert} the linearized problem
\begin{equation*}
  \label{eq:NLSlinw}
  \begin{split}
  i\partial_t w(t,\cdot)&+\Delta_{\R^d}w(t,\cdot)
  -\alpha^2 w(t,\cdot)-V_d w(t,\cdot) \\
  &+p\varphi_{d,p}|R_\alpha(\cdot)|^{p-1}\Re
  w(t,\cdot)+i\varphi_{d,p}|R_\alpha(\cdot)|^{p-1}\Im w(t,\cdot)=0.
  \end{split}
\end{equation*}
Finally, we rescale by setting $w(t,x)=\widetilde w(\alpha^2 t,\alpha
x)$. This yields
\begin{equation}
  \label{eq:NLSlin}
  \begin{split}
  i\partial_t \widetilde w(t,x)&+\Delta_{\R^d,x}\widetilde w(t,x)
  -\widetilde w(t,x)-\alpha^{-2}V_d(\alpha^{-1}x) \widetilde w(t,x) \\
  &+p\varphi_{d,p}(\alpha^{-1}x)\left |\widetilde R_\alpha(x)\right |^{p-1}\Re
  \widetilde w(t,x)+i\varphi_{d,p}(\alpha^{-1}x)\left |\widetilde R_\alpha(x)\right |^{p-1}\Im
  \widetilde w(t,\cdot)=0
\end{split}
\end{equation}
with $\widetilde R_\alpha(x)=\alpha^{-\frac{2}{p-1}}R_\alpha(\alpha^{-1}x)$.
Eq.~\eqref{eq:NLSlin} is equivalent to the system
\begin{equation}
  \label{eq:NLSlinsys}
  \partial_t
  \begin{pmatrix}
    \Re \widetilde w(t,\cdot) \\
    \Im \widetilde w(t,\cdot)
  \end{pmatrix}
  =
  \mc L_\alpha
  \begin{pmatrix}
    \Re \widetilde w(t,\cdot) \\
    \Im \widetilde w(t,\cdot)
  \end{pmatrix}
\end{equation}
with the spatial differential operator
\[ \mc L_\alpha:=
  \begin{pmatrix}
    0 & \mc L_{\alpha,-} \\
    -\mc L_{\alpha,+} & 0
  \end{pmatrix},
\]
where
\begin{align*}
\mc  L_{\alpha,-}f(x)&:=-\Delta_{\R^d}f(x)+f(x)-\varphi_{d,p}(\alpha^{-1}x)\left
                    |\widetilde
                    R_\alpha(x)\right
                  |^{p-1}f(x)+\alpha^{-2}V_d(\alpha^{-1}x)f(x) \\
\mc L_{\alpha,+}f(x)&:=-\Delta_{\R^d}f(x)+f(x)-p\varphi_{d,p}(\alpha^{-1}x)\left
                    |\widetilde
                    R_\alpha(x)\right
                  |^{p-1}f(x)+\alpha^{-2}V_d(\alpha^{-1}x)f(x).
\end{align*}
Consequently, (linear) stability properties of the soliton
$v_\alpha^*$ are encoded in the spectrum of the operator $\mc
L_{\alpha}$, which we consider on the space
$L^2_\mathrm{rad}(\R^d,\C^2)$.
This is a natural choice since
the operators $\mc L_{\alpha,\pm}$ are self-adjoint on
$L^2_{\mathrm{rad}}(\R^d)$.
Formally at least, in the limit $\alpha\to\infty$, $\mc L_\alpha$ reduces to its
well-known Euclidean counterpart $\mc L$, given by
\[ \mc L= \begin{pmatrix}
    0 & \mc L_- \\
    -\mc L_+ & 0
  \end{pmatrix} \]
and
\begin{align*}
  \mc
  L_-f(x)&=-\Delta_{\R^d}f(x)+f(x)-\left
                    |Q_{\R^d}(x)\right |^{p-1}f(x)
  \\
\mc L_+f(x)&=-\Delta_{\R^d}f(x)+f(x)-p\left
                    |Q_{\R^d}(x)
                    \right |^{p-1}f(x).
\end{align*}
This suggests a perturbative spectral analysis, based on the
Euclidean situation. 

\subsection{Spectral properties in the Euclidean case}
Our base case will be the Euclidean operator $\mc L$ which was
extensively studied in the literature, see e.g.~\cite{Wei83, Wei85,
  Gri88, ChaGusNakTsa07}. Nevertheless, there are still a
number of substantial questions that remain unanswered.
We summarize some of the known results but restrict ourselves to the
radial case. Since we will be dealing with spectra of nonself-adjoint
operators, there are some ambiguities that need to be clarified first.

\begin{definition}
  \label{def:essspec}
Let $T$ be a closed operator on a Banach space
$X$. We define the \emph{essential spectrum}
$\sigma_e(T)$ of $T$ by
\[ \sigma_e(T):=\bigcap_{K\in \mc K(X)}\sigma(T+K), \]
where $\mc K(X)$ denotes the set of all compact operators on $X$.
Furthermore, $\sigma_p(T)$ is the set of all eigenvalues of $T$.
\end{definition}

\begin{remark}
  \label{rem:essspec}
  There are other meaningful definitions of essential spectra for
  nonself-adjoint operators in the
  literature, see e.g.~\cite{GusWei69, EdmEva87, HunLee07} for a
  discussion on this. The
  choice we made is the largest possible that is invariant under
  relatively compact perturbations. 
  However,
  for the particular class of operators we will be concerned with, all the usual
  definitions turn out to be equivalent, see \cite{HunLee07}.
\end{remark}

\begin{theorem}[\cite{Wei83, Wei85,
  Gri88, ChaGusNakTsa07, Sch09}]
  \label{thm:L}
  The operator $\mc L: H^2_\mathrm{rad}(\R^d,\C^2)\subset
  L^2_\mathrm{rad}(\R^d,\C^2)
  \to L^2_\mathrm{rad}(\R^d,\C^2)$ is
  closed and has the following properties:
  \begin{itemize}
  \item The spectrum $\sigma(\mc L)$ is a subset of $\R\cup i\R$.
    \item If $\lambda\in \sigma(\mc L)$ then $-\lambda\in \sigma(\mc L)$.
  \item The essential spectrum of $\mc L$ is given by
    \[ \sigma_e(\mc L)=\{z\in \C: \Re z=0, |\Im z|\geq 1\}. \]
  \item The set $\sigma(\mc L)\setminus \sigma_e(\mc L)$ is free of
    accumulation points and consists of
    eigenvalues with finite algebraic multiplicities.
  \item We have $0\in \sigma_p(\mc L)$ and
    \[ \ker\mc L=\left \langle\begin{pmatrix}0 \\
          Q_{\R^d}\end{pmatrix}\right \rangle. \]
  \item For the kernels of powers of $\mc L$ we have
    \[ \dim\ker(\mc L^2)=\dim\ker(\mc L^3)=\begin{cases}
        2 & \mbox{if }p\not= 1+\frac{4}{d} \\
        4 & \mbox{if }p=1+\frac{4}{d}
      \end{cases}. \]
    In particular, the algebraic multiplicity of the eigenvalue
    $0\in\sigma_p(\mc L)$ equals $4$ in the $L^2$-critical case
    $p=1+\frac{4}{d}$ and $2$ otherwise.
    \item In the $L^2$-subcritical case $p<1+\frac{4}{d}$, $\mc L$ has
      no positive eigenvalues. In the $L^2$-supercritical
      case $p>1+\frac{4}{d}$, $\mc L$ has precisely one positive
      eigenvalue $\lambda$ and the eigenvalues $\pm\lambda$ 
are simple.
  \end{itemize}
\end{theorem}

\begin{remark}
  The picture one has in mind is as follows. Starting from the
  supercritical case $p>1+\frac{4}{d}$, the two nonzero real
  eigenvalues move towards
  the origin as $p$ decreases. Precisely when $p=1+\frac{4}{d}$, the
  two eigenvalues merge and the algebraic multiplicity of $0\in
  \sigma_p(\mc L)$
  increases by two. If one decreases $p$ further into the
  subcritical regime $p<1+\frac{4}{d}$, a pair of purely imaginary
  eigenvalues emerges from
  $0$. In particular, the ground state is linearly stable in the
  subcritical case and unstable in the
  supercritical case. These linear stability properties are reflected in the
  nonlinear theory. Indeed, in the subcritical case the ground
  state is orbitally stable and in the supercritical case it is
  unstable. The critical case $p=1+\frac{4}{d}$ is more delicate as
  there is spectral stability (that is to say, no spectrum away from
  the imaginary axis) but quite strong instability in the
  nonlinear theory. 
\end{remark}

\begin{remark}
Important issues that remain unsolved concern the existence of
eigenvalues and/or resonances embedded in the essential spectrum and
the ``gap property''. The latter refers to the absence of eigenvalues
on the imaginary axis between $0$ and $i$ in the
supercritical case. These spectral properties are important for the
(nonlinear) asymptotic stability theory of the ground state. Some of
them have
been verified numerically or even proved rigorously in special cases,
see
e.g.~\cite{DemSch06, KriSch06, ChaGusNakTsa07, CosHuaSch12}, but there is
no systematic theoretical understanding so far.
\end{remark}

\subsection{Spectral properties in the curved geometry}
To begin, we show that the structural properties of the spectrum in the curved
case are the same as in the Euclidean case. An important prerequisite
is the nonnegativity of $\mc L_{\alpha,-}$, which we establish first.

\begin{proposition}
  \label{prop:L-nonneg}
  There exists an $\alpha_0>0$ such that, for all $\alpha\geq\alpha_0$,
  $\mc L_{\alpha,-}: H^2_\mathrm{rad}(\R^d)\subset
  L^2_\mathrm{rad}(\R^d)\to L^2_\mathrm{rad}(\R^d)$ is self-adjoint
  with the following properties:
\begin{itemize}
\item The essential spectrum of $\mc L_{\alpha,-}$ is given by
  $\sigma_e(\mc L_{\alpha,-})=[1+V_{0,d}\alpha^{-2},\infty)$.
\item We have  $0\in \sigma_p(\mc L_{\alpha,-})$ and
$\ker\mc L_{\alpha,-}=\langle Q_{\R^d}+\rho_\alpha
\rangle$.
\item The operator $\mc L_{\alpha,-}$ satisfies
  \[ \left (\mc L_{\alpha,-}f|f \right)_{L^2(\R^d)}\gtrsim
\|f\|_{L^2(\R^d)}^2 \]
  for all $f\in \langle Q_{\R^d}+\rho_\alpha\rangle^\perp\cap H^2_\mathrm{rad}(\R^d)$ and
  all $\alpha\geq\alpha_0$.
  \end{itemize}
\end{proposition}

\begin{proof}
 We define $\mc L_{\alpha,0}: H^2_\mathrm{rad}(\R^d)\subset
 L^2_\mathrm{rad}(\R^d)\to L^2_\mathrm{rad}(\R^d)$ by $\mc L_{\alpha,0}
 f:=-\Delta_{\R^d}f+(1+V_{0,d}\alpha^{-2})f$ and set
 \[ W_\alpha(x):=-\varphi_{d,p}(\alpha^{-1}x)\left
     |Q_{\R^d}(x)+\rho_\alpha(x)\right|^{p-1}+\alpha^{-2}\left [V_d(\alpha^{-1}x)-V_{0,d}\right].
 \]
 Then we have $\mc L_{\alpha,-}f=\mc L_{\alpha,0} f+W_\alpha f$. By Fourier
 analysis it follows that
 \[ \sigma(\mc L_{\alpha,0})=\sigma_e(\mc L_{\alpha,0})=[1+V_{0,d}\alpha^{-2},\infty). \]
 Furthermore,
 by Hypothesis \ref{hyp:A}, Lemma \ref{lem:boundsrhslin}, Proposition \ref{prop:exrho}, and
 Lemma \ref{lem:Strauss}, we have $W_\alpha\in L^\infty(\R^d)\cap
 C(\R^d)$ and
 \[ \lim_{|x|\to\infty}W_\alpha(x)=0. \]
 Consequently, $f\mapsto W_\alpha f: L^2_\mathrm{rad}(\R^d)\to
 L^2_\mathrm{rad}(\R^d)$ is bounded and the Kato-Rellich theorem (see
 e.g.~\cite{Tes14}, p.~159, Theorem 6.4) shows that $\mc L_{\alpha,-}$
 is self-adjoint. In particular, $\sigma(\mc L_{\alpha,-})\subset \R$.
 Furthermore, by \cite{Tes14}, p.~258, Theorem 10.2, the operator
 $f\mapsto W_\alpha f: L^2_\mathrm{rad}(\R^d)\to
 L^2_\mathrm{rad}(\R^d)$
 is relatively compact with respect to $\mc
 L_{\alpha,0}$ and Weyl's theorem (see e.g.~\cite{Tes14}, p.~171,
 Theorem 6.19) implies that $\sigma_e(\mc
 L_{\alpha,-})=[1+V_{0,d}\alpha^{-2},\infty)$. As a consequence, $\sigma(\mc
 L_{\alpha,-})\setminus \sigma_e(\mc L_{\alpha,-})$ consists of
 isolated eigenvalues only. The same is true for the limiting operator
 $\mc L_-$, i.e., $\sigma(\mc L_-)\setminus \sigma_e(\mc L_-)$
 consists of isolated eigenvalues only, where $\sigma_e(\mc L_-)=[1,\infty)$.

 Next, we show that there exists a constant $\mu>0$ such that
 \begin{equation}
   \label{eq:infmu}
   (-\infty,-\mu)\subset \rho(\mc L_{\alpha,-})
 \end{equation}
 for all $\alpha\geq \alpha_0$.
 To this end we use the resolvent bound
 $\|(\lambda-\mc L_{0,\alpha})^{-1}\|_{L^2(\R^d)}\leq
 |\lambda|^{-1}$, valid for all $\lambda<0$, which is a consequence of
 the self-adjointness of $\mc L_{\alpha,0}$ and $\sigma(\mc
 L_{\alpha,0})\subset [1+V_{0,d}\alpha^{-2},\infty)$. Furthermore, we
 note that the operator $\mc B_\alpha f:=W_\alpha f$ satisfies
\[ \|\mc B_\alpha f\|_{L^2(\R^d)}\leq
  \|W_\alpha\|_{L^\infty(\R^d)}\|f\|_{L^2(\R^d)}
\lesssim \|f\|_{L^2(\R^d)} \]
for all $\alpha\geq\alpha_0$ and $f\in L^2_\mathrm{rad}(\R^d)$.
Consequently, if $\mu>0$ is sufficiently large, the operator $1-\mc
B_\alpha(\lambda-\mc L_{\alpha,0})^{-1}$ is invertible for all
$\lambda<-\mu$ by a Neumann
series argument and the identity, 
\[ \lambda-\mc L_{\alpha,-}=\left [1-\mc B_\alpha(\lambda-\mc
  L_{\alpha,0})^{-1}\right ](\lambda-\mc L_{\alpha,0}), \]
proves \eqref{eq:infmu}.

Now we turn to the computation of $\ker\mc L_{\alpha,-}$.
Obviously, $0\in\sigma_p(\mc L_{\alpha,-})$ since $\widetilde
R_\alpha=Q_{\R^d}+\rho_\alpha\in H^2_\mathrm{rad}(\R^d)\cap C^2(\R^d)$ by Lemma
\ref{lem:rhoC2} and
\begin{align*}
  \mc L_{\alpha,-}\widetilde R_\alpha(x)
  &=-\Delta_{\R^d}\widetilde
  R_\alpha(x)+\widetilde
  R_{\alpha}(x)-\varphi_{d,p}(\alpha^{-1}x)\left |\widetilde
    R_\alpha(x)\right |^{p-1}\widetilde
R_\alpha(x)+\alpha^{-2}V_d(\alpha^{-1}x)\widetilde R_\alpha(x) \\
&=0.
\end{align*}
In particular, $\langle\widetilde R_\alpha\rangle\subset \ker(\mc
L_{\alpha,-})$.
To prove the reverse inclusion, suppose $f\in
H^2_\mathrm{rad}(\R^d)\setminus\{0\}$ satisfies $\mc
L_{\alpha,-}f=0$. By the one-dimensional Sobolev embedding we have
$f\in C(\R^d\setminus\{0\})$ and thus, $W_\alpha f\in
L^2_\mathrm{rad}(\R^d)\cap C(\R^d\setminus\{0\})$.
Consequently, Lemma \ref{lem:raddist} implies that $\widehat
f(r):=r^{\frac{d-1}{2}}f(re_1)$ belongs to $C^2(0,\infty)$ and
satisfies
\[ \widehat f''(r)-\frac{(d-1)(d-3)}{4r^2}\widehat f(r)-[1+V_{0,d}\alpha^{-2}]\widehat
  f(r)=W_\alpha(re_1)\widehat f(r) \]
for all $r>0$.
According to Lemma \ref{lem:phi0phiinf}, there exist constants $a,b\in \C$ such
that $\widehat f(r)=a\phi_0(r)+b\psi_0(r)$, where $|\phi_0(r)|\simeq
r^{\frac{d-1}{2}}$, $|\phi_0'(r)|\simeq r^{\frac{d-3}{2}}$, and
\begin{align*}
  |\psi_0(r)|&\simeq
  \begin{cases}
    r^\frac12 |\log r| & d=2 \\
    r^{-\frac{d-3}{2}} & d\not= 2
  \end{cases},
& |\psi_0'(r)|&\simeq
  \begin{cases}
    r^{-\frac12} |\log r| & d=2 \\
    r^{-\frac{d-1}{2}} & d\not= 2
  \end{cases}                          
\end{align*}                         
for $r\in (0,\frac12]$. Since $f\in H^1_\mathrm{rad}(\R^d)$, we must have
$b=0$ and this shows that the kernel of $\mc L_{\alpha,-}$ is one-dimensional. Consequently,
 $\ker\mc L_{\alpha,-}=\langle\widetilde R_\alpha \rangle$, as
 claimed.

 Now we define an operator $\mc C_\alpha: L_\mathrm{rad}^2(\R^d)\to L^2_\mathrm{rad}(\R^d)$ such that $\mc
 L_{\alpha,-}=\mc L_-+\mc C_\alpha$, i.e.,
 \begin{align*}
   \mc C_\alpha f(x)
   &=\mc L_{\alpha,-}f(x)-\mc L_- f(x) \\
   &=-\varphi_{d,p}(\alpha^{-1}x)\left
     |Q_{\R^d}(x)+\rho_\alpha(x)\right |^{p-1}f(x)+
     |Q_{\R^d}(x)|^{p-1}f(x)+\alpha^{-2}V_d(\alpha^{-1}x)f(x) \\
   &=:U_\alpha(x)f(x).
 \end{align*}
 We have
 \begin{align*}
   |U_\alpha(x)|
   &\lesssim \left | \left
     |Q_{\R^d}(x)+\rho_\alpha(x)\right|^{p-1}-|Q_{\R^d}(x)|^{p-1}\right
     |
     +|q_\alpha(x)|\left |Q_{\R^d}(x)+\rho_\alpha(x)\right |^{p-1} \\
   &\quad +\alpha^{-2}|V_d(\alpha^{-1}x)| \\
  &\lesssim
    |\rho_\alpha(x)|^{p-1}+|\rho_\alpha(x)|+|q_\alpha(x)F_p'(Q_{\R^d}(x))|
    +|q_\alpha(x)||\rho_\alpha(x)|^{p-1}+\alpha^{-2} \\
   &\lesssim \alpha^{-(p-1)}+\alpha^{-1}+\alpha^{-2}
 \end{align*}
 for all $x\in \R^d$ by Lemma \ref{lem:boundsrhslin} and Proposition
 \ref{prop:exrho}. Here we have used the elementary estimates
 \eqref{eq:el1} and \eqref{eq:el2} from the proof of Lemma
 \ref{lem:boundsrhsnlin}.
 Consequently, $\|U_\alpha\|_{L^\infty(\R^d)}\to 0$ as
 $\alpha\to\infty$ and this shows that the operator $\mc C_\alpha$
 converges to $0$ in norm as $\alpha\to\infty$.
 Recall that $\mc L_-$ is nonnegative. This is a consequence of
 $Q_{\R^d}>0$, $\mc L_- Q_{\R^d}=0$, and the Sturm oscillation
 theorem.
Let $d_0:=\dist(0,\sigma(\mc L_-)\setminus\{0\})$. Since
$0$ is an isolated eigenvalue of $\mc L_-$, we have $d_0>0$.
Let $\gamma: [0,2\pi]\to \C$ be a simple, closed, smooth curve that
encircles the interval $[-1-\mu,\frac{1}{4}d_0]$ and such that
$\gamma(t)\cap [\frac{3}{4}d_0,\infty)=\emptyset$ for all $t\in
[0,2\pi]$. By construction, $\gamma(t)\in \rho(\mc L_-)$ for all $t\in
[0,2\pi]$ and thus, the spectral projection
\[ \mc P:=\frac{1}{2\pi i}\int_\gamma (z-\mc L_-)^{-1}dz \]
is well defined. By the self-adjointness of $\mc L_-$, we have
$\rg\mc P=\ker\mc L_-=\langle Q_{\R^d}\rangle$
since $0$ is the only spectral point of $\mc L_-$ inside of
$\gamma$.
Recall that $\mc C_\alpha \to 0$ in norm as $\alpha\to\infty$ and
thus, $\gamma(t)\in\rho(\mc L_{\alpha,-})$ for all $t\in [0,2\pi]$ and $\alpha\geq\alpha_0$,
provided $\alpha_0>0$ is sufficiently large. This follows immediately
from the identity
\[ z-\mc L_{\alpha,-}=[1-\mc C_\alpha(z-\mc L_-)^{-1}](z-\mc
  L_-), \]
valid for all $z\in \rho(\mc L_-)$. Thus, the spectral projection
\[ \mc P_\alpha:=\frac{1}{2\pi i}\int_\gamma (z-\mc
  L_{\alpha,-})^{-1}dz \]
is well defined, and we have $\mc P_\alpha\to\mc P$ in norm as
$\alpha\to\infty$.
Consequently, by \cite{Kat95}, p.~34, Lemma 4.10, it follows that
$\dim \rg \mc P_\alpha=\dim\rg \mc P=1$
for all $\alpha\geq \alpha_0$.  Since $0\in \sigma_p(\mc L_{\alpha,-})$,
we conclude that $0$ is the only spectral point of $\mc L_{\alpha,-}$
in the interval $[-1-\mu,\frac14 d_0]$. Finally, with \eqref{eq:infmu}, we
infer that $(-\infty,\frac14 d_0]\cap\sigma(\mc L_{\alpha,-})=\{0\}$
for all $\alpha\geq\alpha_0$.
In particular, $\mc L_{\alpha,-}$ is nonnegative, and this finishes the proof.
\end{proof}

We also note the following simple observation concerning the operator
$\mc L_{\alpha,+}$.

\begin{lemma}
  \label{lem:L+inv}
  The operator $\mc L_{\alpha,+}: H^2_\mathrm{rad}(\R^d)\subset
  L^2_\mathrm{rad}(\R^d)\to L^2_\mathrm{rad}(\R^d)$ is self-adjoint
  and invertible.
\end{lemma}

\begin{proof}
  We write $\mc L_{\alpha,+}=\mc L_++\mc B_\alpha$ with $\mc B_\alpha
  f(x)=W_\alpha(x)f(x)$ and
  \[ W_\alpha(x)=-p\left [\varphi_{d,p}(\alpha^{-1}x)\left |Q_{\R^d}(x)+\rho_\alpha(x)\right
    |^{p-1}-|Q_{\R^d}(x)|^{p-1}\right
  ]+\alpha^{-2}V_d(\alpha^{-1}x). \]
We have $\|W_\alpha\|_{L^\infty(\R^d)}\to 0$ as $\alpha\to\infty$
(cf.~the proof of Proposition \ref{prop:L-nonneg}) and thus, $\mc
B_\alpha$ is a bounded symmetric operator on $L^2_\mathrm{rad}(\R^d)$
that converges to $0$ in norm as $\alpha\to\infty$. Consequently, the
Kato-Rellich theorem implies that $\mc L_{\alpha,+}$ is self-adjoint.
Since $0\in \rho(\mc L_+)$, it follows from the identity
\[ \mc L_{\alpha,+}=[1+\mc B_\alpha \mc L_+^{-1}]\mc L_+, \]
and a Neumann series argument, that
 $\mc L_{\alpha,+}$ is invertible for all
$\alpha\geq\alpha_0$, provided $\alpha_0>0$ is sufficiently large. 
\end{proof}

Based on the results on $\mc L_{\alpha,\pm}$, we can now establish
some basic structural properties concerning the spectrum of $\mc L_\alpha$.

\begin{lemma}
  \label{lem:basicLalpha}
  There exists an $\alpha_0>0$ such that for all $\alpha\geq\alpha_0$
  the operator
  $\mc L_\alpha: H^2_\mathrm{rad}(\R^d, \C^2)\subset
  L^2_\mathrm{rad}(\R^d, \C^2)\to L^2_\mathrm{rad}(\R^d, \C^2)$ is
  closed and the following holds:
  \begin{itemize}
  \item The spectrum of $\mc L_\alpha$ is a subset of $\R\cup i\R$.
    \item If $\lambda\in \sigma(\mc L_\alpha)$ then $-\lambda\in
      \sigma(\mc L_\alpha)$.
    \item The essential spectrum of $\mc L_\alpha$ is given by
      \[ \sigma_e(\mc L_\alpha)=\{z\in \C: \Re z=0, |\Im z|\geq
        1+V_{0,d}\alpha^{-2}\}. \]
\item There exists a $\mu>0$ (independent of $\alpha$) such that $(-\infty,-\mu)\cup
  (\mu,\infty)\subset \rho(\mc L_\alpha)$.
        \item The set $\sigma(\mc L_\alpha)\setminus \sigma_e(\mc
          L_\alpha)$ is free of accumulation points and consists of
    eigenvalues with finite algebraic multiplicities.
  \item We have $0\in \sigma_p(\mc L_\alpha)$ and
    \[ \ker\mc L_\alpha=\left \langle\begin{pmatrix}0 \\
          Q_{\R^d}+\rho_\alpha\end{pmatrix}\right \rangle. \]
  \end{itemize}
\end{lemma}

\begin{proof}
First of all, $\mc L_{\alpha,\pm}$ are self-adjoint and hence
closed. This implies the closedness of $\mc L_\alpha$. Now consider the
unitary operator $\mc U: L^2(\R^d,\C^2)\to L^2(\R^d,\C^2)$ given by
\[ \mc U:=\frac{1}{\sqrt 2}
  \begin{pmatrix}
    1 & i \\ 1 & -i
  \end{pmatrix}, \]
and set $\mc H_\alpha:=i\,\mc U \mc L_\alpha \mc U^*$.
Explicitly, we have
\begin{align*}
  \mc H_\alpha
  &=i\,\mc U\mc L_\alpha \mc U^*
    =\frac{1}{2}
    \begin{pmatrix}
      \mc L_{\alpha,-}+\mc L_{\alpha,+} & -\mc L_{\alpha,-}+\mc L_{\alpha,+} \\
\mc L_{\alpha,-}-\mc L_{\alpha,+} & -\mc L_{\alpha,-}-\mc
L_{\alpha,+}
    \end{pmatrix}
=\mc H_{0,\alpha}+\mc H_\alpha', 
\end{align*}
where
\begin{align*} 
\mc H_{0,\alpha}
&:=
  \begin{pmatrix}
  -\Delta_{\R^d}+1+V_{0,d}\alpha^{-2}  & 0 \\
0 & \Delta_{\R^d}-1-V_{0,d}\alpha^{-2}
  \end{pmatrix}, \\
\mc H_\alpha'
\begin{pmatrix}
  f_1 \\ f_2
\end{pmatrix}
&:=
\begin{pmatrix}
  U_\alpha & W_\alpha \\
-W_\alpha & -U_\alpha
\end{pmatrix}
\begin{pmatrix}
  f_1 \\ f_2
\end{pmatrix},
\end{align*} 
and
\begin{align*}
  U_\alpha(x)
  &:=-\tfrac{p+1}{2}\varphi_{d,p}(\alpha^{-1}x)\left
    |Q_{\R^d}(x)+\rho_\alpha(x)\right |^{p-1}
    +\alpha^{-2}[V_d(\alpha^{-1}x)-V_{0,d}], \\
  W_\alpha(x)
  &:=-\tfrac{p-1}{2}\varphi_{d,p}(\alpha^{-1}x)\left
    |Q_{\R^d}(x)+\rho_\alpha(x)\right |^{p-1}.
\end{align*}
Evidently, $\mc H_{0,\alpha}$ is self-adjoint, and
\[ \sigma(\mc
H_{0,\alpha})=\sigma_e(\mc
H_{0,\alpha})=(-\infty,-1-V_{0,d}\alpha^{-2}]\cup
[1+V_{0,d}\alpha^{-2},\infty). \]
Furthermore, $\mc H_\alpha'$ is bounded, and $U_\alpha, W_\alpha\in
L^\infty(\R^d)\cap C(\R^d)$ with
\[ \lim_{|x|\to\infty}U_\alpha(x)=\lim_{|x|\to\infty}W_\alpha(x)=0 \]
by Hypothesis \ref{hyp:A}, Proposition \ref{prop:exrho}, and Lemma
\ref{lem:Strauss}. By \cite{Tes14}, p.~201, Lemma 7.21,
we see that $\mc H_\alpha'(z-\mc H_{0,\alpha})^{-1}$ is compact for
any $z\in \rho(\mc H_{0,\alpha})$.  In other words, $\mc H_\alpha'$
is relatively compact with respect to $\mc H_{0,\alpha}$.
Consequently, by \cite{Sch02}, p.~173, Theorem 7.28,
\[ \sigma_e(\mc H_\alpha)=\sigma_e(\mc H_{0,\alpha}+\mc
  H_\alpha')=\sigma_e(\mc
  H_{0,\alpha})=(-\infty,-1-V_{0,d}\alpha^{-2}]\cup
[1+V_{0,d}\alpha^{-2},\infty), \]
and, since $\mc H_\alpha$ is
unitarily equivalent to $i\mc L_\alpha$, the statement on
$\sigma_e(\mc L_\alpha)$ follows.

From the identity
\[ z-\mc H_\alpha=\left [1-\mc H_\alpha'(z-\mc
    H_{0,\alpha})^{-1}\right ](z-\mc H_{0,\alpha}),\quad z\in \rho(\mc
  H_{0,\alpha}), \]
we infer that $z-\mc H_\alpha$ is invertible 
for $z\in \rho(\mc H_{0,\alpha})$ if and only if $1-\mc H_\alpha'(z-\mc H_{0,\alpha})^{-1}$
is invertible. By the self-adjointness of $\mc H_{0,\alpha}$ we have
the bound $\|(z-\mc H_{0,\alpha})^{-1}\|_{L^2(\R^d,\C^2)}\leq
|\Im z|^{-1}$ and thus, $z-\mc H_\alpha$ is certainly invertible
for all $z$ sufficiently far away from the real axis. Furthermore,
$\|\mc H_\alpha'\|_{L^2(\R^d,\C^2)}\lesssim 1$ for all
$\alpha\geq\alpha_0$, and thus,
there exists a $\mu>0$ such that
\[ \{z\in \C: \Re z=0, |\Im z|> \mu\}\subset \rho(\mc H_\alpha) \]
for all $\alpha\geq \alpha_0$.
Consequently, the analytic Fredholm
theorem (see e.g.~\cite{Sim15}, p.~194, Theorem 3.14.3) applied to $\mc
H_\alpha'(z-\mc H_{0,\alpha})^{-1}$ shows that $\sigma(\mc
H_\alpha)\setminus \sigma_e(\mc H_\alpha)$ consists of isolated
eigenvalues of finite algebraic multiplicities which do not accumulate
at any point outside of $\sigma_e(\mc H_\alpha)$.

Next, we turn to the proof that $\sigma(\mc L_\alpha)\subset \R\cup i\R$. Since
$\sigma_e(\mc L_\alpha)\subset i\R$ and $\sigma(\mc
L_\alpha)\setminus\sigma_e(\mc L_\alpha)$ consists of eigenvalues
only, it suffices to prove that $\sigma_p(\mc L_\alpha)\subset \R\cup
i\R$.
Furthermore, we may restrict ourselves to nonzero eigenvalues.
Let $\lambda\in \sigma_p(\mc L_{\alpha})\setminus \{0\}$ with eigenfunction $f=(f_1,f_2)\in H^2(\R^d,\C^2)$.
The eigenvalue equation $(\lambda-\mc L_\alpha)f=0$ is equivalent to
\begin{equation}
  \label{eq:evL}
  \begin{cases}
    \mc L_{\alpha,-} f_2=\lambda f_1, \\
    \mc L_{\alpha,+}f_1=-\lambda f_2.
    \end{cases}
  \end{equation}
  Let $\mc P_\alpha: L^2_\mathrm{rad}(\R^d)\to L^2_\mathrm{rad}(\R^d)$ be the orthogonal projection onto $\langle
  Q_{\R^d}+\rho_\alpha\rangle$ and set $\mc P_\alpha^\perp:=1-\mc
  P_\alpha$.
  Note that we must have $\mc P_\alpha^\perp f_2\not= 0$ because otherwise,
  \[ f_1=\tfrac{1}{\lambda}\mc L_{\alpha,-}(\mc P_\alpha f_2+\mc
    P_\alpha^\perp f_2)=\tfrac{1}{\lambda}\mc L_{\alpha,-}\mc P_\alpha
    f_2=0, \] by Proposition \ref{prop:L-nonneg}, and from the second
  equation in \eqref{eq:evL} we infer that $f_2=0$. This is a
  contradiction to $f=(f_1,f_2)$ being an eigenfunction.  Note further
  that
  $\mc P_\alpha$ is the spectral projection associated to the
  eigenvalue $0\in \sigma_p(\mc L_{\alpha,-})$ and thus,
  $\mc P_\alpha$ commutes with $\mc L_{\alpha,-}$. From the first
  equation in \eqref{eq:evL} and Proposition \ref{prop:L-nonneg} we
  obtain
 \[ 0\not= (\mc L_{\alpha,-}\mc P_\alpha^\perp f_2|\mc P_\alpha^\perp
  f_2)_{L^2(\R^d)}=(\mc L_{\alpha,-} f_2|f_2)_{L^2(\R^d)}=\lambda
  (f_1|f_2)_{L^2(\R^d)}, \]
and the second equation in \eqref{eq:evL} yields
\[ (\mc
  L_{\alpha,+}f_1|f_1)_{L^2(\R^d)}=-\lambda(f_2|f_1)_{L^2(\R^d)}=-\lambda\overline{(f_1|f_2)_{L^2(\R^d)}}. \]
Consequently, since $(f_1|f_2)_{L^2(\R^d)}\not= 0$,
\[ \lambda^2=-\frac{(\mc L_{\alpha,-}f_2|f_2)_{L^2(\R^d)}
    (\mc
    L_{\alpha,+}f_1|f_1)_{L^2(\R^d)}}{|(f_1|f_2)_{L^2(\R^d)}|^2}\in
  \R, \]
which implies that $\lambda\in \R\cup i\R$. From Eq.~\eqref{eq:evL} it
is also evident that $-\lambda\in \sigma_p(\mc L_\alpha)$.

Finally, by setting $\lambda=0$ in Eq.~\eqref{eq:evL}, we obtain from
Proposition \ref{prop:L-nonneg} and Lemma \ref{lem:L+inv} that
\[ \ker\mc L_\alpha=\left \langle \begin{pmatrix}0 \\
    Q_{\R^d}+\rho_\alpha\end{pmatrix}\right \rangle. \]
In particular, $0\in \sigma_p(\mc L_\alpha)$.
\end{proof}

Now we can show that the linear stability of the soliton in the curved
geometry is determined by the stability of the Euclidean ground state, at least if $p\not=1+\frac{4}{d}$.

\begin{lemma}
  \label{lem:linstab}
If $p\not=1+\frac{4}{d}$ then there exists an $\alpha_0>0$ such
 that for all
 $\alpha\geq\alpha_0$ the following holds.
 \begin{itemize}
   \item The algebraic multiplicity of $0\in \sigma_p(\mc L_\alpha)$
     equals $2$.
  \item If $p<1+\frac{4}{d}$, there are no positive eigenvalues of $\mc
    L_\alpha$.
    \item If $p>1+\frac{4}{d}$, there exists precisely one
      positive eigenvalue $\lambda_\alpha\in \sigma_p(\mc L_\alpha)$
      and the eigenvalues $\pm \lambda_\alpha\in \sigma_p(\mc L_\alpha)$ are simple.
  \end{itemize}
\end{lemma}

\begin{proof}
  Acoording to Lemma \ref{lem:basicLalpha}, there exists a $\mu>0$ such
  that $(-\infty,-\mu)\cup (\mu,\infty)\subset \rho(\mc L_\alpha)$ for
  all $\alpha\geq\alpha_0$.
  Let $\gamma: [0,1]\to\C$ be a simple, closed, smooth curve such that
  $\gamma(t)\in \rho(\mc L)$ for all $t\in [0,1]$ and $\gamma$
  encircles the interval $[-\mu-1,\mu+1]$ in such a way that 
  only real eigenvalues of $\mc L$ lie inside of $\gamma$. This is
  possible since $0\in \sigma_p(\mc L)$ is isolated. Let
  \[ \mc P:=\frac{1}{2\pi i}\int_\gamma (z-\mc L)^{-1}dz. \]
  Since $\mc L_\alpha-\mc L$ is bounded and converges to $0$ in norm
  as $\alpha\to\infty$ (see the proofs of Proposition
  \ref{prop:L-nonneg} and Lemma \ref{lem:L+inv}), $\gamma(t)\in \rho(\mc L_\alpha)$ for all
  $t\in [0,1]$ and all
  $\alpha\geq \alpha_0$, provided $\alpha_0>0$ is sufficiently large.
  Consequently,
  \[ \mc P_\alpha:=\frac{1}{2\pi i}\int_\gamma (z-\mc L_\alpha)^{-1}dz \]
  is well defined, and $\mc P_\alpha\to \mc P$ in norm as
  $\alpha\to\infty$. This implies
  \[ \dim\rg \mc P_\alpha=\dim\rg \mc P=
    \begin{cases}
      2, & p<1+\frac{4}{d}, \\
      4, & p>1+\frac{4}{d},
    \end{cases}
  \]
  by Theorem \ref{thm:L}.
Suppose now that $p<1+\frac{4}{d}$ and there exists a positive eigenvalue
$\lambda_\alpha\in \sigma_p(\mc L_\alpha)$. Then, by Lemma
\ref{lem:basicLalpha}, $-\lambda_\alpha\in \sigma_p(\mc L_\alpha)$
and, since $0\in \sigma_p(\mc L_\alpha)$, we must have
$\dim\rg\mc P_\alpha\geq 3$. This contradicts $\dim\rg\mc P_\alpha=2$ and thus, there
can be no positive eigenvalue of $\mc L_\alpha$ in the case $p<1+\frac{4}{d}$.
If $p>1+\frac{4}{d}$, there exists a unique positive eigenvalue
$\lambda\in \sigma_p(\mc L)$ with algebraic multiplicity $1$ (Theorem
\ref{thm:L}).
Let $\widetilde \gamma: [0,1]\to \rho(\mc L)\subset \C$ be a simple, closed, smooth curve
that encircles the interval $[\frac{\lambda}{2},\mu+1]$ and such that
$\lambda$ is the only spectral point of $\mc L$ that lies inside of $\widetilde\gamma$.
Set
\[ \widetilde{\mc P}:=\frac{1}{2\pi i}\int_{\widetilde \gamma}(z-\mc
  L)^{-1}dz. \]
As above,
\[ \widetilde{\mc P}_\alpha:=\frac{1}{2\pi i}\int_{\widetilde
    \gamma}(z-\mc L_\alpha)^{-1}dz \]
is well defined for sufficiently large $\alpha$ and
$\dim\rg\widetilde{\mc P}_\alpha=\dim\rg\widetilde{\mc P}=1$.
Consequently, there exists a positive simple eigenvalue $\lambda_\alpha\in
\sigma_p(\mc L_\alpha)$ and by Lemma \ref{lem:basicLalpha},
$-\lambda_\alpha\in \sigma_p(\mc L_\alpha)$. Furthermore, by symmetry,
$-\lambda_\alpha\in \sigma_p(\mc L_\alpha)$ must be simple, too. Since $\dim\rg\mc
P_\alpha=4$ and $0\in \sigma_p(\mc L_\alpha)$, there can be no other
nonzero eigenvalues in $[-\mu-1,\mu+1]$ as they would have to
come in pairs. Since
$(\mu,\infty)\subset \rho(\mc L_\alpha)$, it follows
that there exists a unique simple positive eigenvalue
$\lambda_\alpha\in \sigma_p(\mc L_\alpha)$.
In particular,  the algebraic multiplicity of $0\in
\sigma_p(\mc L_\alpha)$ must equal $2$.
\end{proof}

\section{Spectral stability in the critical case}
\noindent In the critical case $p=1+\frac{4}{d}$, the situation is subtle and
the stability of the soliton
depends on the fine structure of the geometry. 

\subsection{Refined properties of $\mc L_{\alpha,+}$}
\begin{lemma}
  \label{lem:L+spec}
There exists an $\alpha_0>0$ such that for all $\alpha\geq \alpha_0$
the following holds. We have
\[ \sigma_e(\mc L_{\alpha,+})=[1+V_{0,d}\alpha^{-2},\infty), \]
$\mc L_{\alpha,+}$ has precisely one negative eigenvalue
$\lambda_\alpha^*<0$, and this eigenvalue is
simple. Furthermore, if $f_\alpha^*\in H^2_\mathrm{rad}(\R^d)\setminus \{0\}$ is an
associated eigenfunction, i.e., $\mc L_{\alpha,+}f_\alpha^*=\lambda_\alpha^*
f_\alpha^*$, then we have
\[ \big (f_\alpha^* \big | Q_{\R^d}+\rho_\alpha\big )_{L^2(\R^d)}\not=
  0. \]
\end{lemma}

\begin{proof}
  As in the proof of Proposition \ref{prop:L-nonneg}, we write $\mc
  L_{\alpha,+}f=\mc L_{\alpha,0}f+W_\alpha f$, where $\mc
  L_{\alpha,0}: H^2_\mathrm{rad}(\R^d)\subset
  L^2_\mathrm{rad}(\R^d)\to L^2_\mathrm{rad}(\R^d)$ is given by $\mc
  L_{\alpha,0}f=-\Delta_{\R^d}f+(1+V_{0,d}\alpha^{-2})f$, and
  \[ W_\alpha(x):=-p\varphi_{d,p}(\alpha^{-1}x)\left
      |Q_{\R^d}(x)+\rho_\alpha(x)\right
    |^{p-1}+\alpha^{-2}\left [V_d(\alpha^{-1}x)-V_{0,d}\right]. \]
  Thus, by repeating the argument from the proof of Proposition
  \ref{prop:L-nonneg}, we find $\sigma_e(\mc
  L_{\alpha,+})=[1+V_{0,d}\alpha^{-2},\infty)$, and there exists a
  $\mu>0$ such that $(-\infty,-\mu)\subset \rho(\mc L_{\alpha,+})$ for
  all $\alpha$ sufficiently large.

We define $\mc C_\alpha: L^2_\mathrm{rad}(\R^d)\to L^2_\mathrm{rad}(\R^d)$ by
\begin{align*}
 \mc C_\alpha f(x)
&=\mc L_{\alpha,+}f(x)-\mc L_+ f(x) \\
&=\left [-p\varphi_{d,p}(\alpha^{-1}x)\left
                                         |Q_{\R^d}(x)+\rho_\alpha(x)\right
                                         |^{p-1}+p|Q_{\R^d}(x)|^{p-1}+\alpha^{-2}V_d(\alpha^{-1}x)\right ]f(x),
\end{align*}
As in the proof of Proposition \ref{prop:L-nonneg}, we infer that
$\mc C_\alpha\to 0$ in norm as $\alpha\to\infty$.
Let $\gamma: [0,1]\to \C$ be given by $\gamma(t)=-\mu+\mu e^{2\pi i
  t}$. Then, by Lemma \ref{lem:L+inv} and the above, $\gamma(t)\in \rho(\mc
L_{\alpha,+})$ for all $t\in [0,1)$ and all $\alpha\geq \alpha_0$,
provided $\alpha_0>0$ is sufficiently large. Define the spectral
projections
\[ \mc P:=\frac{1}{2\pi i}\int_\gamma (z-\mc L_+)^{-1}dz,\qquad \mc
  P_\alpha:=\frac{1}{2\pi i}\int_\gamma (z-\mc
  L_{\alpha,+})^{-1}dz. \]
Then $\mc P_\alpha\to \mc P$ in norm as
$\alpha\to\infty$. Furthermore, $\mc L_+$ has precisely one simple
eigenvalue inside of $\gamma$ \cite{Wei85, ChaGusNakTsa07}, which implies that $\dim\rg\mc P=1$.
Consequently,
from \cite{Kat95}, p.~34, Lemma 4.10, we conclude that $\dim\rg
\mc P_\alpha=\dim\rg\mc P=1$ for all $\alpha\geq\alpha_0$. In
conjunction with $(-\infty,-\mu)\subset \rho(\mc L_{\alpha,+})$ and
the self-adjointness of $\mc L_{\alpha,+}$ (Lemma \ref{lem:L+inv}), this
means that $\mc L_{\alpha,+}$ has precisely one negative eigenvalue $\lambda_\alpha^*<0$,
and this eigenvalue is simple.

Let $f^*\in \rg \mc P$ with $\|f^*\|_{L^2(\R^d)}=1$ and set $f_\alpha^*:=\mc P_\alpha
f^*$.
Then $f^*$ is an eigenfunction of $\mc L_+$ to the eigenvalue
$\lambda^*<0$. Furthermore,
\[ \|f_\alpha^*-f^*\|_{L^2(\R^d)}=\|(\mc P_\alpha-\mc
  P)f^*\|_{L^2(\R^d)}\to 0 \]
as $\alpha\to\infty$ and thus, $f_\alpha^*\not= 0$ for all
$\alpha\geq\alpha_0$ if $\alpha_0>0$ is sufficiently large.
As a consequence, $f_\alpha^*$ is an eigenfunction of $\mc
L_{\alpha,+}$ with eigenvalue $\lambda_\alpha^*$ and any other
eigenfunction to this eigenvalue is a multiple of $f_\alpha^*$.
Since $\lambda^*$ is the only negative eigenvalue of $\mc L_+$, it
follows by Sturm oscillation theory that $f^*$ does not have zeros. In
particular, $(f^*|Q_{\R^d})_{L^2(\R^d)}\not= 0$.  The fact that
\[ (f_\alpha^*|Q_{\R^d}+\rho_\alpha)_{L^2(\R^d)}\to
  (f^*|Q_{\R^d})_{L^2(\R^d)} \]
as $\alpha\to\infty$ thus implies that
$(f_\alpha^*|Q_{\R^d}+\rho_\alpha)_{L^2(\R^d)}\not= 0$ for all
$\alpha\geq\alpha_0$, provided $\alpha_0>0$ is sufficiently large.
\end{proof}

\begin{lemma}
  \label{lem:PLPsa}
  Let $\alpha>0$ be sufficiently large and denote by $\mc
  P_\alpha^\perp$ the orthogonal projection onto $\langle
  Q_{\R^d}+\rho_\alpha\rangle^\perp$, i.e.,
  \[ \mc P_\alpha^\perp
    f:=f-\frac{(f|Q_{\R^d}+\rho_\alpha)_{L^2(\R^d)}}{\|Q_{\R^d}+\rho_\alpha\|_{L^2(\R^d)}^2}
    (Q_{\R^d}+\rho_\alpha). \]
  Then the operator $\mc P_\alpha^\perp \mc L_{\alpha,+}\mc
  P_\alpha^\perp: H^2_\mathrm{rad}(\R^d)\subset 
  L_\mathrm{rad}^2(\R^d)\to L^2_\mathrm{rad}(\R^d)$ is self-adjoint, and 
  \[ \sigma_e(\mc P_\alpha^\perp \mc L_{\alpha,+}\mc
    P_\alpha^\perp)=[1+V_{0,d}\alpha^{-2},\infty). \]
\end{lemma}

\begin{proof}
The proof is based on the standard trick (see e.g.~\cite{Kat95},
p.~246) of using the decomposition
\[ \mc P_\alpha^\perp \mc L_{\alpha,+}\mc P_\alpha^\perp=\mc
  L_{\alpha,+}+(\mc P_\alpha^\perp-1) \mc L_{\alpha,+}\mc
  P_\alpha^\perp+\mc L_{\alpha,+}(\mc P_\alpha^\perp-1)=:\mc
  L_{\alpha,+}+\mc K_\alpha. \]
Since $\dim\rg(\mc P_\alpha^\perp-1)=1$, the operator $\mc K_\alpha:
H_\mathrm{rad}^2(\R^d)\subset L_\mathrm{rad}^2(\R^d)\to
L_\mathrm{rad}^2(\R^d)$ has finite rank.  The estimate,
\begin{align*}
  \|\mc K_\alpha f\|_{L^2(\R^d)}
  &\leq \frac{|(\mc L_{\alpha,+}\mc P_\alpha^\perp f
    |\widetilde R_\alpha)_{L^2(\R^d)}|}{\|\widetilde R_\alpha\|_{L^2(\R^d)}^2} \|\widetilde
    R_\alpha\|_{L^2(\R^d)}+\frac{|(f|\widetilde
    R_\alpha)_{L^2(\R^d)}|}{\|\widetilde
    R_\alpha\|_{L^2(\R^d)}^2}\|\mc L_{\alpha,+}\widetilde
    R_\alpha\|_{L^2(\R^d)} \\
  &\leq \frac{(f|\mc P_\alpha^\perp \mc L_{\alpha,+}\widetilde
    R_\alpha)_{L^2(\R^d)}}{\|\widetilde R_\alpha\|_{L^2(\R^d)}}
    +\frac{\|\mc L_{\alpha,+}\widetilde R_\alpha
    \|_{L^2(\R^d)}}{\|\widetilde
    R_\alpha\|_{L^2(\R^d)}}\|f\|_{L^2(\R^d)} \\
  &\lesssim \frac{\|\mc L_{\alpha,+}\widetilde R_\alpha
    \|_{L^2(\R^d)}}{\|\widetilde
    R_\alpha\|_{L^2(\R^d)}}\|f\|_{L^2(\R^d)}
\end{align*}
for all $f\in H^2_\mathrm{rad}(\R^d)$, where $\widetilde
R_\alpha:=Q_{\R^d}+\rho_\alpha$, shows that $\mc K_\alpha$ extends to
a bounded operator $\mc K_\alpha: L^2_\mathrm{rad}(\R^d)\to
L^2_\mathrm{rad}(\R^d)$ of finite rank.
In particular, $\mc K_\alpha$ is compact.
Furthermore,
\[ \mc K_\alpha^*=\mc P_\alpha^\perp \mc L_{\alpha,+}(\mc
  P_\alpha^\perp-1)
  +(\mc P_\alpha^\perp-1)\mc L_{\alpha,+}
=\mc P_\alpha^\perp \mc L_{\alpha,+}\mc P_\alpha^\perp-\mc
L_{\alpha,+}=\mc K_\alpha, \]
and thus $\mc K_\alpha$ is self-adjoint. By the Kato-Rellich
theorem (see e.g.~\cite{Tes14}, p,~159, Theorem 10.2) it follows that
$\mc P_\alpha^\perp \mc L_{\alpha,+}\mc P_\alpha^\perp$ is
self-adjoint.  Weyl's theorem (see e.g.~\cite{Tes14}, p.~171,
Theorem 6.19), in conjunction with Lemma \ref{lem:L+spec},
yields the statement on the essential spectrum.
\end{proof}

Next, we establish a crucial dichotomy for $\mc L_{\alpha,+}$.

\begin{proposition}
  \label{prop:L+nonneg}
  Let $\alpha>0$ be sufficiently large.
  \begin{itemize}
    \item If $(\mc L_{\alpha,+}^{-1}(Q_{\R^d}+\rho_\alpha)
        |Q_{\R^d}+\rho_\alpha
     )_{L^2(\R^d)}>0$ then there exists an $f_\alpha\in \langle
     Q_{\R^d}+\rho_\alpha\rangle^\perp\cap H^2_{\mathrm{rad}}(\R^d)$ such that
     \[ (\mc L_{\alpha,+}f_\alpha|f_\alpha)_{L^2(\R^d)}<0. \]
  \item If
    $(\mc L_{\alpha,+}^{-1}(Q_{\R^d}+\rho_\alpha)
        |Q_{\R^d}+\rho_\alpha
     )_{L^2(\R^d)}\leq 0$ then
\[ (\mc L_{\alpha,+}f|f)_{L^2(\R^d)}\geq 0 \]
for all $f\in \langle Q_{\R^d}+\rho_\alpha\rangle^\perp \cap H^2_\mathrm{rad}(\R^d)$.
\end{itemize}
\end{proposition}

\begin{proof}
  We first assume that
  \[\left. \left (\mc L_{\alpha,+}^{-1}\widetilde R_\alpha
    \right |\widetilde R_\alpha \right )_{L^2(\R^d)}>0, \]
  where $\widetilde
        R_\alpha=Q_{\R^d}+\rho_\alpha$.
Let $\mc P_\alpha^\perp$ be the orthogonal projection on $\langle
\widetilde R_\alpha\rangle^\perp$, i.e.,
\[ \mc P_{\alpha}^\perp f=f-\frac{(f|\widetilde
    R_\alpha)_{L^2(\R^d)}}{\|\widetilde
    R_\alpha\|_{L^2(\R^d)}^2}\widetilde R_\alpha. \]
We set $f_\alpha:=\mc P_\alpha^\perp \mc L_{\alpha,+}^{-1}\widetilde
R_\alpha\in \langle \widetilde R_\alpha\rangle^\perp\cap H^2_\mathrm{rad}(\R^d)$.
Then we have
\begin{align*}
  (\mc L_{\alpha,+}f_\alpha|f_\alpha)_{L^2(\R^d)}
&=\Big (\mc L_{\alpha,+}\mc P_\alpha^\perp \mc L_{\alpha,+}^{-1}\widetilde
R_\alpha\Big |\mc P_\alpha^\perp \mc L_{\alpha,+}^{-1}\widetilde
R_\alpha\Big )_{L^2(\R^d)}\\
&= \Big (\widetilde R_\alpha \Big | \mc L_{\alpha,+}^{-1}\widetilde
                                 R_\alpha\Big )_{L^2(\R^d)}
-\frac{(\mc L_{\alpha,+}^{-1}\widetilde R_\alpha|\widetilde
                                 R_\alpha)_{L^2(\R^d)}}{\|\widetilde
                                 R_\alpha\|_{L^2(\R^d)}^2}
\Big (\mc L_{\alpha,+}\widetilde R_\alpha \Big | \mc
                                 L_{\alpha,+}^{-1}\widetilde R_\alpha
                                 \Big )_{L^2(\R^d)} \\
&\quad -\frac{(\mc L_{\alpha,+}^{-1}\widetilde R_\alpha | 
\widetilde R_\alpha)_{L^2(\R^d)}}{\|\widetilde
R_\alpha\|_{L^2(\R^d)}^2}
\Big (\widetilde R_\alpha \Big | \widetilde R_\alpha\Big )_{L^2(\R^d)} \\
&\quad +\frac{(\mc L_{\alpha,+}^{-1}\widetilde R_\alpha | \widetilde
R_\alpha )_{L^2(\R^d)}^2}{\|\widetilde R_\alpha\|_{L^2(\R^d)}^4}
\Big (\mc L_{\alpha,+}\widetilde R_\alpha \Big | \widetilde
                                                         R_\alpha \Big
                                                                            )_{L^2(\R^d)}
  \\
&=-\Big ( \mc L_{\alpha,+}^{-1}\widetilde
                                 R_\alpha
       \Big | \widetilde R_\alpha \Big )_{L^2(\R^d)}
       +\frac{(\mc L_{\alpha,+}^{-1}\widetilde R_\alpha | \widetilde
R_\alpha )_{L^2(\R^d)}^2}{\|\widetilde R_\alpha\|_{L^2(\R^d)}^4}
\Big (\mc L_{\alpha,+}\widetilde R_\alpha \Big | \widetilde
                                                         R_\alpha \Big
       )_{L^2(\R^d)} \\
&<\frac{(\mc L_{\alpha,+}^{-1}\widetilde R_\alpha | \widetilde
R_\alpha )_{L^2(\R^d)}^2}{\|\widetilde R_\alpha\|_{L^2(\R^d)}^4}
\Big (\mc L_{\alpha,+}\widetilde R_\alpha \Big | \widetilde
                                                         R_\alpha \Big
                                                                            )_{L^2(\R^d)}.       
\end{align*}
Thus, it suffices to show that $(\mc L_{\alpha,+}\widetilde
R_\alpha|\widetilde R_\alpha)_{L^2(\R^d)}\leq 0$. Explicitly, we have
\begin{align*}
  \mc L_{\alpha,+}\widetilde R_\alpha(x)
  &=\mc L_{\alpha,-}\widetilde
  R_\alpha(x)
   -(p-1)\varphi_{d,p}(\alpha^{-1}x)\left |\widetilde R_\alpha(x)\right
    |^{p-1}\widetilde R_\alpha(x) \\
  &=-(p-1)\varphi_{d,p}(\alpha^{-1}x)\left |\widetilde R_\alpha(x)\right
    |^{p-1}\widetilde R_\alpha(x),
\end{align*}
and thus,
\[ \Big (\mc L_{\alpha,+}\widetilde R_\alpha \Big |
  \widetilde R_\alpha \Big
  )_{L^2(\R^d)}=-(p-1)\int_{\R^d}\varphi_{d,p}(\alpha^{-1}x)
  \left |\widetilde R_\alpha(x)\right |^{p+1} dx< 0, \]
since $\varphi_{d,p}\geq 0$ by Hypothesis \ref{hyp:A}. In summary,
$(\mc L_{\alpha,+}f_\alpha|f_\alpha)_{L^2(\R^d)}<0$, as claimed.

Next, we assume that
\[ \Big (\mc L_{\alpha,+}^{-1}\widetilde R_\alpha \Big | \widetilde R_\alpha
  \Big )_{L^2(\R^d)}\leq 0. \]
Suppose there exists an $f_\alpha\in \langle \widetilde
R_\alpha\rangle^\perp \cap H^2_\mathrm{rad}(\R^d)$ such that $(\mc
L_{\alpha,+}f_\alpha | f_\alpha)_{L^2(\R^d)}<0$.
Consider the operator $\mc P_\alpha^\perp \mc L_{\alpha,+}\mc
P_\alpha^\perp$. 
By assumption, we have
\[ 0>\Big (\mc L_{\alpha,+}f_\alpha \Big | f_\alpha \Big)_{L^2(\R^d)}
  =\Big (\mc L_{\alpha,+}\mc P_\alpha^\perp f_\alpha \Big | \mc
  P_\alpha^\perp f_\alpha \Big)_{L^2(\R^d)}
=\Big (\mc P_\alpha^\perp \mc L_{\alpha,+}\mc P_\alpha^\perp f_\alpha \Big | 
f_\alpha \Big)_{L^2(\R^d)}, \]
and thus, by Lemma \ref{lem:PLPsa}, 
$\mc P_\alpha^\perp \mc L_{\alpha,+}\mc P_\alpha^\perp$
must have a negative eigenvalue $\lambda_\alpha<0$. In other words, there
exists a nontrivial $g_\alpha\in \langle \widetilde
R_\alpha\rangle^\perp \cap H^2_\mathrm{rad}(\R^d)$
such that $\mc P_\alpha^\perp \mc L_{\alpha,+}g_\alpha=\lambda_\alpha g_\alpha$.
This means that there exists a $c_\alpha\in \C$ such that
\[ \mc L_{\alpha,+}g_\alpha=\lambda_\alpha g_\alpha+c_\alpha\widetilde
  R_\alpha. \]
We claim that $c_\alpha\not= 0$.
To see this, recall that $\mc L_{\alpha,+}$ has a unique negative eigenvalue
$\lambda_\alpha^*<0$ (which is simple) and if $f_\alpha^*$ is an associated 
eigenfunction, we have $(f_\alpha^* | \widetilde R_\alpha)_{L^2(\R^d)}\not= 0$, see Lemma \ref{lem:L+spec}.
Suppose now that $c_\alpha=0$. Then $g_\alpha$ is an eigenfunction
of $\mc L_{\alpha,+}$ with negative eigenvalue $\lambda_\alpha$, and thus,
$\lambda_\alpha=\lambda_\alpha^*$ and $g_\alpha$ must be a multiple of
$f_\alpha^*$. This, however, contradicts $(g_\alpha | \widetilde
R_\alpha)_{L^2(\R^d)}=0$, and the claim $c_\alpha\not= 0$
follows. 
Note further that $\lambda_\alpha\not=\lambda_\alpha^*$ because
otherwise we would arrive at the contradiction
\[ 0=\big (g_\alpha \big | (\mc L_{\alpha,+}-\lambda_\alpha^*)f_\alpha^*\big
  )_{L^2(\R^d)}=\big ((\mc L_{\alpha,+}-\lambda_\alpha^*)g_\alpha
  \big | f_\alpha^*\big )_{L^2(\R^d)}=c_\alpha \big (\widetilde
  R_\alpha \big | f_\alpha^*\big )_{L^2(\R^d)}\not= 0. \]
Consequently, we 
have
\[ (\mc
  L_{\alpha,+}-\lambda_\alpha)^{-1}\widetilde R_\alpha=\tfrac{1}{c_\alpha}g_\alpha.
\]
Furthermore,
\[ \Big ((\mc L_{\alpha,+}-\lambda_\alpha)^{-1}\widetilde R_\alpha
  \Big | \widetilde R_\alpha \Big )_{L^2(\R^d)}=\tfrac{1}{c_\alpha}(g_\alpha
  | \widetilde R_\alpha)_{L^2(\R^d)}=0. \]
Now we define a function $\phi_\alpha: (-\infty,0]\setminus
\{\lambda_\alpha^*\}\to \R$ by
\[ \phi_\alpha(\lambda):=\Big ((\mc L_{\alpha,+}-\lambda)^{-1}\widetilde R_\alpha
  \Big | \widetilde R_\alpha \Big )_{L^2(\R^d)}. \]
Note that $\phi_\alpha$ is differentiable and
\begin{align*} \phi_\alpha'(\lambda)
  &=\Big ((\mc
  L_{\alpha,+}-\lambda)^{-2}\widetilde R_\alpha \Big | \widetilde
  R_\alpha \Big )_{L^2(\R^d)}
=\Big ((\mc
L_{\alpha,+}-\lambda)^{-1}\widetilde R_\alpha \Big |
(\mc
  L_{\alpha,+}-\lambda)^{-1}\widetilde
    R_\alpha \Big )_{L^2(\R^d)} \\
  &=\Big \|(\mc L_{\alpha,+}-\lambda)^{-1}\widetilde R_\alpha\Big \|_{L^2(\R^d)}^2>0
\end{align*}
for all $\lambda\in (-\infty,0]\setminus \{\lambda_\alpha^*\}$.
By assumption,
\[ \phi_\alpha(0)=\Big (\mc L_{\alpha,+}^{-1}\widetilde R_\alpha \Big | \widetilde
  R_\alpha \Big )_{L^2(\R^d)}\leq 0 \]
and $\phi_\alpha(\lambda_\alpha)=0$. Thus, we must have
$\lambda_\alpha<\lambda_\alpha^*$ since otherwise, we would arrive at the
contradiction
\[ 0\geq \phi_\alpha(0)=\int_{\lambda_\alpha}^0
  \phi_\alpha'(\lambda)d\lambda+\underbrace{\phi_\alpha(\lambda_\alpha)}_{=0}>0. \]
However, $\lambda_\alpha<\lambda_\alpha^*$ is also impossible since it
leads to the contradiction
\begin{align*} 0
  &=\phi_\alpha(\lambda_\alpha)=\Big ((\mc
  L_{\alpha,+}-\lambda_\alpha)^{-1}\widetilde R_\alpha \Big |
    \widetilde R_\alpha \Big )_{L^2(\R^d)}
    =\Big ((\mc L_{\alpha,+}-\lambda_\alpha)\widetilde S_\alpha \Big |
    \widetilde S_\alpha \Big )_{L^2(\R^d)} \\
  &=\Big ((\mc L_{\alpha,+}-\lambda_\alpha^*)\widetilde S_\alpha \Big
    |
    \widetilde S_\alpha\Big
    )_{L^2(\R^d)}+(\lambda_\alpha^*-\lambda_\alpha)\big \|\widetilde
    S_\alpha\big \|_{L^2(\R^d)}^2 \\
  &>0,
\end{align*}
where $\widetilde S_\alpha:=(\mc
L_{\alpha,+}-\lambda_\alpha)^{-1}\widetilde R_\alpha$ and we have used
the fact that $\mc L_{\alpha,+}-\lambda_\alpha^*$ is nonnegative, see
Lemma \ref{lem:L+spec}. In summary, we see that there cannot exist an
$f_\alpha\in \langle \widetilde R_\alpha\rangle^\perp\cap
H^2_\mathrm{rad}(\R^d)$ with $(\mc L_{\alpha,+}f_\alpha |
f_\alpha)_{L^2(\R^d)}<0$, and this finishes the proof.
\end{proof}

\subsection{The auxiliary operator $\mc L_{\alpha,-}^\frac12 \mc
  L_{\alpha,+}\mc L_{\alpha,-}^\frac12$}

By Proposition \ref{prop:L-nonneg}, $\mc L_{\alpha,-}$ is nonnegative
and thus, the square root $\mc L_{\alpha,-}^\frac12$ is well defined
either via the functional calculus for self-adjoint operators or by the
Dunford-Taylor integral, see e.g.~\cite{Kat95}, p.~281, Theorem
3.35. Furthermore, since $\mc L_{\alpha,-}^\frac12$ is self-adjoint,
we have
\[ \ker\mc L_{\alpha,-}^\frac12=\ker \mc L_{\alpha,-}=\left
    \langle Q_{\R^d}+\rho_\alpha\right\rangle \]
and $\rg \mc
L_{\alpha,-}^\frac12=\langle Q_{\R^d}+\rho_\alpha\rangle^\perp$ by Proposition \ref{prop:L-nonneg}.
As expected from the Euclidean case, the auxiliary operator
$\mc L_{\alpha,-}^\frac12
  \mc L_{\alpha,+}\mc L_{\alpha,-}^\frac12$ plays a crucial role.

\begin{definition}
  \label{def:J}
  Let $\alpha>0$ be sufficiently large.
  We define an operator
  \[ \mc J_\alpha: \mc D(\mc J_\alpha)\subset \langle
  Q_{\R^d}+\rho_\alpha\rangle^\perp \to \langle Q_{\R^d}+\rho_\alpha
  \rangle^\perp \] by
\[  \mc D(\mc J_\alpha):=\left \{f\in \mc D(\mc L_{\alpha,-}^\frac12)\cap
    \langle Q_{\R^d}+\rho_\alpha\rangle^\perp: \mc
    L_{\alpha,-}^\frac12 f\in \mc D(\mc L_{\alpha,+}) \mbox{ and }\mc
    L_{\alpha,+}\mc L_{\alpha,-}^\frac12 f\in \mc D(\mc
    L_{\alpha,-}^\frac12)\right \} \]
and
$\mc J_\alpha f:=\mc L_{\alpha,-}^\frac12 \mc L_{\alpha,+}\mc
  L_{\alpha,-}^\frac12 f$.
\end{definition}

It is not immediately obvious that $\mc J_\alpha$ is densely
defined. Thus, we first establish this fact using the following simple
property of maximally defined products.

\begin{lemma}
  \label{lem:AB}
  Let $(X,\|\cdot|_X)$, $(Y,\|\cdot\|_Y)$, and $(Z,\|\cdot\|_Z)$ be
  Banach spaces. Furthermore, let $A: \mc D(A)\subset Y\to Z$ and $B:
  \mc D(B)\subset X\to Y$ be densely defined linear operators and
  assume that $B$ is bounded invertible.
Then the maximally defined operator\footnote{That is to
      say, $\mc D(AB):=\{x\in \mc D(B): Bx\in \mc D(A)\}$.} $AB: \mc
    D(AB)\subset X\to Z$ is densely
    defined and $\mc D(AB)$ is a core for $B$.
\end{lemma}

\begin{proof}
   Let $x\in X$ and $\epsilon>0$ be arbitrary. Since $\mc D(B)$
    is dense in $X$, we can find an $x'\in \mc D(B)$ such that
    $\|x-x'\|_X<\frac{\epsilon}{2}$. By the density of $\mc D(A)$ in
    $Y$, there exists a $\widetilde y\in \mc D(A)$ such that
    $\|Bx'-\widetilde y\|_Y<\frac{\epsilon}{2}\|B^{-1}\|_{\mc B(X,Y)}^{-1}$.
    Set $\widetilde x:=B^{-1}\widetilde y$. By definition, $\widetilde
    x\in \mc D(AB)$ and
    \begin{align*}
      \|x-\widetilde x\|_X
      &=\|x-x'\|_X+\|x'-\widetilde
        x\|_X<\tfrac{\epsilon}{2}+\|B^{-1}(Bx'-B\widetilde x)\|_X \\
      &\leq
     \tfrac{\epsilon}{2}+ \|B^{-1}\|_{\mc B(X,Y)}\|Bx'-\widetilde
        y\|_Y <\epsilon.
    \end{align*}

To prove the second assertion, let $x\in \mc D(B)$. We have to show
that there exists a sequence $(x_n)_{n\in\N}\subset \mc D(AB)$ such
that $x_n\to x$ in $X$ and $Bx_n\to Bx$ in $Y$ as $n\to\infty$.
Since
 $\mc D(A)$ is dense in $Y$, there exists a sequence
$(y_n)_{n\in\N}\subset \mc D(A)$ such that $y_n\to Bx$ in $Y$ as
$n\to\infty$.
We set $x_n:=B^{-1}y_n$. Then $(x_n)_{n\in\N}\subset \mc D(AB)$ and we
have $Bx_n\to Bx$ in $Y$ as well as
\[ \|x_n-x\|_X=\|B^{-1}(Bx_n-Bx)\|_X\lesssim \|y_n-Bx\|_Y\to 0 \]
as $n\to\infty$.
\end{proof}

\begin{lemma}
  \label{lem:Jdense}
  Let $\alpha>0$ be sufficiently large.
  Then the operator $\mc J_\alpha$
is densely defined.
\end{lemma}

\begin{proof}
  To begin with, we define an auxiliary operator $\mc A_\alpha: \mc
  D(\mc A_\alpha)\subset L^2_\mathrm{rad}(\R^d)\to \langle
  Q_{\R^d}+\rho_\alpha\rangle^\perp$ by
  \[ \mc D(\mc A_\alpha):=\left \{ f\in \mc D(\mc
      L_{\alpha,+})=H^2_\mathrm{rad}(\R^d): \mc L_{\alpha,+}f\in \mc
      D(\mc L_{\alpha,-}^\frac12)\right \} \]
  and $\mc A_\alpha f:=\mc L_{\alpha,-}^\frac12 \mc L_{\alpha,+}f$.
  Since $0\notin \sigma(\mc L_{\alpha,+})$ by Lemma \ref{lem:L+inv},
  Lemma \ref{lem:AB} shows that $\mc A_\alpha$ is densely defined.
Next, we define another auxiliary operator $\mc B_\alpha: \mc D(\mc B_\alpha)\subset \langle
Q_{\R^d}+\rho_\alpha\rangle^\perp \to \langle
Q_{\R^d}+\rho_\alpha\rangle^\perp$
by
\[ \mc D(\mc B_\alpha):=\mc D(\mc L_{\alpha,-}^\frac12)
    \cap \langle Q_{\R^d}+\rho_\alpha\rangle^\perp \]
  and $\mc B_\alpha f:=\mc L_{\alpha,-}^\frac12 f$.
  Obviously, $\mc B_\alpha$ is densely defined and,
  since $\ker \mc L_{\alpha,-}^\frac12=\langle Q_{\R^d}+\rho_\alpha
  \rangle$, it follows that $\mc B_\alpha$ is injective.
  Furthermore, by the self-adjointness of $\mc L_{\alpha,-}^\frac12$,
  $\rg \mc L_{\alpha,-}^\frac12=\langle
  Q_{\R^d}+\rho_\alpha\rangle^\perp$ and thus, for any $g\in \langle
  Q_{\R^d}+\rho_\alpha\rangle^\perp$ we can find an $f\in \mc D(\mc
  L_{\alpha,-}^\frac12)$ such that $\mc L_{\alpha,-}^\frac12 \mc P_\alpha^\perp
  f=\mc L_{\alpha,-}^\frac12 f=g$, where $\mc P_\alpha^\perp$ denotes the
  orthogonal projection on $\langle
  Q_{\R^d}+\rho_\alpha\rangle^\perp$. 
Consequently, $\mc P_\alpha^\perp f\in \mc D(\mc B_\alpha)$ and $\mc
B_\alpha \mc P_\alpha^\perp f=g$. This shows that $\mc B_\alpha$ is
surjective. From the closedness of $\mc L_{\alpha,-}^\frac12$ it follows
immediately that $\mc B_\alpha$ is closed and the closed graph
theorem implies that $\mc
B_\alpha$ is bounded invertible. Now observe that $\mc J_\alpha=\mc
A_\alpha \mc B_\alpha$, where the product $\mc A_\alpha \mc B_\alpha$
is maximally defined. Consequently, Lemma \ref{lem:AB} implies that
$\mc J_\alpha$ is densely defined. 
\end{proof}

\begin{remark}
  \label{rem:Jcore}
  Lemma \ref{lem:AB} also shows that $\mc D(\mc J_\alpha)$ is a core for
  the operator $\mc B_\alpha$ defined in the proof of Lemma \ref{lem:Jdense}.
\end{remark}

The importance of $\mc J_\alpha$
derives from the following observation.

\begin{lemma}
  \label{lem:Jinjsur}
  Let $\alpha>0$ be sufficiently large and $\lambda\in \C\setminus
  \{0\}$. Then we have the following implications.
  \begin{itemize}
  \item If $\lambda \in \rho(\mc L_\alpha)$ then 
    $\lambda^2+\mc J_\alpha$ is surjective.
    \item The operator $\lambda-\mc L_\alpha$ is injective if and only if
      $\lambda^2+\mc J_\alpha$ is injective.
  \end{itemize}
\end{lemma}

\begin{proof}
Let $\lambda\in \rho(\mc L_\alpha)$ and
 $g\in
  \langle Q_{\R^d}+\rho_\alpha\rangle^\perp$. We have to show that there exists an $f\in
  \mc D(\mc J_\alpha)$ such that $(\lambda^2+\mc J_\alpha) f=g$.
  By the self-adjointness of $\mc L_{\alpha,-}^\frac12$, we have
  \[ \rg\mc
L_{\alpha,-}^\frac12=(\ker \mc L_{\alpha,-}^\frac12)^\perp \]
 and thus, there
  exists an $g_2 \in \mc D(\mc L_{\alpha,-}^\frac12)$ such that $\lambda \mc
  L_{\alpha,-}^\frac12 g_2=g$.   Since
  $\lambda\in \rho(\mc L_\alpha)$,
 there exists $(f_1,f_2)\in H^2(\R^d,\C^2)$
  such that
  \[ (\lambda -\mc L_\alpha)
    \begin{pmatrix}
      f_1 \\ f_2
    \end{pmatrix}
=
\begin{pmatrix}
  0 \\ g_2
\end{pmatrix}. \]
Equivalently,
\begin{equation}
  \label{eq:specLexplicit}
  \begin{cases}
    \lambda f_1-\mc L_{\alpha,-}f_2=0, \\
    \mc L_{\alpha,+}f_1+\lambda f_2=g_2.
  \end{cases}
\end{equation}
By inserting the first equation into the second one, we find
\[ \mc L_{\alpha,+}\mc L_{\alpha,-}f_2=-\lambda^2 f_2+\lambda g_2\in
  \mc D(\mc L_{\alpha,-}^\frac12), \]
and applying $\mc L_{\alpha,-}^\frac12$ yields
\[ \mc L_{\alpha,-}^\frac12 \mc L_{\alpha,+}\mc L_{\alpha,-}^\frac12
  f=-\lambda^2 f+\lambda \mc L_{\alpha,-}^\frac12 g_2
=-\lambda^2 f+g \]
with $f:=\mc L_{\alpha,-}^\frac12 f_2\in \mc D(\mc J_\alpha)$.

To prove the second assertion, we first assume that $\lambda^2+\mc
J_\alpha$ is injective.
Suppose
\[ (\lambda-\mc L_\alpha)
  \begin{pmatrix}
    f_1 \\ f_2
  \end{pmatrix}
  =
  \begin{pmatrix}
    0 \\ 0
  \end{pmatrix}. \]
Then, by setting $g=g_2=0$ in the above computation, we find
$(\lambda^2+\mc J_\alpha)f=0$ for $f=\mc L_{\alpha,-}^\frac12 f_2$.
This shows that $f_2\in \ker \mc L_{\alpha,-}^\frac12=\ker \mc
L_{\alpha,-}$ and the first equation in \eqref{eq:specLexplicit} implies
that $f_1=0$. Subsequently, the second equation in
\eqref{eq:specLexplicit} with $g_2=0$ shows that $f_2=0$ as well.

It remains to prove the reverse implication, i.e., we assume that $\lambda-
\mc L_\alpha$ is injective and show that $\lambda^2+\mc
J_\alpha$ is injective. Consider the equation $(\lambda^2+\mc
J_\alpha)f=0$ for an arbitrary $f\in \mc D(\mc J_\alpha)$. We have to
show that $f=0$.
Set $f_1:=\mc L_{\alpha,-}^\frac12 f$.
From $f\in \mc D(\mc J_\alpha)$ we infer that $f_1\in \mc D(\mc
L_{\alpha,-}^\frac12 \mc L_{\alpha,+})\subset \mc D(\mc L_{\alpha,+})
=H^2_\mathrm{rad}(\R^d)$.
Furthermore, we define
$f_2:=-\tfrac{1}{\lambda}\mc L_{\alpha,+}f_1=-\tfrac{1}{\lambda}\mc
L_{\alpha,+}\mc L_{\alpha,-}^\frac12 f\in \mc D(\mc L_{\alpha,-}^\frac12)$. Then we have 
\[ \mc L_{\alpha,-}^\frac12 f_2=-\tfrac{1}{\lambda}\mc J_\alpha
  f=\lambda f \in \mc D(\mc J_\alpha)\subset \mc D(\mc
  L_{\alpha,-}^\frac12), \]
which shows that $f_2\in \mc D(\mc L_{\alpha,-})=H^2_\mathrm{rad}(\R^d)$.
Consequently,
\begin{align*}
  (\lambda-\mc L_\alpha)
  \begin{pmatrix}
    f_1 \\ f_2
  \end{pmatrix}
  &=
    \begin{pmatrix}
      \lambda f_1-\mc L_{\alpha,-}f_2 \\
      \mc L_{\alpha,+}f_1+\lambda f_2
    \end{pmatrix}
  =
  \begin{pmatrix}
    \lambda\mc L_{\alpha,-}^\frac12 f-\lambda\mc L_{\alpha,-}^\frac12 f \\ 0
  \end{pmatrix}
=
  \begin{pmatrix}
    0 \\ 0
  \end{pmatrix}
\end{align*}
and it follows that $f_1=f_2=0$ by the injectivity of $\lambda-\mc
L_\alpha$.
Since $f_1=\mc L_{\alpha,-}^\frac12 f$, we infer that $f\in \ker \mc
L_{\alpha,-}^\frac12=\langle Q_{\R^d}+\rho_\alpha\rangle$.  Together with
$f\in \mc D(\mc J_\alpha)$, this implies that $f=0$.
\end{proof}

A consequence of Lemma \ref{lem:Jinjsur} is the
self-adjointness of $\mc J_\alpha$.

\begin{lemma}
  \label{lem:Jsa}
  Let $\alpha>0$ be sufficiently large. Then the operator
$\mc J_\alpha$ is self-adjoint.
\end{lemma}

\begin{proof}
  By the self-adjointness of $\mc L_{\alpha,-}^\frac12$, $\mc
  L_{\alpha,+}$ and Lemma \ref{lem:Jdense}, it follows that $\mc
  J_\alpha$ is symmetric.
  In other words, $\mc J_\alpha\subset \mc J_\alpha^*$ and, since $\mc
  J_\alpha^*$ is closed, $\mc J_\alpha$ is closable and its closure
  $\overline{\mc J_\alpha}$ is symmetric, too.
  Now consider the operators $\pm i+\mc J_\alpha=\mu_\pm^2+\mc J_\alpha$,
  where $\mu_\pm:=\frac{1}{\sqrt 2}\pm \frac{i}{\sqrt 2}$.
  By Lemma
\ref{lem:basicLalpha}, $\mu_\pm\in \rho(\mc L_\alpha)$ and thus, Lemma
\ref{lem:Jinjsur} implies that $\pm i+\mc J_\alpha$ is
surjective. Consequently, $\pm i+\overline{\mc J_\alpha}$ is
surjective and therefore, $\overline{\mc J_\alpha}$ is self-adjoint
(see e.g.~\cite{Kat95}, p.~271, Theorem 3.16).
Let $g\in \mc D(\overline{\mc J_\alpha})$ be arbitrary. By the
surjectivity of $i+\mc J_\alpha$, there exists an $f\in \mc D(\mc
J_\alpha)$ such that $(i+\mc J_\alpha)f=(i+\overline{\mc J_\alpha})g$
and $\mc J_\alpha\subset \overline{\mc J_\alpha}$ implies that
$(i+\overline{\mc J_\alpha})(f-g)=0$. Since $\sigma_p(\overline{\mc
  J_\alpha})\subset \R$, we must have $f-g=0$ and therefore, $g\in \mc
D(\mc J_\alpha)$. Thus, we have proved that $\mc D(\overline{\mc
  J_\alpha})\subset \mc D(\mc J_\alpha)$ and this shows that $\mc
J_\alpha=\overline{\mc J_\alpha}$.
\end{proof}

We need one last technical result.

\begin{lemma}
  \label{lem:Jcore}
  Let $\alpha>0$ be sufficiently large and define $\mc B_\alpha: \mc
  D(\mc B_\alpha)\subset \langle Q_{\R^d}+\rho_\alpha\rangle^\perp \to
  \langle Q_{\R^d}+\rho_\alpha\rangle^\perp$ by $\mc D(\mc B_\alpha):=\mc
  D(\mc L_{\alpha,-}^\frac12)\cap \langle
  Q_{\R^d}+\rho_\alpha\rangle^\perp$ and $\mc B_\alpha f:=\mc
  L_{\alpha,-}^\frac12 f$. Then the (maximally defined) operator $\mc
  L_{\alpha,+}\mc B_\alpha$ is densely defined, closed, and $\mc D(\mc J_\alpha)$ is a
  core for $\mc L_{\alpha,+}\mc B_\alpha$.
\end{lemma}

\begin{proof}
  Recall from the proof of Lemma \ref{lem:Jdense} that $\mc B_\alpha$
  is closed and bounded invertible. As a consequence, Lemma
  \ref{lem:AB} shows that $\mc
  L_{\alpha,+}\mc B_\alpha$ is densely defined. Let $(f_n)_{n\in\N}\subset \mc D(\mc
  L_{\alpha,+}\mc B_\alpha)\subset \mc D(\mc B_\alpha)$ with $f_n\to f$ and $\mc L_{\alpha,+}\mc
  B_\alpha f_n\to h$ as $n\to\infty$. Then we have
\[ \left \|\mc B_\alpha f_n-\mc L_{\alpha,+}^{-1}h \right
  \|_{L^2(\R^d)}
=\left \|\mc L_{\alpha,+}^{-1}\left (\mc L_{\alpha,+}\mc B_\alpha
    f_n-h\right )\right \|_{L^2(\R^d)}
\lesssim \left \|\mc L_{\alpha,+}\mc B_\alpha
    f_n-h\right \|_{L^2(\R^d)}\to 0 \]
as $n\to\infty$ and the closedness of $\mc B_\alpha$
implies that $f\in \mc D(\mc B_\alpha)$ and $\mc B_\alpha f=\mc
L_{\alpha,+}^{-1}h \in \mc D(\mc L_{\alpha,+})$. Consequently, $f\in
\mc D(\mc L_{\alpha,+}\mc B_\alpha)$ and $\mc L_{\alpha,+}\mc B_\alpha
f=\mc L_{\alpha,+}\mc L_{\alpha,+}^{-1}h=h$. This proves the
closedness of $\mc L_{\alpha,+}\mc B_\alpha$. 

Next, we claim that $\mc L_{\alpha,+}\mc B_\alpha$ has closed
range. Indeed, let $(h_n)_{n\in\N}\subset \rg(\mc L_{\alpha,+}\mc
B_\alpha)$
with $h_n\to h$ as $n\to\infty$. Then there exists a sequence
$(f_n)_{n\in\N}\subset \mc D(\mc L_{\alpha,+}\mc B_\alpha)$ such that
$\mc L_{\alpha,+}\mc B_\alpha f_n=h_n$. In other words, $f_n=\mc
B_\alpha^{-1}\mc L_{\alpha,+}^{-1}h_n$ and thus, $f_n\to f$ as $n\to\infty$
for some $f\in \langle Q_{\R^d}+\rho_\alpha\rangle^\perp$. By the
closedness of $\mc L_{\alpha,+}\mc B_\alpha$, we infer that $f\in \mc
D(\mc L_{\alpha,+}\mc B_\alpha)$ and $\mc L_{\alpha,+}\mc B_\alpha
f=h$, which shows that $h\in \rg(\mc L_{\alpha,+}\mc B_\alpha)$.

Now we define an auxiliary operator $\mc A_\alpha: \mc D(\mc
A_\alpha)\subset \langle Q_{\R^d}+\rho_\alpha\rangle^\perp \to \rg(\mc
L_{\alpha,+}\mc B_\alpha)$ by $\mc D(\mc A_\alpha):=\mc D(\mc
L_{\alpha,+}\mc B_\alpha)$ and $\mc A_\alpha f:=\mc L_{\alpha,+}\mc
B_\alpha f$. By the above, $\mc A_\alpha$ is densely defined, closed,
and bijective. Thus, the closed graph theorem shows that $\mc
A_\alpha$ is bounded invertible. 
By definition, $\mc D(\mc J_\alpha)=\mc D(\mc L_{\alpha,-}^\frac12 \mc
A_\alpha)$.
Lemma \ref{lem:AB} implies that $\mc
D(\mc J_\alpha)$ is a core for $\mc A_\alpha$ and hence for $\mc
L_{\alpha,+}\mc B_\alpha$.
\end{proof}

\subsection{Spectral stability in the critical case}

Now we can establish a stability criterion also in the critical case
$p=1+\frac{4}{d}$.

\begin{lemma}
  \label{lem:linstabcrit}
  If $p=1+\frac{4}{d}$ then there exists an $\alpha_0>0$ such that for
  all $\alpha\geq \alpha_0$ the following holds.
  \begin{itemize}
  \item If $(\mc
    L_{\alpha,+}^{-1}(Q_{\R^d}+\rho_\alpha)|Q_{\R^d}+\rho_\alpha)_{L^2(\R^d)}>0$
    then $\mc L_\alpha$ has precisely one positive eigenvalue
    $\lambda_\alpha$ and the eigenvalues
    $\pm \lambda_\alpha\in \sigma_p(\mc L_\alpha)$ are simple.
 \item If $(\mc
    L_{\alpha,+}^{-1}(Q_{\R^d}+\rho_\alpha)|Q_{\R^d}+\rho_\alpha)_{L^2(\R^d)}\leq
    0$ then $\mc L_\alpha$ has no positive eigenvalues.
  \end{itemize}
\end{lemma}

\begin{proof}
  Let $\mc P$ and $\mc P_\alpha$ be the spectral projections from the
  proof of Lemma \ref{lem:linstab}. By Theorem \ref{thm:L} and
  \cite{Kat95}, p.~34, Lemma 4.10, we have
  $\dim\rg\mc P_\alpha=\dim\rg\mc P=4$ and thus, by Lemma
  \ref{lem:basicLalpha}, there can be at most one positive eigenvalue
  $\lambda_\alpha>0$ and if so, the eigenvalues $\pm \lambda_\alpha\in
  \sigma_p(\mc L_\alpha)$ will be simple since $0\in
  \sigma_p(\mc L_\alpha)$.

  Now assume that $(\mc
  L_{\alpha,+}^{-1}(Q_{\R^d}+\rho_\alpha)|Q_{\R^d}+\rho_\alpha)_{L^2(\R^d)}>0$.
  Then, by Proposition \ref{prop:L+nonneg}, we can find an $f_\alpha\in \langle
  Q_{\R^d}+\rho_\alpha\rangle^\perp\cap H^2_\mathrm{rad}(\R^d)$ such
  that $(\mc L_{\alpha,+}f_\alpha|f_\alpha)_{L^2(\R^d)}<0$. From the
  self-adjointness of $\mc L_{\alpha,-}$, we have
  \[ \rg \mc
    L_{\alpha,-}^\frac12=(\ker \mc L_{\alpha,-}^\frac12)^\perp=
    \ker(\mc L_{\alpha,-})^\perp=\langle
    Q_{\R^d}+\rho_\alpha\rangle^\perp. \]
 Thus, since $f_\alpha\perp Q_{\R^d}+\rho_\alpha$,
 there exists a
  $\widetilde g_\alpha\in \mc D(\mc L_{\alpha,-}^\frac12)$ such that $\mc
  L_{\alpha,-}^\frac12 \widetilde g_\alpha=f_\alpha$. Set
  $g_\alpha:=\mc P_\alpha^\perp \widetilde g_\alpha$, where $\mc
  P_\alpha^\perp$ is the orthogonal projection onto $\langle
  Q_{\R^d}+\rho_\alpha\rangle^\perp$. Then we have $g_\alpha\in \mc
  D(\mc B_\alpha)$ and $\mc B_\alpha
  g_\alpha=\mc L_{\alpha,-}^\frac12 \mc P_\alpha^\perp \widetilde
  g_\alpha=\mc L_{\alpha,-}^\frac12 \widetilde g_\alpha=f_\alpha$,
  where $\mc B_\alpha$ is the operator defined in Lemma
  \ref{lem:Jcore}. By construction, 
  \[ \Big (\mc L_{\alpha,+}\mc B_\alpha g_\alpha \Big | \mc
   B_\alpha g_\alpha\Big )_{L^2(\R^d)}=(\mc
    L_{\alpha,+}f_\alpha|f_\alpha)_{L^2(\R^d)}<0. \]
Since $\mc D(\mc J_\alpha)$ is a core for $\mc
L_{\alpha,+}\mc B_\alpha$ (Lemma
\ref{lem:Jcore}), we can
find for any given $\epsilon>0$ 
an $\widetilde f_\alpha\in \mc
D(\mc J_\alpha)$ such that $\|\mc L_{\alpha,+}\mc B_\alpha \widetilde
f_\alpha - \mc L_{\alpha,+}\mc B_\alpha
g_\alpha\|_{L^2(\R^d)}<\epsilon$ and
\[ \|\mc B_\alpha \widetilde f_\alpha-\mc B_\alpha
  g_\alpha\|_{L^2(\R^d)}
  =\|\mc L_{\alpha,+}^{-1}[\mc L_{\alpha,+}\mc B_\alpha \widetilde
  f_\alpha
  -\mc L_{\alpha,+}\mc B_\alpha g_\alpha]\|_{L^2(\R^d)}\lesssim
  \epsilon. \]
Consequently, by choosing $\epsilon>0$ sufficiently small, we find
\begin{align*}
  0&>\Big (\mc L_{\alpha,+}\mc B_\alpha \widetilde f_\alpha \Big | \mc
     B_\alpha \widetilde f_\alpha \Big )_{L^2(\R^d)}
  =\Big (\mc L_{\alpha,+}\mc L_{\alpha,-}^\frac12 \widetilde f_\alpha
    \Big | \mc L_{\alpha,-}^\frac12 \widetilde f_\alpha \Big
     )_{L^2(\R^d)}
  =\Big (\mc L_{\alpha,-}^\frac12 \mc
  L_{\alpha,+}\mc L_{\alpha,-}^\frac12 \widetilde f_\alpha \Big |
  \widetilde f_\alpha \Big )_{L^2(\R^d)} \\   
 &= \Big (\mc J_\alpha \widetilde f_\alpha \Big | \widetilde f_\alpha
  \Big )_{L^2(\R^d)}.
  \end{align*}
Lemma \ref{lem:Jsa} therefore implies 
that $\mc J_\alpha$ has negative spectrum, i.e.,
there exists a $\lambda_\alpha>0$ such that $-\lambda_\alpha^2\in \sigma(\mc J_\alpha)$.
If $-\lambda_\alpha^2-\mc J_\alpha=-(\lambda_\alpha^2+\mc J_\alpha)$ is not
surjective, then, by Lemma \ref{lem:Jinjsur}, $\lambda_\alpha\in \sigma(\mc
L_\alpha)$ and by Lemma \ref{lem:basicLalpha}, $\lambda_\alpha \in
\sigma_p(\mc L_\alpha)$. If $-(\lambda_\alpha^2+\mc J_\alpha)$ is not
injective, Lemma \ref{lem:Jinjsur} implies that $\lambda_\alpha\in
\sigma_p(\mc L_\alpha)$.

If, on the other hand, $(\mc
L_{\alpha,+}^{-1}(Q_{\R^d}+\rho_\alpha)|Q_{\R^d}+\rho_\alpha)_{L^2(\R^d)}\leq
0$, we obtain
\[ (\mc J_\alpha f | f)_{L^2(\R^d)}=\Big (\mc L_{\alpha,+}\mc
  L_{\alpha,-}^\frac12 f \Big | \mc L_{\alpha,-}^\frac12 f \Big
  )_{L^2(\R^d)}\geq 0 \]
for all $f\in \mc D(\mc J_\alpha)$, by Proposition \ref{prop:L+nonneg}.
Thus, Lemma \ref{lem:Jsa} implies that $\sigma(\mc
J_\alpha)\subset [0,\infty)$ and from Lemma \ref{lem:Jinjsur} we infer
that $\lambda-\mc L_\alpha$ is injective for any
$\lambda>0$. Consequently, Lemma \ref{lem:basicLalpha} shows that
$\sigma(\mc L_\alpha)\cap \R=\{0\}$.
\end{proof}

\begin{proof}[Proof of Theorems \ref{thm:spec}, \ref{thm:stabnoncrit},
  and \ref{thm:stabcrit}]
Consider the map $\mc V_\alpha: L^2_\mathrm{rad}(\M^d)\to
L^2_\mathrm{rad}(\R^d)$ given by
\[ \mc V_\alpha f(x):=\alpha^{-\frac{d}{2}}\left
    (\frac{A(\alpha^{-1}|x|)}{\alpha^{-1}|x|}\right
  )^{\frac{d-1}{2}}f(\alpha^{-1}|x|,y), \]
with inverse
\[ \mc V_\alpha^{-1}f(r,y)=\alpha^\frac{d}{2}\left
    (\frac{r}{A(r)}\right )^{\frac{d-1}{2}}f(\alpha r e_1). \]
We have
\begin{align*}
  \|\mc V_\alpha
  f\|_{L^2(\R^d)}^2
  &=\alpha^{-d}|\S^{d-1}|\int_0^\infty \left
    (\frac{A(\alpha^{-1}r)}{\alpha^{-1}r}\right)^{d-1}
    |f(\alpha^{-1}r,y)|^2 r^{d-1} dr \\
  &=|\S^{d-1}|\int_0^\infty |f(r,y)|^2 A(r)^{d-1}dr \\
  &=\|f\|_{L^2(\M^d)}^2,
\end{align*}
and thus, $\mc V_\alpha$ is unitary for any $\alpha>0$.
Furthermore, recall that
  \[ Q_{\M^d,\alpha}(r,y)=\alpha^\frac{2}{p-1}\left
      (\frac{r}{A(r)}\right)^{\frac{d-1}{2}}\left [Q_{\R^d}(\alpha
      r e_1)+\rho_\alpha(\alpha r e_1)\right ], \]
  and thus, for any radial $f\in C^\infty_c(\M^d)$, we have
\[ \widetilde{\mc L}_{\M^d,\alpha,\pm} f=\alpha^2 \mc V_\alpha^{-1}\mc
  L_{\alpha,\pm}\mc V_\alpha f. \]
Consequently, the closure $\mc L_{\M^d,\alpha}$ of $\widetilde{\mc
  L}_{\M^d,\alpha}$ is given by
\[ \mc L_{\M^d,\alpha}=\alpha^2\begin{pmatrix}
    0 & \mc V_\alpha^{-1}\mc L_{\alpha,-}\mc V_\alpha \\
    - \mc V_\alpha^{-1} \mc L_{\alpha,+}\mc V_\alpha & 0
    \end{pmatrix},
\]
and
$\mc L_{\M^d,\alpha}$ is unitarily equivalent to $\alpha^2 \mc
L_\alpha$. This implies the claimed statements.
\end{proof}

\section{Stability and curvature}
\label{sec:stabcurv}

\noindent From \cite{BanDuy15}, we know that in negative curvature there is blow-up instability for sufficiently high energy.  
In this last section we give numerical evidence of how this instability manifests in the bifurcation theory from the Euclidean situation. 
The soliton may become linearly unstable in
the curved geometry if the curvature is strictly negative
everywhere or otherwise.
More precisely, 
we consider the model case of a warping function
$A(r)=r+c_1r^3+c_2r^5$, in the critical case 
$d=2$, $p=3$. The sectional curvatures of the 
manifold $\M^2$ are given by
\[ -\frac{A''(r)}{A(r)}=-\frac{6c_1+20c_2r^2}{1+c_1r^2+c_2r^4},\qquad
  \frac{1-A'(r)^2}{A(r)^2}=-\frac{(3c_1+5c_2r^2)(2+3c_1r^2+5c_2r^4)}{(1+c_1r^2+c_2r^4)^2}. \]

\subsection{A formal expansion}
As before, we write $\widetilde R_\alpha=Q_{\R^2}+\rho_\alpha$ with
$\rho_\alpha$ from Proposition \ref{prop:exrho}.
By Lemma \ref{lem:rhoC2}, $\widetilde R_\alpha\in C^2(\R^2)$ and
\[ \Delta_{\R^2}\widetilde R_\alpha(x)-\widetilde
  R_\alpha(x)-\alpha^{-2}V_2(\alpha^{-1}x)\widetilde
  R_\alpha(x)+\varphi_{2,3}(\alpha^{-1}x)\widetilde R_\alpha(x)^3=0 \]
for all $x\in \R^2$. When written out explicitly for our model case, this reads
\begin{equation}
  \label{eq:R32}
  \begin{split}
(1&+c_1\alpha^{-2}r^2+c_2\alpha^{-4}r^4)^2(\Delta_{\R^2}-1)\widetilde
R_\alpha \\
&-\alpha^{-2}\left
  [2c_1+(c_1^2+8c_2)\alpha^{-2}r^2+6c_1c_2\alpha^{-4}r^4+4c_2^2\alpha^{-6}r^6\right]\widetilde
R_\alpha \\
&+(1+c_1\alpha^{-2}r^2+c_2\alpha^{-4}r^4)\widetilde R_\alpha^3=0,
\end{split}
  \end{equation}
where $r(x)=|x|$. Now we assume an asymptotic expansion of the form
\begin{equation}
  \label{eq:asymexp}
  \widetilde
  R_\alpha(x)=Q_{\R^2}(x)+\alpha^{-2}Q_1(x)+\alpha^{-4}Q_2(x)+\alpha^{-6}Q_E(x,\alpha),
  \end{equation}
where $\|Q_E(\cdot,\alpha)\|_{L^2(\R^2)}\lesssim 1$ and
$\|\partial_\alpha Q_E(\cdot,\alpha)\|_{L^2(\R^2)}\lesssim \alpha^{-1}$ for all
$\alpha\gg 1$.
Then, in view of Theorem \ref{thm:ex}, the soliton profile on $\M^2$ is given by
\[ Q_{\M^2,\alpha}(r,y)=\alpha\left (\frac{r}{A(r)}\right)^\frac12
  \left [Q_{\R^2}(\alpha r e_1)+\alpha^{-2}Q_1(\alpha r
    e_1)+\alpha^{-4}Q_2(\alpha r e_1)+\alpha^{-6}Q_E(\alpha r
    e_1,\alpha)\right ] \]
and thus,
\begin{align*}
  \|Q_{\M^2,\alpha}\|_{L^2(\M^2)}^2
  &=\int_0^\infty\int_\R Q_{\M^2,\alpha}(r,y)^2 A(r)\frac{2}{y^2+1}dy
    dr=2\pi\int_0^\infty Q_{\M^2,\alpha}(r,y)^2 A(r) dr \\
   &=2\pi\int_0^\infty \left [Q_{\R^2}(r e_1)+\alpha^{-2}Q_1(r
    e_1)+\alpha^{-4}Q_2( r e_1)+\alpha^{-6}Q_E(r
     e_1,\alpha)\right ]^2rdr \\
  &=\|Q_{\R^2}\|_{L^2(\R^2)}^2+2\alpha^{-2}(Q_{\R^2}|Q_1)_{L^2(\R^2)}+\alpha^{-4}\left
    [\|Q_1\|_{L^2(\R^2)}^2+2(Q_{\R^2}|Q_2)_{L^2(\R^2)}\right ] \\
  &\quad + O(\alpha^{-6}).
\end{align*}
In order to compute the profiles $Q_1$ and $Q_2$, we plug the ansatz
\eqref{eq:asymexp} into Eq.~\eqref{eq:R32} and solve
order by order in $\alpha$. This yields
\begin{align*}
  \alpha^0&: \Delta_{\R^2}Q_{\R^2}-Q_{\R^2}+Q_{\R^2}^3=0, \\
  \alpha^{-2}&:
               \mc L_+ Q_1=-c_1(2Q_{\R^2}+r^2Q_{\R^2}^3),
  \\
  \alpha^{-4}&: \mc L_+ Q_2=
               -2c_1Q_1+(3c_1^2-8c_2)r^2Q_{\R^2}
               -3c_1r^2Q_{\R^2}^2Q_1+3Q_{\R^2}Q_1^2
               +(c_1^2-c_2)r^4 Q_{\R^2}^3,
\end{align*}
with $\mc L_+=-\Delta_{\R^2}+1-3Q_{\R^2}^2$.
By definition, $Q_{\R^2,\alpha}(x)=\alpha Q_{\R^2}(\alpha x)$
satisfies
\[ \Delta_{\R^2}Q_{\R^2,\alpha}-\alpha^2
  Q_{\R^2,\alpha}+Q_{\R^2,\alpha}^3=0. \]
By differentiating this equation with respect to $\alpha$,
we see that
\[ S_0(x):=\partial_\alpha
  Q_{\R^2,\alpha}(x)|_{\alpha=1}=x^j\partial_j Q_{\R^2}(x)+Q_{\R^2}(x) \]
satisfies
\[ \mc L_+ S_0=-\Delta_{\R^2}S_0+S_0-3Q_{\R^2}^2S_0=-2Q_{\R^2}. \]
Consequently,
\begin{align*}
  (Q_{\R^2}|Q_1)_{L^2(\R^2)}
  &=-\tfrac12 (\mc L_+ S_0|Q_1)_{L^2(\R^2)}
    =-\tfrac12 (S_0|\mc L_+ Q_1)_{L^2(\R^2)} \\
    &=c_1(S_0|Q_{\R^2})_{L^2(\R^2)}+\tfrac{1}{2}c_1(S_0|r^2
      Q_{\R^2}^3)_{L^2(\R^2)} \\
  &=0,
  \end{align*}
  since
  \[ 0=\partial_\alpha \|Q_{\R^2}\|_{L^2(\R^2)}^2=\partial_\alpha
    \|Q_{\R^2,\alpha}\|_{L^2(\R^2)}^2
    =2(\partial_\alpha Q_{\R^2,\alpha}|Q_{\R^2,\alpha})_{L^2(\R^2)} \]
  and
  \[ 0=\partial_\alpha \|rQ_{\R^2}^2\|_{L^2(\R^2)}^2=\partial_\alpha
    \|rQ_{\R^2,\alpha}^2\|_{L^2(\R^2)}^2=4(\partial_\alpha
    Q_{\R^2,\alpha}|r^2Q_{\R^2,\alpha}^3)_{L^2(\R^2)} \]
  which, when evaluated at $\alpha=1$, reads
  \[ 0=(S_0|Q_{\R^2})_{L^2(\R^2)}=(S_0|r^2Q_{\R^2}^3)_{L^2(\R^2)}. \]
  This implies that
  \[
    \|Q_{\M^2,\alpha}\|_{L^2(\M^2)}^2=\|Q_{\R^2}\|_{L^2(\R^2)}^2+\alpha^{-4}\left
        [\|Q_1\|_{L^2(\R^2)}^2+2(Q_{\R^2}|Q_2)_{L^2(\R^2)}\right
      ] +O(\alpha^{-6}). \]
The sign of $\partial_\alpha \|Q_{\M^2,\alpha}\|_{L^2(\R^2)}^2$ is
to leading order determined by the sign of
\[ \kappa:=\|Q_1\|_{L^2(\R^2)}^2+2(Q_{\R^2}|Q_2)_{L^2(\R^2)}. \]
More precisely, we have
\[ \partial_\alpha
  \|Q_{\M^2,\alpha}\|_{L^2(\M^2)}^2=-4\alpha^{-5}\kappa+O(\alpha^{-7}) \]
and the soliton is linearly unstable for sufficiently large
$\alpha$ if $\kappa>0$, see Theorem \ref{thm:stabcrit} and Remark
\ref{rem:stabcrit}.

\subsection{Stability}
\label{sec:numstab}
By using the defining equation for $Q_2$, we find the expression
\begin{align*}
\kappa&=\|Q_1\|_{L^2(\R^2)}^2-(\mc L_+ S_0|Q_2)_{L^2(\R^2)}
        =\|Q_1\|_{L^2(\R^2)}^2-(S_0|\mc L_+ Q_2)_{L^2(\R^2)} \\
      &=\|Q_1\|_{L^2(\R^2)}^2\\
  &\quad +\big(S_0\big|
    2c_1Q_1-(3c_1^2-8c_2)r^2Q_{\R^2}
    +3c_1r^2Q_{\R^2}^2Q_1-3Q_{\R^2}Q_1^2-(c_1^2-c_2)r^4 Q_{\R^2}^3\big)_{L^2(\R^2)}.
  \end{align*}
It is convenient to introduce the function $\widehat Q_1$, defined as
the unique solution (in $H^2_\mathrm{rad}(\R^2)$) of the equation
\[ \mc L_+ \widehat Q_1=-2Q_{\R^2}-r^2Q_{\R^2}^3. \]
Then we have $Q_1=c_1\widehat Q_1$ and we arrive at
$\kappa=c_1^2 b_1+c_2 b_2$ with
\begin{align*}
  b_1&=\|\widehat Q_1\|_{L^2(\R^2)}^2+\big (S_0 \big |
  2\widehat Q_1-3r^2Q_{\R^2}+3r^2Q_{\R^2}^2\widehat
          Q_1-3Q_{\R^2}\widehat Q_1^2-r^4Q_{\R^2}^3\big )_{L^2(\R^2)}, \\
b_2&=\big (S_0|8r^2Q_{\R^2}+r^4Q_{\R^2}^3\big )_{L^2(\R^2)}.
\end{align*}
Consequently, the issue is to determine the signs of $b_1$ and $b_2$
(which depend only on the Euclidean profile $Q_{\R^2}$).
An integration by parts yields
\[
  b_2=-8(Q_{\R^2}|r^2Q_{\R^2})_{L^2(\R^2)}-\tfrac12(Q_{\R^2}|r^4Q_{\R^2}^3)_{L^2(\R^2)}<0 \]
and numerical evaluation shows, somewhat surprisingly, that $b_1\geq 14\pi$, see
Appendix \ref{sec:num}. This means that the simple choice $c_1=1$ and
$c_2=0$ provides a negatively curved metric that makes the soliton
linearly unstable.  In addition, we see that there are values of $c_1 \ll c_2$ such that the mass condition for stability is possibly true.
Of course, to establish orbital stability, further analysis is required on such a manifold, for which the metric expansion 
is far from standard examples.

\begin{appendix}

\section{Background material}
\label{sec:back}

\noindent For the convenience of the reader and to fix notation, we compile some
background material on radial distributions and distributional
solutions of Poisson's equation.

\subsection{Radial distributions}
As usual, for $U\subset \R^d$ open, we denote by $\mc
D(U)=C^\infty_c(U)$ the set of test functions. For
$(\varphi_n)_{n\in\N}\subset \mc D(U)$ and $\varphi\in \mc D(U)$, we say that
$\lim_{n\to\infty}\varphi_n=\varphi$ in $\mc D(U)$ if there exists a
compact $K\subset U$ such that $\supp\varphi_n\subset K$ for all $n\in
\N$ and for any $k\in \N_0$,
$\|\varphi_n-\varphi\|_{W^{k,\infty}(U)}\to 0$ as $n\to\infty$. Here,
\[ \|\varphi\|_{W^{k,\infty}(U)}=\sum_{|\beta|\leq k}\|\partial^\beta
  \varphi\|_{L^\infty(U)} \]
with the usual multi-index notation.
This notion of convergence defines a topology on $\mc D(U)$ and the
space $\mc D'(U)$ of continuous linear functionals on $\mc D(U)$ is
called the space of distributions.

In order to define radial distributions, 
we start with a test function $f\in \mc D(\R^d)$ and define its spherical mean
$Mf$ by
\[
  Mf(x):=\frac{1}{|\S^{d-1}|}\int_{\S^{d-1}}f(|x|\omega)d\sigma(\omega), \]
where $\sigma$ is the standard surface measure on the sphere
$\S^{d-1}$. Clearly, $Mf=f$ if and only if $f$ is radial.
The most important properties are summarized in the next lemma.

\begin{lemma}
  \label{lem:M}
  We have $\Delta_{\R^d}Mf=M\Delta_{\R^d}f$ for all $f\in
  C^\infty_c(\R^d)$. Furthermore, for any $s\geq 0$ we have
  \[ \|Mf\|_{H^s(\R^d)}\lesssim \|f\|_{H^s(\R^d)} \]
  for all $f\in C^\infty_c(\R^d)$. Finally, $M$ extends to
  a self-adjoint operator on $L^2(\R^d)$. 
\end{lemma}

\begin{proof}
  We use polar coordinates $x=r\omega'$ defined by $r=|x|$,
  $\omega'=\frac{x}{|x|}$ for $x\in\R^d\backslash\{0\}$. Since $Mf$ is radial, we obtain
  \begin{align*}
    \Delta_{\R^d} Mf(x)
    &=\left
      (\partial_r^2+\frac{d-1}{r}\partial_r\right )Mf(r\omega')
      =\frac{1}{|\S^{d-1}|}\left
      (\partial_r^2+\frac{d-1}{r}\partial_r\right
      )\int_{\S^{d-1}}f(r\omega)d\sigma(\omega) \\
    &=\frac{1}{|\S^{d-1}|}\int_{\S^{d-1}}\left
      (\partial_r^2+\frac{d-1}{r}\partial_r\right
      )f(r\omega)d\sigma(\omega) \\
    &=\frac{1}{|\S^{d-1}|}\int_{\S^{d-1}}\left
      (\partial_r^2+\frac{d-1}{r}\partial_r+\frac{1}{r^2}\Delta_{\S^{d-1},\omega}\right
      )f(r\omega)d\sigma(\omega) \\
       &=\frac{1}{|\S^{d-1}|}\int_{\S^{d-1}}\Delta_{\R^d}f(r\omega)d\sigma(\omega)
    \\
    &=M\Delta_{\R^d} f(x)
  \end{align*}
  for any $x\in \R^d\backslash\{0\}$. Next, by Cauchy-Schwarz,
  \begin{align*}
    \|Mf\|_{L^2(\R^d)}^2
    &=\int_0^\infty \int_{\S^{d-1}}|Mf(r\omega)|^2
      d\sigma(\omega)r^{d-1}dr \\
    &\lesssim \int_0^\infty
      \int_{\S^{d-1}}\int_{\S^{d-1}}|f(r\omega')|^2d\sigma(\omega')d\sigma(\omega)r^{d-1}dr
    \\
    &\lesssim \|f\|_{L^2(\R^d)}^2
  \end{align*}
  Since $M$ commutes with $\Delta_{\R^d}$, we can also estimate
  \begin{align*}
    \|Mf\|_{H^{2k}(\R^d)}
    &\simeq \|\Delta_{\R^d}^k Mf\|_{L^2(\R^d)}+\|Mf\|_{L^2(\R^d)}
      =\|M\Delta_{\R^d}^kf\|_{L^2(\R^d)}+\|Mf\|_{L^2(\R^d)} \\
    &\lesssim \|\Delta_{\R^d}^k f\|_{L^2(\R^d)}+\|f\|_{L^2(\R^d)} \\
    &\simeq \|f\|_{H^{2k}(\R^d)}
  \end{align*}
  for any $k\in \N_0$. By interpolation, we obtain the claimed
  estimate. Clearly, $Mf$ has compact support if $f\in
  C^\infty_c(\R^d)$ and the Sobolev embedding theorem shows that in fact $Mf\in
  C^\infty_c(\R^d)$. In particular,
  $\Delta_{\R^d}Mf(x)=M\Delta_{\R^d}f(x)$ holds for all $x\in \R^d$ by
  continuity.
  
  By density, $M$ extends to a bounded operator on $L^2(\R^d)$, and we
  have
  \begin{align*}
    (Mf|g)_{L^2(\R^d)}
    &=\int_{\R^d}Mf(x)\overline{g(x)}dx
      =\int_0^\infty\int_{\S^{d-1}}Mf(r\omega)\overline{g(r\omega)}d\sigma(\omega)r^{d-1}dr
    \\
    &=\frac{1}{|\S^{d-1}|}\int_0^\infty\int_{\S^{d-1}}\int_{\S^{d-1}}
      f(r\omega')d\sigma(\omega')\overline{g(r\omega)}d\sigma(\omega)
      r^{d-1}dr \\
    &=\frac{1}{|\S^{d-1}|}\int_0^\infty\int_{\S^{d-1}}f(r\omega')\int_{\S^{d-1}}\overline{g(r\omega)}d\sigma(\omega)d\sigma(\omega')r^{d-1}dr
    \\
    &=\int_0^\infty
      f(r\omega')\overline{Mg(r\omega')}d\sigma(\omega')r^{d-1}dr \\
    &=(f|Mg)_{L^2(\R^d)}
  \end{align*}
  for all $f,g\in C^\infty_c(\R^d)$ by Fubini. Consequently, $M$
  extends to a self-adjoint operator on $L^2(\R^d)$.
\end{proof}

We use the same symbol $M$ to denote the extension of the spherical
mean to $L^2(\R^d)$.
For $s\geq 0$ we define the closed subspace $H^s_\mathrm{rad}(\R^d)\subset H^s(\R^d)$ of
radial functions in $H^s(\R^d)$ by
\[ H^s_\mathrm{rad}(\R^d):=\{f \in H^s(\R^d): Mf=f\}. \]

It is now straightforward to further extend $M$ to
distributions. Indeed, for $u\in \mc D'(\R^d)$ we define $\widehat Mu$ by
\[ (\widehat Mu)(\varphi):=u(M\varphi) \]
for all $\varphi\in \mc D(\R^d)$. Obviously, $\widehat Mu$ is a linear form on
$\mc D(\R^d)$ and, for any $K\subset \R^d$ compact, we can find a $k\in
\N_0$ such that
\begin{align*}
  |(\widehat Mu)(\varphi)|
  &=|u(M\varphi)|\lesssim \|M\varphi\|_{W^{k,\infty}(\R^d)}\lesssim
    \|M\varphi\|_{H^{k+d}(\R^d)}
    \lesssim \|\varphi\|_{H^{k+d}(\R^d)}\lesssim \|\varphi\|_{W^{k+d,\infty}(K)}
\end{align*}
for all $\varphi\in C^\infty_c(K)$ by Sobolev embedding and Lemma
\ref{lem:M}.
This estimate shows that $\widehat Mu\in \mc D'(\R^d)$ and by the self-adjointness of $M$ on $L^2(\R^d)$, the
operator $\widehat M: \mc D'(\R^d)\to \mc D'(\R^d)$ is an extension of
$M$ to the space of distributions. Consequently, it is justified to
simplify notation by writing $M$ instead of $\widehat M$.
Accordingly, a distribution $u\in \mc D'(\R^d)$ is said to be radial
if $Mu=u$. Note that by Lemma \ref{lem:M}, $\Delta_{\R^d}$ maps
radial distributions to radial distributions.

\subsection{Regularity results}
We state and prove two regularity results for radial distributional solutions
of Poisson's equation.
It is
convenient to introduce the following notation.

\begin{definition}
\label{def:sharp}
  Let $U\subset \R^d$ be open and $f\in
L^1_\mathrm{loc}(U)$. Then we define the distribution $f^\sharp\in \mc
D'(U)$ by
\[ f^\sharp(\varphi):=\int_U f(x)\varphi(x)dx \]
for $\varphi\in \mc D(U)$.
\end{definition}

\begin{lemma}
  \label{lem:raddist}
  Let $f,g\in C(\R^d\setminus\{0\})\cap L^1_\mathrm{loc}(\R^d)$ be
  radial and suppose $f$ satisfies
    \[ \Delta_{\R^d}f^\sharp=g^\sharp \mbox{ in }\mc D'(\R^d). \]
  Then the function $\widehat f: (0,\infty)\to\R$, defined by
  $\widehat f(r):=r^\frac{d-1}{2}f(re_1)$, belongs to $C^2(0,\infty)$ and
  satisfies
  \[ \widehat f''(r)-\frac{(d-1)(d-3)}{4r^2}\widehat
    f(r)=r^\frac{d-1}{2}g(re_1) \]
  for all $r>0$.
\end{lemma}

\begin{proof}
Let
  \[ \mc
    D_d\psi(r):=\psi''(r)-\frac{(d-1)(d-3)}{4r^2}\psi(r). \]
  The operator $\mc D_d$ maps $\mc D(0,\infty)$ to $\mc D(0,\infty)$
  continuously and is formally self-adjoint on $L^2(0,\infty)$. Thus,
  $\mc D_d$ extends to $\mc D'(0,\infty)$ by setting $\mc D_d
  v(\psi):=v(\mc D_d \psi)$ for $v\in \mc D'(0,\infty)$ and $\psi\in
  \mc D(0,\infty)$. Furthermore, we have the identity
 \[ \Delta_{\R^d}\left
        (|\cdot|^{-\frac{d-1}{2}}\psi(|\cdot|)\right)(x)
      =|x|^{-\frac{d-1}{2}}(\mc D_d \psi)(|x|) \] 
    for all $x\in \R^d$.
  Now note that $\psi\in C^\infty_c(0,\infty)$ implies
  $|\cdot|^{-\frac{d-1}{2}}\psi(|\cdot|)\in C^\infty_c(\R^d)$ and
  thus, every distribution $u\in \mc D'(\R^d)$ defines a distribution
 $\widehat u\in \mc D'(0,\infty)$ by setting
  \[ \widehat u(\psi):=u\left
      (|\cdot|^{-\frac{d-1}{2}}\psi(|\cdot|)\right ) \]
  for $\psi\in \mc D(0,\infty)$. Then we have
  \begin{align*}
    \Delta_{\R^d}u
    \left (|\cdot|^{-\frac{d-1}{2}}\psi(|\cdot|)\right )
    &=
      u\left (\Delta_{\R^d}\left
      (|\cdot|^{-\frac{d-1}{2}}\psi(|\cdot|)\right )\right ) \\
    &=u\left (|\cdot|^{-\frac{d-1}{2}}(\mc D_d \psi)(|\cdot|)\right
      )=\widehat u(\mc D_d \psi) \\
    &=\mc D_d \widehat u(\psi)
  \end{align*}
  for all $\psi\in \mc D(0,\infty)$,
and the equation $\Delta_{\R^d}f^\sharp=g^\sharp$ in
  $\mc D'(\R^d)$ implies that
  \[ \mc D_d \widehat{f^\sharp}=\widehat{g^\sharp} \mbox{ in }\mc
    D'(0,\infty). \]
  Explicitly, we have
  \begin{align*} \widehat{f^\sharp}(\psi)
    &=f^\sharp\left
      (|\cdot|^{-\frac{d-1}{2}}\psi(|\cdot|)\right )
      =\int_{\R^d}f(x)|x|^{-\frac{d-1}{2}}\psi(|x|)dx
      =|\S^{d-1}|\int_0^\infty r^\frac{d-1}{2}f(re_1)\psi(r)dr \\
  \end{align*}
  and thus, $\widehat{f^\sharp}=\hat f^\sharp$ with
  $\hat f(r):=|\S^{d-1}|r^{\frac{d-1}{2}}f(re_1)$.
  This yields
  \[ \mc D_d \hat f^\sharp=\hat g^\sharp \mbox{ in }\mc D'(0,\infty), \]
  and by \cite{Hor03}, p.~58, Corollary 3.1.6, it follows that
  $\hat f\in C^2(0,\infty)$ and
$\mc D_d \hat f=\hat g$ holds in the classical sense.
  \end{proof}

\begin{lemma}
  \label{lem:regLaplace}
  Let $f,g\in C(\R^d)$ be radial and suppose $f$
  satisfies
  \[ \Delta_{\R^d}f^\sharp=g^\sharp \mbox{ in }\mc D'(\R^d). \]
 Then $f\in
  C^2(\R^d)$ and $\Delta_{\R^d}f(x)=g(x)$ for all $x\in \R^d$ in
  the classical sense. 
\end{lemma}

\begin{proof}
 From Lemma \ref{lem:raddist} we know that $\widehat f(r):=r^{\frac{d-1}{2}}f(re_1)$
 belongs to $C^2(0,\infty)$ and satisfies
\[ \widehat f''(r)-\frac{(d-1)(d-3)}{4r^2}\widehat f(r)=r^{\frac{d-1}{2}}g(re_1) \]
for all $r>0$. In particular, $f\in C^2(\R^d\setminus\{0\})$.
We set $\phi(s):=s^{-\frac{d-1}{4}}\widehat f(\sqrt
s)=f(\sqrt s e_1)$. Then $\phi\in C^2(0,\infty)\cap C([0,\infty))$ and
\[   \phi''(s)+\frac{d}{2s}\phi'(s)=\frac{1}{s}h(s) \]
for all $s>0$ and $h(s):=\frac{1}{4}g(\sqrt s e_1)$.
Obviously, $h\in C([0,\infty))$.
  A fundamental system for the homogeneous equation is given by
  $\{1,\psi_0\}$, where $\psi_0(s)=-\frac{2}{d-2}s^{-\frac{d-2}{2}}$
  if $d\geq 3$ and $\psi_0(s)=\log s$ if $d=2$.
  Note that $W(1,\psi_0)(s)=s^{-\frac{d}{2}}$ and thus, 
by the variation of constants formula, $\phi$ can be written as
  \[ \phi(s)=c_0+c_1 \psi_0(s)-\int_0^s
    \psi_0(t)t^{\frac{d}{2}-1}h(t)dt+\psi_0(s)\int_0^s
    t^{\frac{d}{2}-1}h(t)dt \]
  for some constants $c_0,c_1\in \R$. Since $\phi\in C([0,\infty))$, we
  must have $c_1=0$ and therefore,
  \[ \phi'(s)=\psi_0'(s)\int_0^s
    t^{\frac{d}{2}-1}h(t)dt=s^{-\frac{d}{2}}\int_0^s t^{\frac{d}{2}-1}h(t)dt. \]
Consequently, by de l'H\^{o}pital's rule,
  \[ \lim_{s\to 0+}\phi'(s)=\lim_{s\to 0+}\frac{\int_0^s
      t^{\frac{d}{2}-1}h(t)dt}{s^{\frac{d}{2}}}
  =\lim_{s\to
    0+}\frac{s^{\frac{d}{2}-1}h(s)}{\frac{d}{2}s^{\frac{d}{2}-1}}=\tfrac{2}{d}h(0), \]
and we see that $\phi\in C^1([0,\infty))$.
Furthermore,
\[ \phi''(s)=s^{-1}h(s)-\tfrac{d}{2}s^{-\frac{d}{2}-1}\int_0^s
  t^{\frac{d}{2}-1}h(t)dt=s^{-1}h(s)-\tfrac{d}{2}s^{-1}\phi'(s) \]
and thus,
\[ \lim_{s\to 0+}\left [s\phi''(s)\right ]=\lim_{s\to0+}\left
    [h(s)-\tfrac{d}{2}\phi'(s)\right ]= 0. \]
By definition, $f(x)=\phi(|x|^2)$ and thus,
\[ \partial_j\partial_k
  f(x)=4x_jx_k\phi''(|x|^2)+2\phi'(|x|^2)\delta_{jk}. \]
This implies that
\[ \lim_{x\to 0}\partial_j\partial_k f(x)=2\phi'(0)\delta_{jk}, \]
since
\[ |x_jx_k\phi''(|x|^2)|\leq |x|^2|\phi''(|x|^2)|\to 0 \]
as $|x|\to 0$. Consequently, $f\in C^2(\R^d)$.
\end{proof}

    \section{Construction of a fundamental system}

    \begin{lemma}
      \label{lem:phi0phiinf}
      Let $\lambda>0$ and $V\in
      C^\infty([0,\infty))$. Furthermore, suppose that for any $k\in
      \N_0$ there exists a $C_k>0$ such that
      $|V^{(k)}(r)|\leq C_k\langle r\rangle^{-2-k}$ for all $r\geq 0$.
           Then the equation
           \begin{equation}
             \label{eq:radLaplambda}
             \phi''(r)-\frac{(d-1)(d-3)}{4r^2}\phi(r)-V(r)\phi(r)-\lambda^2
             \phi(r)=0
             \end{equation}
      has fundamental systems $\{\phi_0, \psi_0\}$ on $(0,\frac12]$ and
      $\{\phi_\infty,\psi_\infty\}$ on $[\frac14,\infty)$,
      respectively, of
      the form
      \begin{align*}
        \phi_0(r)&=r^{\frac{d-1}{2}}[1+a_0(r)],
        & \psi_0(r)&=
\begin{cases}
  r^\frac12 \log r[1+b_0(r)], & d=2, \\
  -\frac{1}{d-2}r^{-\frac{d-3}{2}}[1+b_0(r)], & d\not= 2, 
\end{cases}  \\
        \phi_\infty(r)&=e^{-\lambda r}[1+a_\infty(r)],
        &
\psi_\infty(r)&=
\tfrac{1}{2\lambda}e^{\lambda r}[1+b_\infty(r)].
      \end{align*}
For any $k\in \N_0$ there exists a $C_k>0$ such that\footnote{The
  bounds on the error functions $a_0$, $b_0$ are not optimal but
  simple to work with and sufficient for our purposes.}
\begin{align*}
  |a_0^{(k)}(r)|
  &\leq C_k r^{1-k},
  & \mbox{for all }r&\in (0,\tfrac12], \\
  |b_0^{(k)}(r)|&\leq C_k
                  \begin{cases}
                    |\log r|^{-1}r^{-k}, & d=2, \\
                    r^{1-k}, & d\not= 2,
                  \end{cases}
& \mbox{for all }r&\in (0,\tfrac12], \\
  |a_\infty^{(k)}(r)|+|b_\infty^{(k)}(r)|
  &\leq
    C_k r^{-1-k},
 & \mbox{for all }
r&\geq \tfrac14.                     
\end{align*}
\end{lemma}

\begin{proof}
  We start with the construction of the solution $\phi_0$. 
  Note that the equation
  \[ f''(r)-\frac{(d-1)(d-3)}{4r^2}f(r)=0 \]
  has the fundamental system $\{f_0,g_0\}$, given by
  \[ f_0(r)=r^{\frac{d-1}{2}},\qquad 
    g_0(r)=\begin{cases}r^\frac12\log r, & \mbox{if }d=2, \\
      -\frac{1}{d-2}r^{-\frac{d-3}{2}}, & \mbox{if }d\not= 2,
    \end{cases} \]
  and $W(f_0,g_0)=1$.
  Thus, in view of the variation of constants formula, $\phi_0$ is
  supposed to solve the equation
  \[ \phi_0(r)=f_0(r)-f_0(r)\int_0^r
    g_0(s)[V(s)+\lambda^2]\phi_0(s)ds+g_0(r)\int_0^r
    f_0(s)[V(s)+\lambda^2]\phi_0(s)ds. \]
  We rewrite this equation in terms of the auxiliary function $h$,
  defined by $\phi_0=f_0 h$. This yields the Volterra equation
  \begin{equation}
    \label{eq:Volterra0}
    h(r)=1+\int_0^r K(r,s)h(s)ds,
    \end{equation}
  with the kernel
  \[ K(r,s):=\left [\frac{g_0(r)}{f_0(r)}f_0(s)^2-f_0(s)g_0(s)\right
    ][V(s)+\lambda^2]. \]
  If $d\geq 3$ we have the bound
  \[ |K(r,s)|\lesssim r^{-d+2}s^{d-1}+s\lesssim s \]
  for all $0\leq s\leq r\leq \frac12$.  If $d=2$, we fix an arbitrary
  $\delta\in (0,1)$ and estimate
  \[ |K(r,s)|\lesssim |\log r| s + s
    |\log
    s|\lesssim s^{1-\delta} \]
  for all $0\leq s\leq r\leq \frac12$. Consequently,
  \[ \int_0^\frac12 \sup_{r\in [s,\frac12]}|K(r,s)|ds\lesssim 1, \]
  and the standard existence result for Volterra equations (see
  e.g.~\cite{SchSofSta10a}, Lemma 2.4) yields the existence of a
  solution $h\in L^\infty(0,\frac12)$ to Eq.~\eqref{eq:Volterra0},
  satisfying the bound
  \[ |h(r)-1|\lesssim \int_0^r |K(r,s)||h(s)|ds\lesssim
    \|h\|_{L^\infty(0,\frac12)}\int_0^r s^{1-\delta}ds\lesssim r^{2-\delta} \]
  for all $r\in [0,\frac12]$.
This proves the existence of $\phi_0(r)=f_0(r) h(r)=f_0(r)[1+a_0(r)]$,
with the bound $|a_0(r)|\lesssim r^{2-\delta}\lesssim r$ for all $r\in [0,\frac12]$.

Next, we turn to the derivative bounds on $a_0$. For any $j\in \N_0$ we have the bound
$|\partial_r^j K(r,s)|\lesssim r^{-j}s^{1-\delta}\lesssim r^{-j}$ for all $0<r\leq s\leq \frac12$.  If
we set $\kappa_j(r):=\partial_r^j K(r,s)|_{s=r}$, then, for $j,k\in \N_0$, we have
$|\kappa_j^{(k)}(r)|\lesssim r^{-j-k}$ for all $r\in (0,\frac12]$.
In terms of $a_0$, Eq.~\eqref{eq:Volterra0} reads
\[ a_0(r)=\int_0^r K(r,s)ds+\int_0^r K(r,s)a_0(s)ds. \]
Thus, 
for $k\in \N$, we obtain
\[ a_0^{(k)}(r)=\kappa_0^{(k-1)}(r)+\sum_{j=0}^{k-1}(\kappa_j
  a_0)^{(k-1-j)}(r)+\int_0^r \partial_r^k K(r,s)a_0(s)ds. \]
Inductively, we find
\[ |a_0^{(k)}(r)|\lesssim r^{1-k}+r^{2-k}\lesssim r^{1-k} \]
for all $r\in (0,\frac12]$, which is the desired bound.

The singular solution $\psi_0$ is constructed via the reduction
formula. Since $\phi_0(r)=r^{\frac{d-1}{2}}[1+O(r)]$, there exists an
$r_0\in (0,\frac12]$ such that $\phi_0(r)>0$ for all $r\in
(0,r_0]$. Consequently,
\[ \psi_0(r):=-\phi_0(r)\int_r^{r_0}\phi_0(s)^{-2}ds \]
is well-defined for all $r\in (0,r_0]$ and provides a solution to
Eq.~\eqref{eq:radLaplambda} on $(0,r_0]$.
We define the function $b_0$ on $(0,r_0]$ by
$\psi_0(r)=g_0(r)[1+b_0(r)]$, i.e.,
\[
  b_0(r)
  :=-\frac{f_0(r)[1+a_0(r)]}{g_0(r)}\int_r^{r_0}f_0(s)^{-2}[1+a_0(s)]^{-2}ds-1.
\]
Observe that
\[ \left (\frac{g_0}{f_0}\right
  )'=\frac{f_0g_0'-f_0'g_0}{f_0^2}=\frac{W(f_0,g_0)}{f_0^2}=f_0^{-2} \]
and thus,
\[
  -\frac{f_0(r)}{g_0(r)}\int_r^{r_0}f_0(s)^{-2}ds=1-c_0\frac{f_0(r)}{g_0(r)},\qquad
  c_0:=\frac{g_0(r_0)}{f_0(r_0)}.
  \]
  Consequently,
  \[  b_0(r)=a_0(r)-c_0\frac{f_0(r)}{g_0(r)}[1+a_0(r)]
    -\frac{f_0(r)}{g_0(r)}[1+a_0(r)]\int_r^{r_0} f_0(s)^{-2}\left
      [(1+a_0(s))^{-2}-1\right ]ds, \]
  and in the case $d\geq 3$ we obtain
  \[ |b_0(r)|\lesssim
    r+r^{d-2}+r^{d-2}\int_r^{r_0}s^{-d+1}s^{2-\delta}ds\lesssim
    r+r^{d-2}+r^{2-\delta}\lesssim r \]
  for all $r\in [0,r_0]$. In the case $d=2$ we have the weaker bound 
  \[ |b_0(r)|\lesssim r+|\log r|^{-1}+|\log r|^{-1}\int_r^{r_0}s^{-1}s^{2-\delta}ds\lesssim
    |\log r|^{-1} \]
  for all $r\in (0,r_0]$.
  The derivative bounds on $b_0$ follow directly by
  differentiating the explicit formula for $b_0$.
  By solving an initial value problem with data at $r=r_0$, we extend
  the solution $\psi_0$ to $(0,\frac12]$ and clearly, $\psi_0\in
  C^\infty((0,\frac12])$ since the coefficients of
  Eq.~\eqref{eq:radLaplambda} are smooth on $(0,\infty)$.

  The solution $\phi_\infty$ is constructed by a similar
  procedure. This time we treat the term $-\frac{(d-1)(d-3)}{4r^2}\phi(r)$
  perturbatively since it is negligible for large $r$. That is to say,
  we first note that the equation
  \[ f''(r)-\lambda^2 f(r)=0 \]
  has the fundamental system $\{f_\infty,g_\infty\}$, where
  $f_\infty(r)=e^{-\lambda r}$ and
  $g_\infty(r)=\frac{1}{2\lambda}e^{\lambda r}$.
  Consequently, we write $\phi_\infty=f_\infty h$ and consider the
  Volterra equation
  \begin{equation}
    \label{eq:Volterrainf}
    h(r)=1+\int_r^\infty K(r,s)h(s)ds
  \end{equation}
  with the kernel
  \[ K(r,s):=\left
      [f_\infty(s)g_\infty(s)-\frac{g_\infty(r)}{f_\infty(r)}f_\infty(s)^2\right
    ]\left [\frac{(d-1)(d-3)}{4s^2}+V(s)\right ]. \]
  We estimate
  \[ |K(r,s)|\lesssim (1+e^{2\lambda r}e^{-2\lambda s})s^{-2}\lesssim
    s^{-2} \]
  for all $\frac14\leq r\leq s$, which yields
  \[ \int_\frac14^\infty \sup_{r\in [\frac14,s]}|K(r,s)|ds\lesssim
    \int_{\frac14}^\infty s^{-2}ds\lesssim 
    1. \]
 The Volterra theorem (see e.g.~\cite{SchSofSta10a}, Lemma 2.4) then
  implies the existence of a solution $h\in
  L^\infty(\frac14,\infty)$. Furthermore,
  \[ |h(r)-1|\leq \int_r^\infty |K(r,s)||h(s)|ds\lesssim
    \|h\|_{L^\infty(\frac14,\infty)}\int_r^\infty s^{-2}ds\lesssim
    r^{-1} \]
  for all $r\geq \frac14$ and thus,
  $\phi_\infty(r)=f_\infty(r)[1+a_\infty(r)]$ with
  $|a_\infty(r)|\lesssim r^{-1}$ for all $r\geq \frac14$.

  For the bounds on the derivatives of $a_\infty$, we rewrite
the Volterra equation for $a_\infty=h-1$ as
\begin{align*}
  a_\infty(r)
  &=\int_r^\infty K(r,s)ds +\int_r^\infty
    K(r,s)a_\infty(s)ds \\
  &=\int_0^\infty K(r,s+r)ds+\int_0^\infty K(r,s+r)a_\infty(s+r)ds.
\end{align*}
Note that
\begin{align*}
  K(r,s+r)
  &=\frac{1}{2\lambda}\left ( 1-\frac{e^{\lambda
        r}}{e^{-\lambda r}}e^{-2\lambda(s+r)}\right )\left [
    \frac{(d-1)(d-3)}{4(s+r)^2}+V(s+r) \right ] \\
  &=\frac{1}{2\lambda}\left (1-e^{-2\lambda s}\right )\left [
    \frac{(d-1)(d-3)}{4(s+r)^2}+V(s+r) \right ]
\end{align*}
and thus, for $j\in \N_0$,
\[ |\partial_r^j K(r,s+r)|\lesssim (s+r)^{-2-j} \]
for all $\frac14\leq r\leq s$.
Now let $k\in \N$ and assume that for any $j\in \N_0$ with $j\leq
k-1$, we have
$|a_\infty^{(j)}(r)|\lesssim r^{-1-j}$ for all $r\geq \frac14$. Then
we obtain
\begin{align*}
  a_\infty^{(k)}(r)
  &=\int_0^\infty \partial_r^k
  K(r,s+r)ds+\int_0^\infty \partial_r^k\left
    [K(r,s+r)a_\infty(s+r)\right ]ds \\
  &=O(r^{-1-k})+\int_r^\infty K(r,s)a_\infty^{(k)}(s)ds
\end{align*}
and thus,
$a_k(r):=r^{1+k}a_\infty^{(k)}(r)$ satisfies the Volterra equation
\[ a_k(r)=O(r^0)+\int_r^\infty K(r,s)r^{1+k}s^{-1-k}a_k(s)ds. \]
Since
\[ \left | K(r,s)r^{1+k}s^{-1-k}\right |\lesssim s^{-2} \]
for all $\frac14\leq r\leq s$, a Volterra iteration yields $a_k\in
L^\infty(\frac14,\infty)$ and we obtain
\[ |a_\infty^{(k)}(r)|=|r^{-1-k}a_k(r)|\leq
  r^{-1-k}\|a_k\|_{L^\infty(\frac14,\infty)}\lesssim r^{-1-k} \]
for all $r\geq \frac14$. Consequently, the stated bounds on the
derivatives of
$a_\infty$ follow inductively.

Finally, for the growing solution $\psi_\infty$, we note that there
exists an $r_1\geq \frac14$ such that $\phi_\infty(r)>0$ for all
$r\geq r_1$ and set
\[ \psi_\infty(r):=\phi_\infty(r)\int_{r_1}^r
  \phi_\infty(s)^{-2}ds. \]
Then $\psi_\infty$ solves Eq.~\eqref{eq:radLaplambda} on
$[r_1,\infty)$ and the function $b_\infty$, defined by
$\psi_\infty=g_\infty(1+b_\infty)$, is given explicitly by
\[
  b_\infty(r)=\frac{f_\infty(r)[1+a_\infty(r)]}{g_\infty(r)}\int_{r_1}^r
  f_\infty(s)^{-2}[1+a_\infty(s)]^{-2}ds-1.
\]
As before,
\[ \frac{f_\infty(r)}{g_\infty(r)}\int_{r_1}^r
  f_\infty(s)^{-2}ds=1-c_1\frac{f_\infty(r)}{g_\infty(r)},\qquad
  c_1:=\frac{g_\infty(r_1)}{f_\infty(r_1)}, \]
and thus,
\[
  b_\infty(r)=a_\infty(r)-c_1\frac{f_\infty(r)}{g_\infty(r)}[1+a_\infty(r)]
+\frac{f_\infty(r)}{g_\infty(r)}[1+a_\infty(r)]\int_{r_1}^r
f_\infty(s)^{-2}[(1+a_\infty(s))^{-2}-1] ds. \]
This yields the bound
\[ |b_\infty(r)|\lesssim r^{-1}+e^{-2\lambda r}+e^{-2\lambda
    r}\int_{r_1}^r e^{2\lambda s}s^{-1}ds\lesssim r^{-1} \]
for all $r\geq \frac14$. The bounds on the derivatives of $b_\infty$
follow in a straightforward manner by differentiating the explicit
expression for $b_\infty$. By solving an initial value problem, the
solution $\psi_\infty$ smoothly extends to all of $[\frac14,\infty)$.
\end{proof}

\section{Numerics}
\label{sec:num}
\subsection{Numerical construction of the soliton profile}
\label{sec:ChebQ}
We would like to obtain a radial solution to
\[ \Delta_{\R^2}Q-Q+Q^3=0. \]
That is to say, we need to solve the radial equation
\begin{equation}
\label{eq:cubicf}
 f''(r)+\frac{1}{r}f'(r)-f(r)+f(r)^3=0 
\end{equation}
for $r\geq 0$. Asymptotically, the nonlinearity is negligible and
thus, we expect the behavior $f(r)\simeq 1$ as $r\to 0+$ and
$f(r)\simeq r^{-\frac12}e^{-r}$ as $r\to\infty$. We encode the
expected asymptotics in the definition of the new variable $g$, given by
\[ f(r)=:(1+r)^{-\frac12} e^{-r}g\left (\frac{r-1}{r+1}\right ). \]
In terms of $g$ and $x:=\frac{r-1}{r+1}$, Eq.~\eqref{eq:cubicf} reads
\begin{equation}
  \label{eq:cubicg}
 \mc R(g):=g''(x)+\frac{3x^2-6x-5}{(1-x)^2(1+x)}g'(x)-\frac{3(3-x)}{4(1-x)^2(1+x)}g(x)+\frac{2}{(1-x)^3}e^{-2\frac{1+x}{1-x}}g(x)^3=0
\end{equation}
for $x\in [-1,1)$.
We compactify the problem \eqref{eq:cubicg} by allowing $x\in [-1,1]$. Evidently, the
endpoints $x=\pm 1$ are singular and this yields the regularity
conditions
\begin{equation}
  \label{eq:regg}
    4g'(-1)-3g(-1)=16 g'(1)+3g(1)=16g''(1)-5g'(1)-3g(1)=0. 
\end{equation}
Note that these conditions are determined by the linear part of the equation
since the coefficient of $g(x)^3$ is not singular at $x=\pm 1$. We solve
Eq.~\eqref{eq:cubicg} by a Chebyshev pseudospectral method. To this
end, we use the basis functions $\phi_n: [-1,1]\to\R$, $n\in \N_0$,
\[ \phi_n(x):=T_n(x)+a_{0,n}+a_{1,n} x+a_{2,n} x^2, \]
where $T_n$ are the standard Chebyshev polynomials and $a_{j,n}$ are chosen in such a way that each $\phi_n$ satisfies the
regularity conditions Eq.~\eqref{eq:regg}. Note that this leads to $\phi_0=\phi_1=\phi_2=0$.
Then we numerically solve the root finding problem  				
\[ \mc R\left (\sum_{n=3}^{25} \beta_n \phi_n \right )(x_k)=0 \]
for $k=0,1,2,\dots,22$, and $x_k \in [-1,1]$ some collocation points.
The expansion coefficients $(\beta_n)_{n=3}^{25}$ are given in Table \ref{tab:Q}.

\begin{table}[ht]
\centering
\caption{Expansion coefficients for the approximate soliton profile} 
\begin{tabular}{|c|c|c|c|c|c|c|}
\hline
$n$ & $3$ & $4$ & $5$ & $6$ & $7$ & $8$\\ 
 $\beta_n$ & $-\frac{2542}{141001}$ & $\frac{8061}{72860}$ & $\frac{23}{25643}$ & $-\frac{17127}{731900}$ & $-\frac{113}{61446}$ & $\frac{407}{88530}$\\[2pt] \hline 
$n$ & $9$ & $10$ & $11$ & $12$ & $13$ & $14$\\ 
 $\beta_n$ & $\frac{80}{79969}$ & $-\frac{195}{296276}$ & $-\frac{167}{607101}$ & $\frac{3}{91531}$ & $\frac{3}{109289}$ & $\frac{1}{42237921}$\\[2pt] \hline  
$n$ & $15$ & $16$ & $17$ & $18$ & $19$ & $20$\\ 
 $\beta_n$ & $\frac{1}{163112}$ & $\frac{1}{171418}$ & $\frac{1}{1839428}$ & $-\frac{1}{412985}$ & $-\frac{1}{693490}$ & $-\frac{1}{3459389}$\\[2pt]  \hline 
$n$ & $21$ & $22$ & $23$ & $24$ & $25$ & \\ 
 $\beta_n$ & $\frac{1}{5641102}$ & $\frac{1}{2626342}$ &
                                                     $\frac{1}{45286837}$
                      & $\frac{1}{10226264}$ & $-\frac{1}{9836273}$ & \\[2pt]
  \hline
 \end{tabular}
 \label{tab:Q}
 \end{table}

\subsection{Numerical construction of $\widehat Q_1$}

The goal is to numerically construct the unique (radial) solution
$\widehat Q_1$ to the
equation
\[ \mc L_+\widehat Q_1=-2Q_{\R^2}-r^2Q_{\R^2}^3. \]
Recall that $S_0(x)=x^j\partial_j Q_{\R^2}(x)+Q_{\R^2}(x)$ satisfies
$\mc L_+S_0=-2Q_{\R^2}$.
Consequently, it suffices to solve $\mc L_+
S_1=-r^2Q_{\R^2}^3$ because then, $\widehat Q_1=S_0+S_1$.
In other words, we need to solve the radial equation
\begin{equation}
  \label{eq:L+f}
  f''(r)+\frac{1}{r}f'(r)-f(r)+3f_0(r)^2f(r)=r^2f_0(r)^3,
  \end{equation}
where $f_0(r)=Q_{\R^2}(re_1)$.
Again, we introduce the auxiliary variable $g$, defined by
\[ f(r)=(1+r)^{-\frac12}e^{-r}g\left (\frac{r-1}{r+1}\right ), \]
which transforms Eq.~\eqref{eq:L+f} into
\begin{equation}
  \label{eq:L+g}
  \begin{split}
  g''(x)&+\frac{3x^2-6x-5}{(1-x)^2(1+x)}g'(x)-\frac{3(3-x)}{4(1-x)^2(1+x)}g(x)
  +\frac{6}{(1-x)^3}e^{-2\frac{1+x}{1-x}}g_0(x)^2g(x) \\
  &=2\frac{(1+x)^2}{(1-x)^5}e^{-2\frac{1+x}{1-x}}g_0(x)^3,
\end{split}
\end{equation}
where $x=\frac{r-1}{r+1}$ and $g_0$ is given by
\[ f_0(r)=(1+r)^{-\frac12}e^{-r}g_0\left (\frac{r-1}{r+1}\right ). \]
We replace $g_0$ by the approximation obtained in Section
\ref{sec:ChebQ} and solve Eq.~\eqref{eq:L+g} by a Chebyshev pseudospectral
method with the basis functions $\phi_n$ from above. This yields an
approximate solution of the form $
\sum_{n=3}^{40} \gamma_n\phi_n$
with the coefficients $(\gamma_n)_{n=3}^{40}$ given in Table
\ref{tab:S1}.

With the numerical approximations to the functions $Q_{\R^2}$ and
$\widehat Q_1$ at hand, it is straightforward to compute (an
approximation to) the constant $b_1$
from Section \ref{sec:numstab}. By numerical integration we find
$\frac{b_1}{2\pi}\approx 7.39$.

 \begin{table}[ht]
\centering
\caption{Expansion coefficients for approximation to $S_1$} 
\begin{tabular}{|c|c|c|c|c|c|c|c|c|}
  \hline
  $n$ & $3$ & $4$ & $5$ & $6$ & $7$ & $8$ & $9$ & $10$\\ 
 $\gamma_n$ & $\frac{54973}{96387}$ & $-\frac{3088}{102021}$ & $-\frac{11563}{65730}$ & $-\frac{622}{123831}$ & $\frac{935}{19694}$ & $\frac{715}{80273}$ & $-\frac{972}{107461}$ & $-\frac{245}{66869}$\\[2pt] \hline 
$n$ & $11$ & $12$ & $13$ & $14$ & $15$ & $16$ & $17$ & $18$\\ 
 $\gamma_n$ & $\frac{43}{75440}$ & $\frac{6}{13097}$ & $\frac{7}{79466}$ & $\frac{23}{138473}$ & $\frac{10}{87071}$ & $-\frac{1}{41044}$ & $-\frac{7}{100544}$ & $-\frac{3}{79736}$\\[2pt] \hline  
$n$ & $19$ & $20$ & $21$ & $22$ & $23$ & $24$ & $25$ & $26$\\ 
 $\gamma_n$ & $-\frac{1}{247350}$ & $\frac{1}{98688}$ & $\frac{1}{104302}$ & $\frac{1}{181864}$ & $\frac{1}{1748151}$ & $-\frac{1}{795239}$ & $-\frac{1}{519650}$ & $-\frac{1}{1141942}$\\[2pt]  \hline 
$n$ & $27$ & $28$ & $29$ & $30$ & $31$ & $32$ & $33$ & $34$\\ 
 $\gamma_n$ & $-\frac{1}{2632970}$ & $\frac{1}{3481458}$ & $\frac{1}{4334802}$ & $\frac{1}{3856839}$ & $\frac{1}{14342913}$ & $-\frac{1}{142634956}$ & $-\frac{1}{42463795}$ & $-\frac{1}{12658667}$\\[2pt]  \hline 
$n$ & $35$ & $36$ & $37$ & $38$ & $39$ & $40$ & & \\ 
 $\gamma_n$ & $\frac{1}{45132528}$ & $-\frac{1}{14926347}$ &
                                                        $\frac{1}{15529718}$ & $-\frac{1}{15419336}$ & $\frac{1}{13135736}$ & $-\frac{1}{36714512}$ &
 & \\[2pt]
  \hline
 \end{tabular}
 \label{tab:S1}
\end{table}

    \end{appendix}
  
\bibliographystyle{plain} \bibliography{nlshyp}

\end{document}